\documentclass[a4paper,11pt]{article}
\usepackage{graphicx}
\usepackage{subfig}
\usepackage{jcappub}
\usepackage[T1]{fontenc} 
\usepackage[normalem]{ulem}

\pdfoutput=1

\newcommand\fverb{\setbox\fverbbox=\hbox\bgroup\verb}
\newcommand\fverbdo{\egroup\medskip\noindent%
			\fbox{\unhbox\fverbbox}\ }
\newcommand\fverbit{\egroup\item[\fbox{\unhbox\fverbbox}]}
\newbox\fverbbox

\newcommand{\be}{\begin{eqnarray}}
\newcommand{\ee}{\end{eqnarray}}

\def\smica{{\tt SMICA}}
\def\nilc{{\tt NILC}}
\def\sevem{{\tt SEVEM}}
\def\commander{\texttt{Commander}}
\def\planck{{\it Planck}}
\def\cromaster{\texttt{cRomaster}}

\title{Planck 2018 constraints on anisotropic birefringence and its cross-correlation with CMB anisotropy}

\author[a,b,1]{A. Gruppuso,\note{Corresponding author.}}
\author[c,d,a,e]{D. Molinari,}
\author[d,a,e]{P. Natoli}
\author[d,e]{and L. Pagano}

\affiliation[a]{Istituto Nazionale di Astrofisica - Osservatorio di Astrofisica e Scienza dello Spazio di Bologna, via Gobetti 101, I-40129 Bologna, Italy}
\affiliation[b]{Istituto Nazionale di Fisica Nucleare, Sezione di Bologna,\\ viale Berti Pichat 6/2, I-40127 Bologna, Italy}
\affiliation[c]{CINECA Interuniversity Consortium,\\ Via Magnanelli 6/3, I-40033 Casalecchio di Reno, Bologna, Italy}
\affiliation[d]{Dipartimento di Fisica e Scienze della Terra, Universit\`a degli Studi di Ferrara,\\ via Saragat 1, I-44122 Ferrara, Italy}
\affiliation[e]{Istituto Nazionale di Fisica Nucleare, Sezione di Ferrara,\\ via Saragat 1, I-44122 Ferrara, Italy}

\emailAdd{alessandro.gruppuso@inaf.it}
\emailAdd{diego.molinari@inaf.it}
\emailAdd{paolo.natoli@unife.it}
\emailAdd{luca.pagano@unife.it}	

\abstract{Parity-violating extensions of standard electromagnetism produce cosmic birefringence, the in vacuo rotation of the linear polarisation direction of a photon during propagation.
We employ {\it Planck} 2018 CMB polarised data to constrain anisotropic birefringence, modeled by its angular power spectrum $C_{\ell}^{\alpha \alpha}$,
and the cross-correlation with CMB temperature maps, $C_{\ell}^{\alpha T}$, at scales larger than $\sim$15 degrees.
We present joint limits on the scale invariant quantity, $A^{\alpha \alpha} \equiv \ell (\ell +1) \, C_{\ell}^{\alpha \alpha} / 2 \pi$, and on the analogous amplitude for the cross-correlation, $A^{\alpha T} \equiv \ell (\ell +1) \, C_{\ell}^{\alpha T} / 2 \pi$.
We find no evidence of birefringence within the 
error budget 
and obtain $A^{\alpha \alpha}  <  0.104 \, \mbox{[deg$^2$]}$ and $A^{\alpha T}=1.50^{+2.41}_{-4.10} \, \mbox{[$\mu$K$\cdot$deg] both at } 95 \% \mbox{ C.L.}$.
The latter bound appears competitive
in constraining  a few early dark energy models recently proposed to alleviate the $H_{0}$ tension. 
Slicing the joint likelihood at $A^{\alpha T}=0$, the bound on $A^{\alpha \alpha}$ becomes tighter at $A^{\alpha \alpha}  <  0.085 \, \mbox{[deg$^2$]}$ at 95$\% \mbox{ C.L.}$.
In addition we recast the constraints on $A^{\alpha \alpha}$ as a bound on the amplitude of primordial magnetic fields responsible for Faraday rotation, finding $B_{\mbox{\tiny 1Mpc}} < 26.9$ nG 
and $B_{\mbox{\tiny 1Mpc}} < 24.3$ nG at 95$\%$ C.L. for the marginalised and sliced case respectively.}
\keywords{CMB, parity symmetry, data analysis}

\begin{document} 
\maketitle
\flushbottom

\section{Introduction}
\label{intro}

A feature of Maxwell's electromagnetism is that it is invariant under the point parity transformation. As a direct consequence, the linear polarisation direction of a photon does not change during propagation {\it in vacuo}. However, parity-violating extensions of the standard electromagnetism have been proposed and exhibit the so-called cosmic birefringence effect, i.e.~the in vacuo rotation of photon polarisation direction during propagation \cite{Carroll:1989vb}. This effect is naturally parameterised by an angle $\alpha$, dubbed the birefringence angle.

The cosmic microwave background (CMB) radiation is linearly polarised at the level of $\sim 1-10\%$ due to Thomson scattering and thus represents a good candidate to perform such 
investigations\footnote{For a more general framework aimed at studying extensions of standard electromagnetism through polarised CMB observations see \cite{Lembo:2020ufn}.}. 
Furthermore, of all photons we receive from the Universe, those from the CMB have travelled the longest journey, a fact that raises, for a given frequency, the chances to detect a rotation. In detail, the effect, which mixes the Q and U Stokes parameters, produces non-null cross correlations between temperature and B-mode polarisation, and between E- and B-mode polarisation, that would be otherwise null in standard electromagnetism \cite{Lue:1998mq,Feng:2004mq,Gluscevic:2010vv}.
If the birefringence effect is in action, these cross-correlations can show up with different amplitudes depending on the direction of observation, $\hat n$, of the CMB anisotropies. In this case the phenomenon is called {\it anisotropic birefringence} effect, otherwise, in case of constant amplitude over the sky, it is denoted as {\it isotropic birefringence}. 
For the latter case, one single angle, $\alpha$, is enough to describe the phenomenon whereas for anisotropic birefringence, in principle, one needs an angle for each available direction, $\alpha(\hat n)$.
Morevoer, irrespectively of the isotropic or anisotropic origin, the mixing of the Q and U Stokes parameters creates spurious B-mode polarisation which has to be taken under control when looking for primordial gravitational waves through CMB polarisation \cite{Mei:2014iaa,Liu:2016dcg,Zhai:2019god}.

Note that, since isotropic and anisotropic birefringence can be disentangled, for example through an expansion of $\alpha(\hat n)$ in spherical harmonics, it is possible to have parity violating effects in action beyond standard electromagnetism even if TB and EB are null {\it on average} over the sky. For instance one can have half of the sky providing a given EB correlation and the other half exactly the opposite of that value: globally the correlation would be null, but not locally \cite{Gubitosi:2011ue}.
This is an example of the case of the {\it purely} anisotropic birefringence effect: one needs to go beyond the two-point correlation function of CMB anisotropies to be sensitive to it \cite{Gluscevic:2012me}.

Current constraints on the isotropic birefringence effect are compatible with no rotation at the level of $\sim 0.5$ deg, see \cite{Molinari:2016xsy,Aghanim:2016fhp,Pogosian:2019jbt}. 
Note that for {\it Planck} the error budget turns out to be dominated by systematics: an uncertainty in the knowledge of the instrumental polarisation direction, turns out to be much larger then the statistical error which is $\sim 0.05$ deg \cite{Aghanim:2016fhp} at 1$\sigma$. 
Future constraints, which critically depend on the ability to keep under control all the systematic effects\footnote{We wish to mention here that a new idea has been set forth recently in \cite{Minami:2019ruj,Minami:2020xfg,Minami:2020fin}, where it is shown how future experiments, as the LiteBIRD satellite \cite{Ishino:2016izb,Sekimoto}, taking into account both foreground and CMB emissions, will be able to jointly constrain isotropic birefringence and a deviation of the direction of instrumental polarisation, in principle without any a priori knowledge of the latter.},
provide improvements of some orders of magnitude depending on the experimental setup considered \cite{Molinari:2016xsy,Pogosian:2019jbt}.

Present constraints on anisotropic birefringence, provided as amplitude of the scale-invariant spectrum of $\alpha(\hat n)$, are compatible with no detection with an uncertainty of $\sim$0.1 deg$^2$ at 95$\%$ C.L., as obtained by {\it Planck} 2015 \cite{Contreras:2017sgi} and Bicep-Keck data \cite{Array:2017rlf} or a fraction between $1/2$ and $1/3$ of that value as recently estimated by the ACTPol 
\cite{Namikawa:2020ffr} and the SPTpol  \cite{Bianchini:2020osu} collaborations.
Other compatible, even though weaker, constraints on this parameter are provided by Polarbear \cite{Ade:2015cao} and WMAP \cite{Gluscevic:2012me} observations.
Again, future CMB observations are expected to improve such a bound by orders of magnitude \cite{Pogosian:2019jbt}.
It is interesting to note that the anisotropic birefringence effect is not affected by the uncertainty of the global instrumental polarisation angle which in fact impacts only on the isotropic birefringence \cite{Gluscevic:2012me,Keating:2012ge}.

In this paper we use polarised {\it Planck} 2018 data to constrain isotropic and anisotropic birefringence on angular scales larger than $\sim$15 degrees. 
Note that this range is complementary to those investigated by the most constraining experiments ACTPol (or SPTpol), whose angular resolution allows to probe the anisotropic effect from 9 deg (or 3.6 deg) to $\sim$5 arcmin. 
For \planck\ 2018 data there has been no critical improvement about the knowledge of the instrumental polarisation angle with respect to the 2015 data, so this systematic effect still dominates the error budget of isotropic birefringence.
Hence, our main focus in this paper is in fact the analysis of anisotropic birefringence and its isotropic part will be evaluated only to check consistency with previous {\it Planck} constraints.
In addition, still employing {\it Planck} 2018 data, we evaluate, for the first time to our knowledge, the cross-correlation between anisotropic birefringence and CMB temperature anisotropies map at the largest angular 
scales\footnote{\label{sptfootnote}The SPTpol work \cite{Bianchini:2020osu} appeared on the arXiv when the present paper was about to be finalised. To our knowledge, the SPTpol and the present are the only articles which provide estimates of the cross-correlation between anisotropic birefringence and CMB temperature map. Interestingly, the two analyses consider multipole ranges which do not overlap.}. 
This is also relevant for early dark energy models which have been recently proposed to alleviate the tension on the Hubble constant $H_0$ \cite{Capparelli:2019rtn}.

In order to estimate the anisotropic birefringence and its cross-correlation with the CMB temperature map, we adopt a brute-force approach which is shortly described here and in more detail in Section \ref{Description} below:
\begin{itemize}
\item From the {\it Planck} 2018 CMB solution T, Q and U maps at high resolution and corresponding realistic simulations,
we build maps of the birefringence angle $\alpha(\hat n)$ at low resolution (allowing analysis at angular range larger than $\sim$15 deg).
\begin{itemize}
	\item This is done by evaluating the CMB angular power spectra (henceforth APS) at high resolution in each of the regions defined by the low resolution pixels.
        \item In each of these regions we have applied suitable estimators to extract the birefringence angle. This is done assuming constant the birefringence angle in that particular region.
\end{itemize}
\item We estimate the APS of the birefringence angle maps,
 i.e. the $C_{\ell}^{\alpha \alpha}$ from $\ell=0$ to $\ell=12$. Note that the monopole term provides constraints on the isotropic birefringence angle. This monopole term is then compared to what found estimating $\alpha$ directly from the whole available sky. 
\item We estimate the APS of the cross-correlation between birefringence angle maps and CMB temperature maps,
 i.e. the $C_{\ell}^{{\alpha} T}$ from $\ell=2$ to $\ell=12$. 
\end{itemize}
The paper is organised as follows: in Section \ref{dataset} we describe the data set and the simulations employed; the validation of the  data-analysis pipeline which provides the maps of $\alpha(\hat n)$ is given in Section \ref{validation}; our null tests analyses are presented in Section \ref{analysis}; constraints on APS and amplitude of scale-invariant $\alpha$-anisotropies and its cross-correlation with the CMB temperature map are given in Section \ref{results} while final considerations are drawn in Section \ref{conclusions}. 

\section{Data set and simulations}
\label{dataset}

We make use of the CMB-cleaned maps provided by the four component separation algorithms employed in \planck, namely  \commander, \nilc, \smica\ and \sevem~\cite{Akrami:2018mcd}. 
These maps, available from the Planck Legacy Archive\footnote{https://www.cosmos.esa.int/web/planck/pla} (PLA), are provided at \texttt{HEALPix}\footnote{http://healpix.sourceforge.net} \cite{Gorski:2004by} resolution $N_{side}=2048$, with a Gaussian beam with $\mathrm{FWHM}=5^\prime$. In addition, we employ Monte Carlo (MC) simulations,
also publicly available from the PLA, which are an updated version of the full focal plane simulations described in \cite{Ade:2015via}, referred as FFP10, see e.g.~\cite{Akrami:2019bkn}.
The FFP10 set contains the most realistic simulations the \planck\ collaboration provides to characterise its 2018 data. They consist of 1000 CMB maps extracted from the current $\Lambda$CDM best-fit model which are beam smeared and contain residuals of beam leakage \cite{Ade:2015via}. These maps are complemented by 300 instrumental noise simulations for each frequency channel which also include residual systematic effects 
as beam leakage again, ADC non linearities, thermal fluctuations (dubbed 4K fluctuations), band-pass mismatch and others \cite{Aghanim:2018fcm}. The latter are processed through the component separation algorithms assuming the same weights employed for the data. 
Following the same procedure adopted in \cite{Akrami:2019bkn}, we have combined the CMB simulations with permutations of the noise 
maps processed through component separation algorithms so to effectively have a MC set of 999 signal and noise 
simulations\footnote{We have removed one of the 1000 CMB realisations because it turned out corrupted.} which do contain also residuals of known systematic effects.

More precisely, we use the half-mission (HM) and the odd-even (OE) version for both data and FFP10 simulations, see again \cite{Akrami:2019bkn} for further details.
These are two different splits of the
\planck\ dataset that are employed to estimate the six CMB spectra in cross-mode: such approach is preferable since it reduces residuals of systematic effects and noise mismatches \cite{Akrami:2018mcd}. 
Moreover, the comparison of the results obtained with the two different data splits, allows one to assess the robustness of our findings.
Following what done in the \planck\ likelihood analysis, we base our main results on the HM data split, which perform better in terms of null tests \cite{Aghanim:2019ame}, and consider the OE data split only for consistency checks.
Hence, we employ the HM splits for all  component separation methods while we consider the OE splits only for \commander\ and \smica.
There is a caveat regarding the use of HM and OE splits: as described above, their CMB signal component also contains residuals of known instrumental, beam related, non-idealities. However, these are computed 
assuming the full mission (FM) set and not the HM or OE splits. In other words, the PLA does not provide a version of such systematic component tailored to either the HM or OE splits.
A potentially relevant consequence is that the cross-mode APS estimation cannot be expected to alleviate these systematic residuals in the simulations as, instead, one would expect to happen for real data. This issue might lead to over-estimation of residuals in the simulations with respect to data. We carefully take into account the latter problem in our pipeline and show that its impact, when measurable, is weak compared to statistical uncertainlty. 
For example, as it will clear below,  it only affects the first few even multipoles in the APS of the birefringence anisotropies.

\section{From CMB polarised maps to $\alpha$-anisotropy maps}
\label{validation}

\subsection{Description and validation of the analysis pipeline}\label{Description}

\begin{figure}[t]
\centering
\includegraphics[width=.45\textwidth]{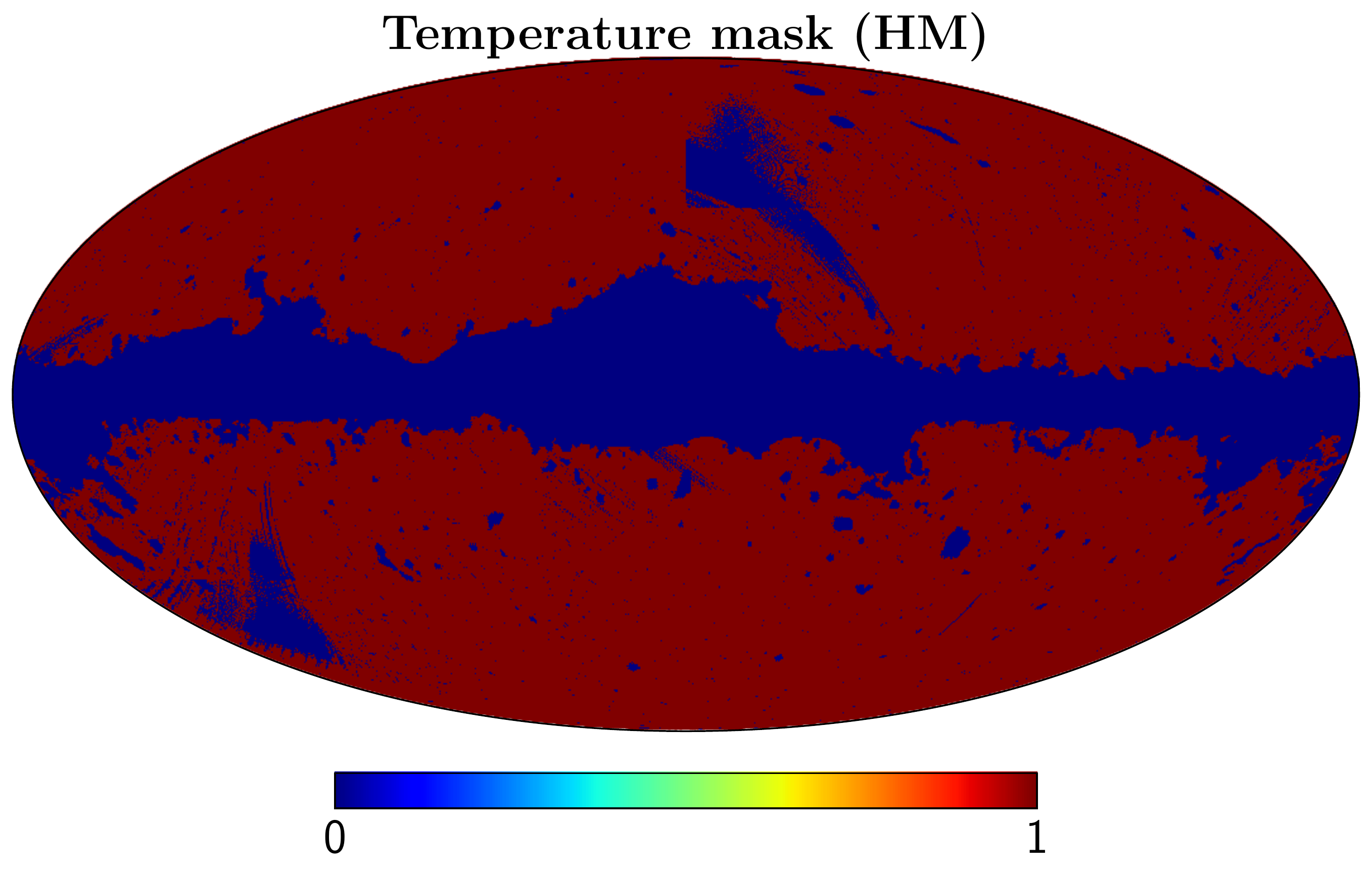}
\includegraphics[width=.45\textwidth]{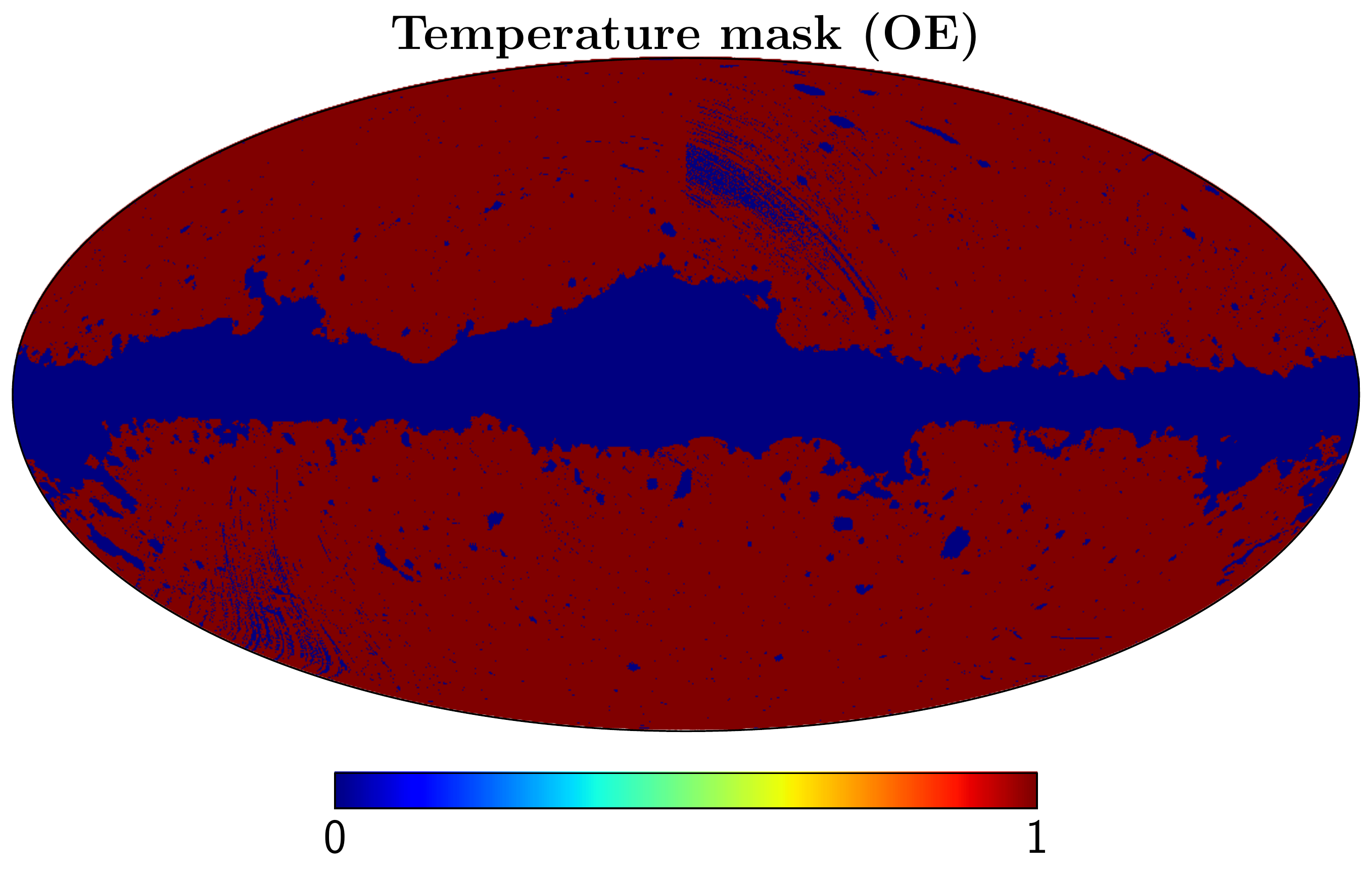}
\includegraphics[width=.45\textwidth]{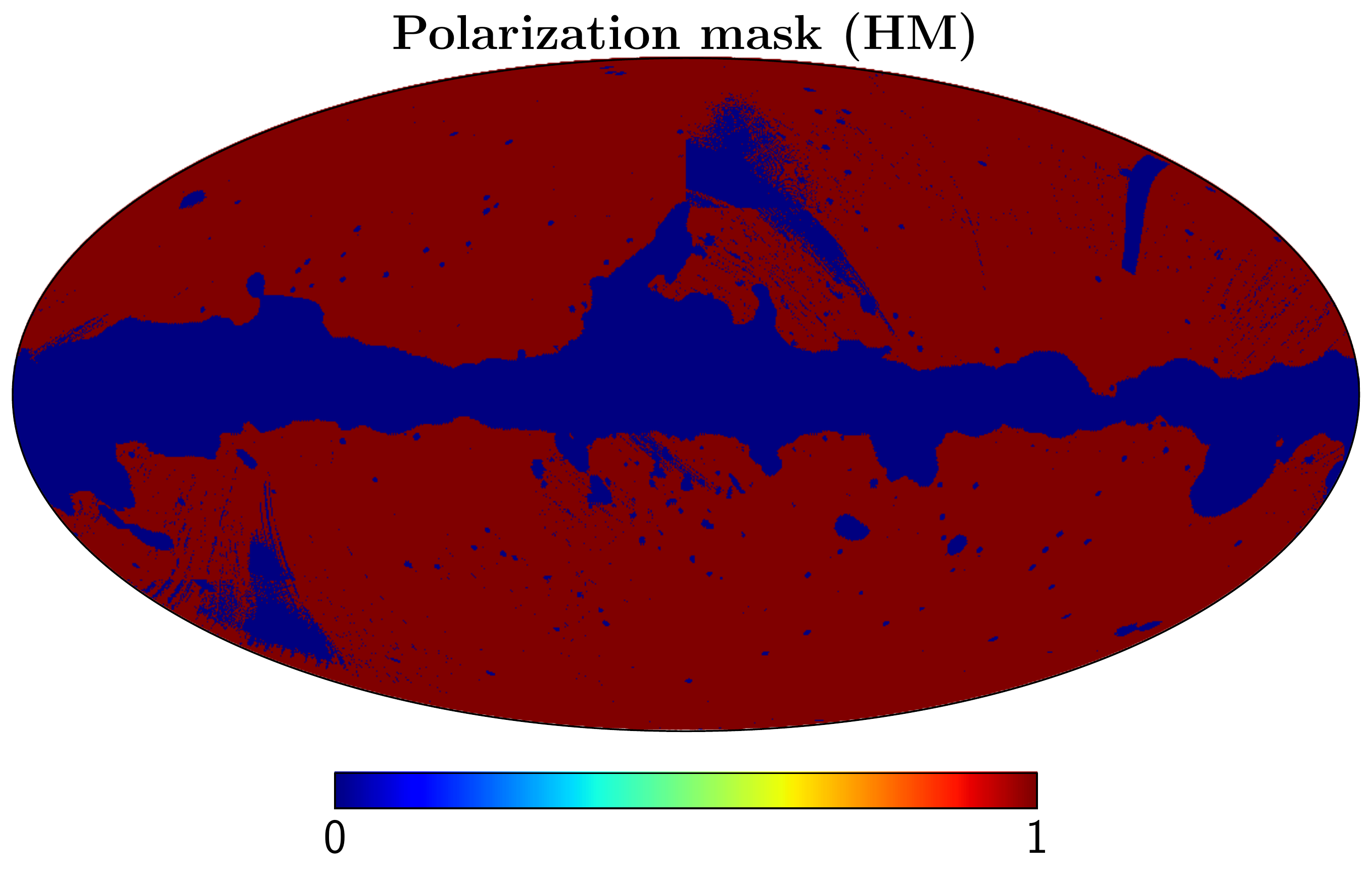}
\includegraphics[width=.45\textwidth]{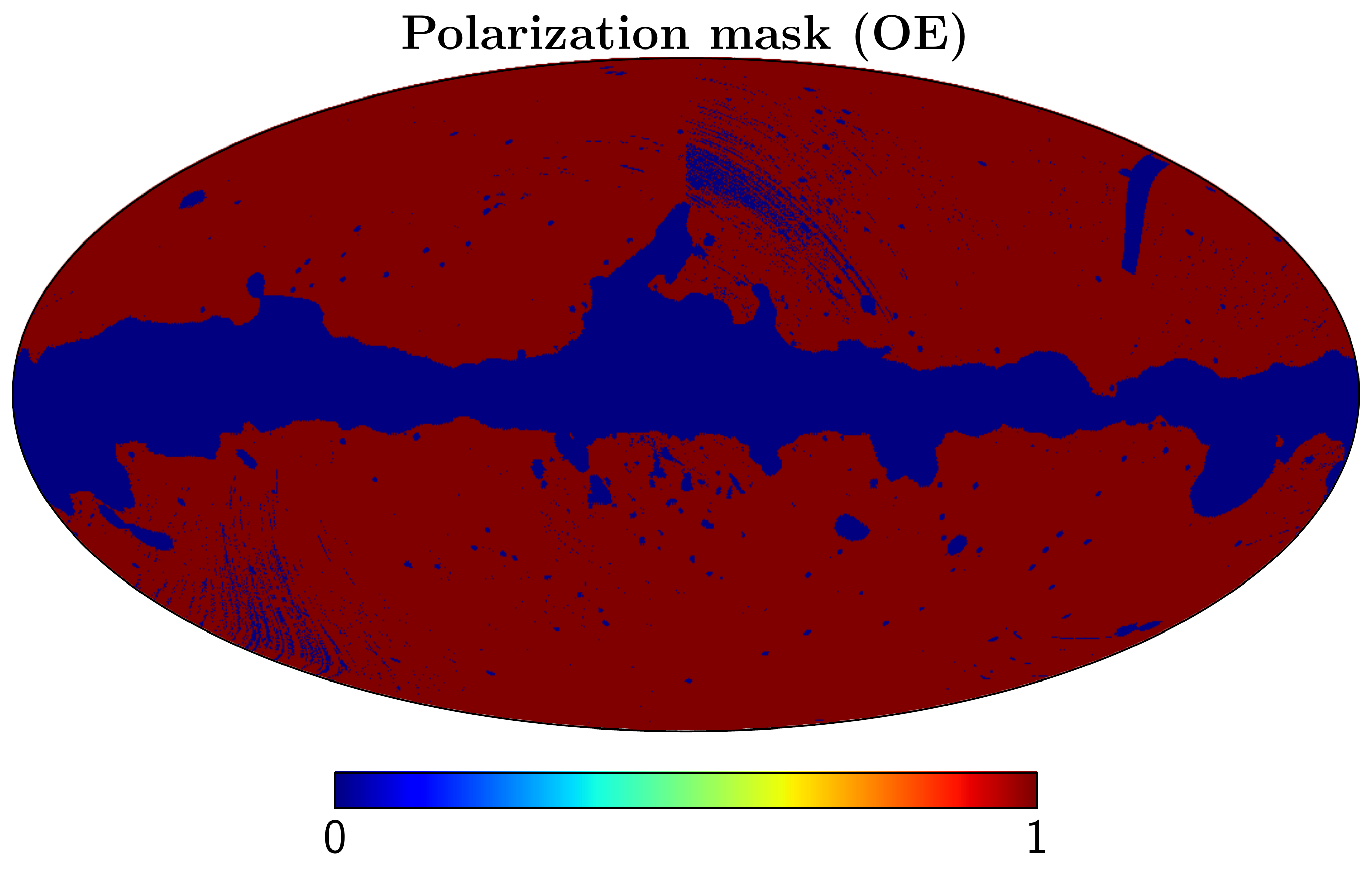}
\caption{Temperature (top row) and polarisation (bottom row) \planck\ common masks used to extract the cross-spectra from HM maps (left column) or OE maps (right column). The differences between the two is due to the masking of missing pixels differently present in the two splits.} \label{fig:common_masks}
\end{figure}

We divide the T, Q and U maps at high resolution, $N_{side}=2048$, in regions corresponding to the pixels at $N_{side}=4$ in the \texttt{HEALPix} pixelation scheme\footnote{The relation between the number of pixels, $N_{pix}$, and $N_{side}$ is given by $N_{pix} = 12 \, N_{side}^2$ \cite{Gorski:2004by}.}.
Within each of these large-scale regions, simply called ``large pixels'' in the 
following\footnote{In other words with ``large pixel'' we mean the high-resolution region defined by the contours of the low resolution pixel.}, we compute the six CMB anisotropy spectra in cross-mode both for the data and the FFP10 simulations processed by the \planck\ component separation algorithms.
To this extent we employ the \cromaster\ code \cite{Polenta:2004qs}, an implementation of the pseudo-C$_{\ell}$ estimator \cite{Hivon:2001jp}.
The masks used for each of these APS estimations, are given by the product of the temperature or polarisation \planck\ common mask \cite{Akrami:2018mcd}, both shown in Figure \ref{fig:common_masks}, 
with the masks which leave as observed each of the large pixels.
Note that the common masks for the HM and OE data splits are (slightly) different because of pixels that have not been observed by some detectors at a given frequency in at least one of the splits or because, even if observed, their polarisation angle coverage is too poor to be decomposed into Q and U polarisation.
These masks are used in the APS estimations without any apodization.

\begin{figure}[t]
\centering
\includegraphics[width=.45\textwidth]{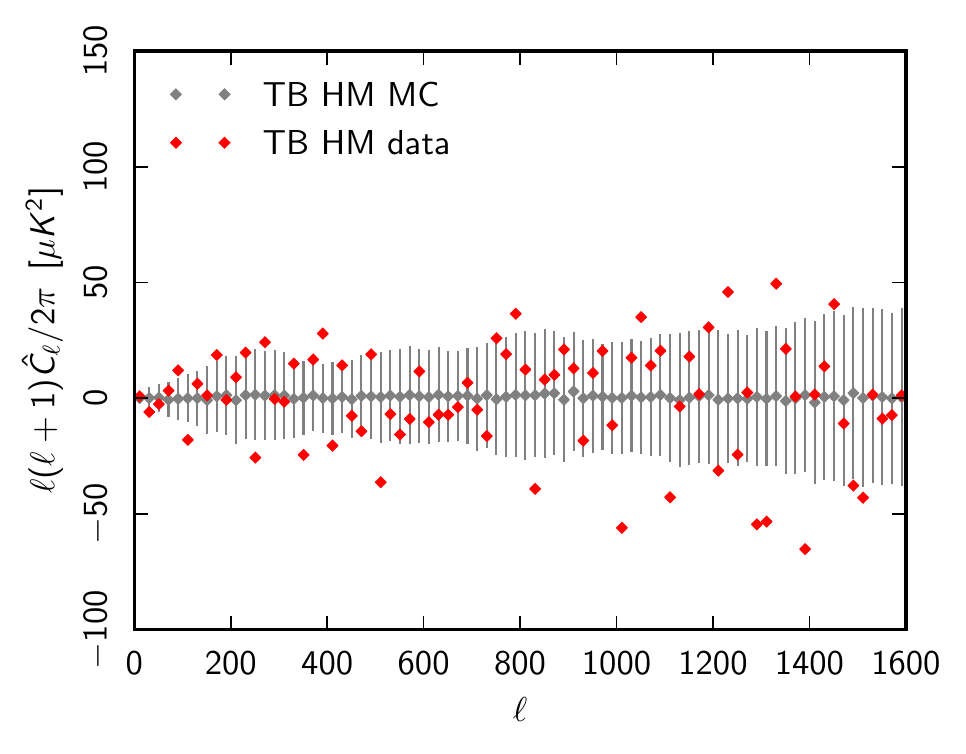}
\includegraphics[width=.45\textwidth]{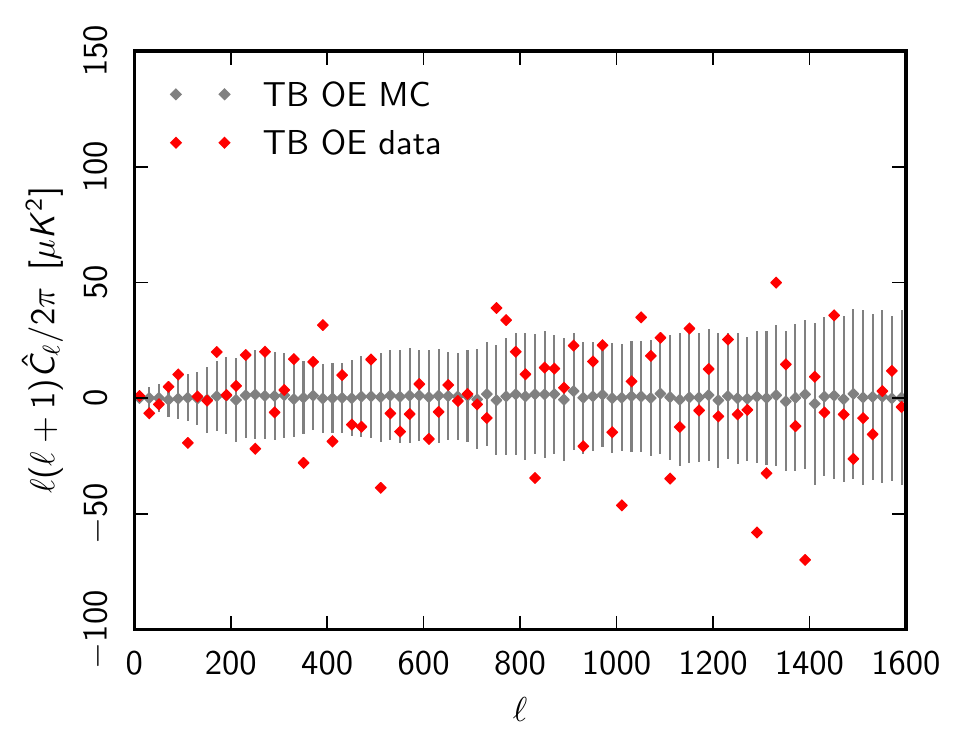}
\includegraphics[width=.45\textwidth]{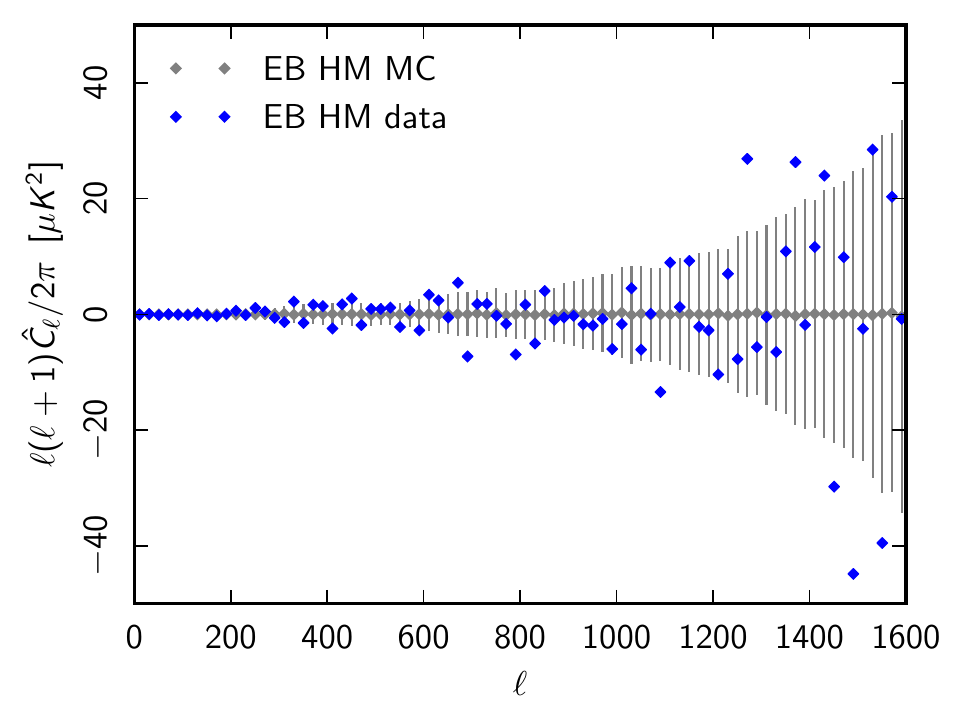}
\includegraphics[width=.45\textwidth]{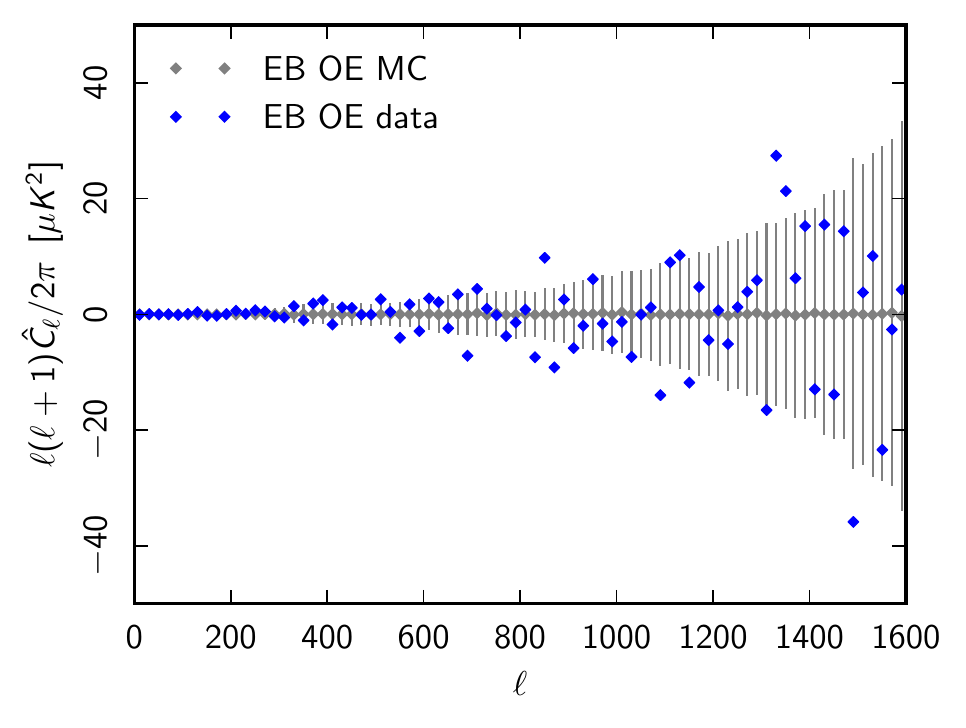}
\caption{TB (upper panels) and EB (lower panels) spectra from HM (left panels) or OE (right panels) data for the pixel number 16 (in ring order) at resolution of $N_{side}=4$.}\label{fig:TBeEBspectra}
\end{figure}

As an example of the extracted APS, in Figure \ref{fig:TBeEBspectra} we show the TB (upper panels) and EB (lower panels) CMB spectra relative to a large pixel at high Galactic latitude, precisely pixel \#16 in the ``ring'' format of the numbering scheme of \texttt{HEALPix} at $N_{side}=4$, for \smica\ HM (left panels) and \smica\ OE data (right panels) compared with the average and the 1$\sigma$ dispersion of the corresponding FFP10 simulations.
The six APS are extracted for each (T,Q,U)-map in the harmonic range  binned with $\Delta_{\ell}=20$ up to a maximum multipole of $\ell = 1980$.
The binning option helps increasing the computational speed of the APS estimation and, at the same time, reduces the off-diagonal correlation of the $C_{\ell}$ induced by the presence of the masks.
The CMB spectra are in fact used up to $\ell_{max}=1511$ to be consistent with tests performed by the \planck\ collaboration on the FFP10 simulations \cite{Akrami:2018mcd}.
In any case, as shown in \cite{Molinari:2016xsy}, the signal-to-noise ratio of the birefringence angle is almost saturated at that multipole and therefore going beyond $\ell_{max} \sim$1500 does not add any crucial information.

In each of the large pixels we suppose constant the birefringence angle, $\alpha$. 
That means that in each of these pixels we can model the impact of the birefringence effect on the CMB spectra as follows \cite{Lue:1998mq,Feng:2006dp}:
\be
\hat C^{TT}_{\ell} &=& C^{TT}_{\ell} \, , \\
\hat C^{EE}_{\ell} &=& C^{EE}_{\ell} \, \cos^2 \left( 2 \alpha \right) + C^{BB}_{\ell} \, \sin^2 \left( 2 \alpha \right) \, , \\
\hat C^{BB}_{\ell} &=& C^{BB}_{\ell} \, \cos^2 \left( 2 \alpha \right) + C^{EE}_{\ell} \, \sin^2 \left( 2 \alpha \right) \, , \\
\hat C^{TE}_{\ell} &=& C^{TE}_{\ell} \, \cos \left( 2 \alpha \right) \, , \\
\hat C^{TB}_{\ell} &=& C^{TB}_{\ell} \, \sin \left( 2 \alpha \right) \, , \\
\hat C^{EB}_{\ell} &=& \frac{1}{2} \left( C^{EE}_{\ell} - C^{BB}_{\ell} \right) \, \sin \left( 4 \alpha \right) \, ,
\ee
where $\hat C_{\ell}^{X}$ and $C_{\ell}^{X}$ with $X=TB, TE, EB, EE, BB$ are the observed and the primordial CMB spectra respectively.
In order to estimate the birefringence angle $\alpha$ from the observed APS, we build the so called D-estimators \cite{Wu:2008qb,Zhao:2015mqa,Gruppuso:2016nhj}, defined as
\begin{eqnarray}
D_{\ell}^{TB}(\alpha) &=& \hat{C_{\ell}}^{TB} \cos(2 \alpha) - \hat{C_{\ell}}^{TE} \sin(2 \alpha) \, ,
\label{estimators1}\\
D_{\ell}^{EB}(\alpha) &=& \hat{C_{\ell}}^{EB} \cos(4 \alpha) - \frac{1}{2}(\hat{C_{\ell}}^{EE}-\hat{C_{\ell}}^{BB}) \sin(4 \alpha) \, .
\label{estimators2}
\end{eqnarray}
It is possible to show, see e.g.\ \cite{Gruppuso:2016nhj}, that the birefringence angle can be evaluated finding $\alpha $ which makes null the expectation value of $D_{\ell}^{Y}$, where $Y=EB, TB$.
This is typically performed through a standard $\chi^2$ minimisation:
\begin{equation}\label{chisq}
\chi_{Y}^2(\alpha) = \sum_{\ell\ell'} D_{\ell}^{Y}M_{\ell\ell'}^{Y Y^{-1}}D_{\ell'}^{Y} \, ,
\end{equation}
where the matrix $M_{\ell\ell'}^{YY}=\langle D_{\ell}^{Y}D_{\ell'}^{Y}\rangle$ is the covariance matrix of the $D_{\ell}^{Y}$ estimator. 
Note that the minimisation of the $\chi^2$ can also be performed jointly for $D_{\ell}^{TB}(\alpha)$ and $D_{\ell}^{EB}(\alpha)$, see \cite{Gruppuso:2016nhj}.
However, since the latter has more constraining power than the former, a joint analysis does not reduce significantly the uncertainty in $\alpha$.
Therefore, we decide to estimate $\alpha$ only through the minimisation of $\chi_{EB}^2$ and consider the companion procedure, given by the minimisation of $\chi_{TB}^2$, 
simply as a consistency test.

\begin{figure}[t]
\centering
\includegraphics[width=.45\textwidth]{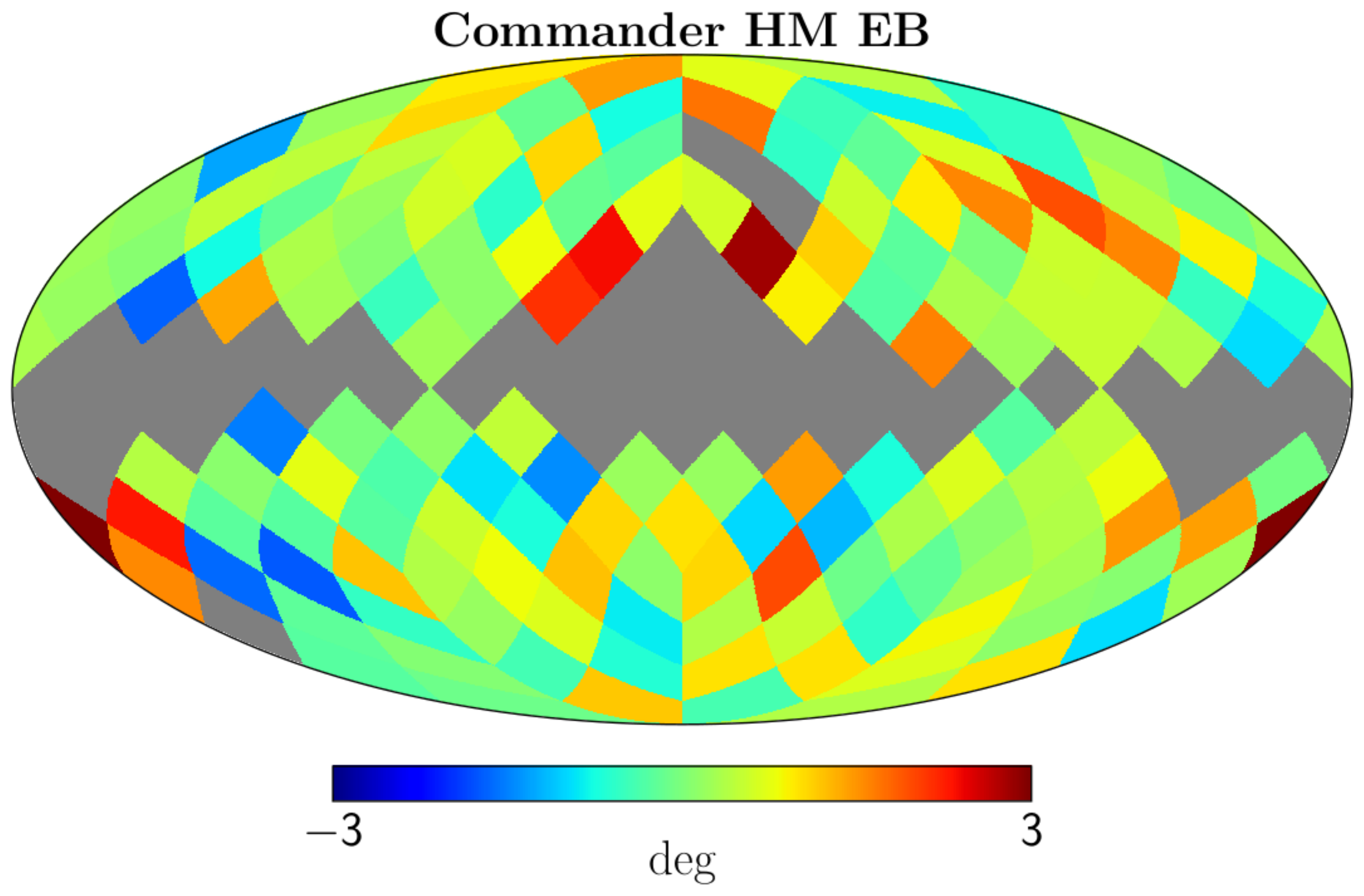}
\includegraphics[width=.45\textwidth]{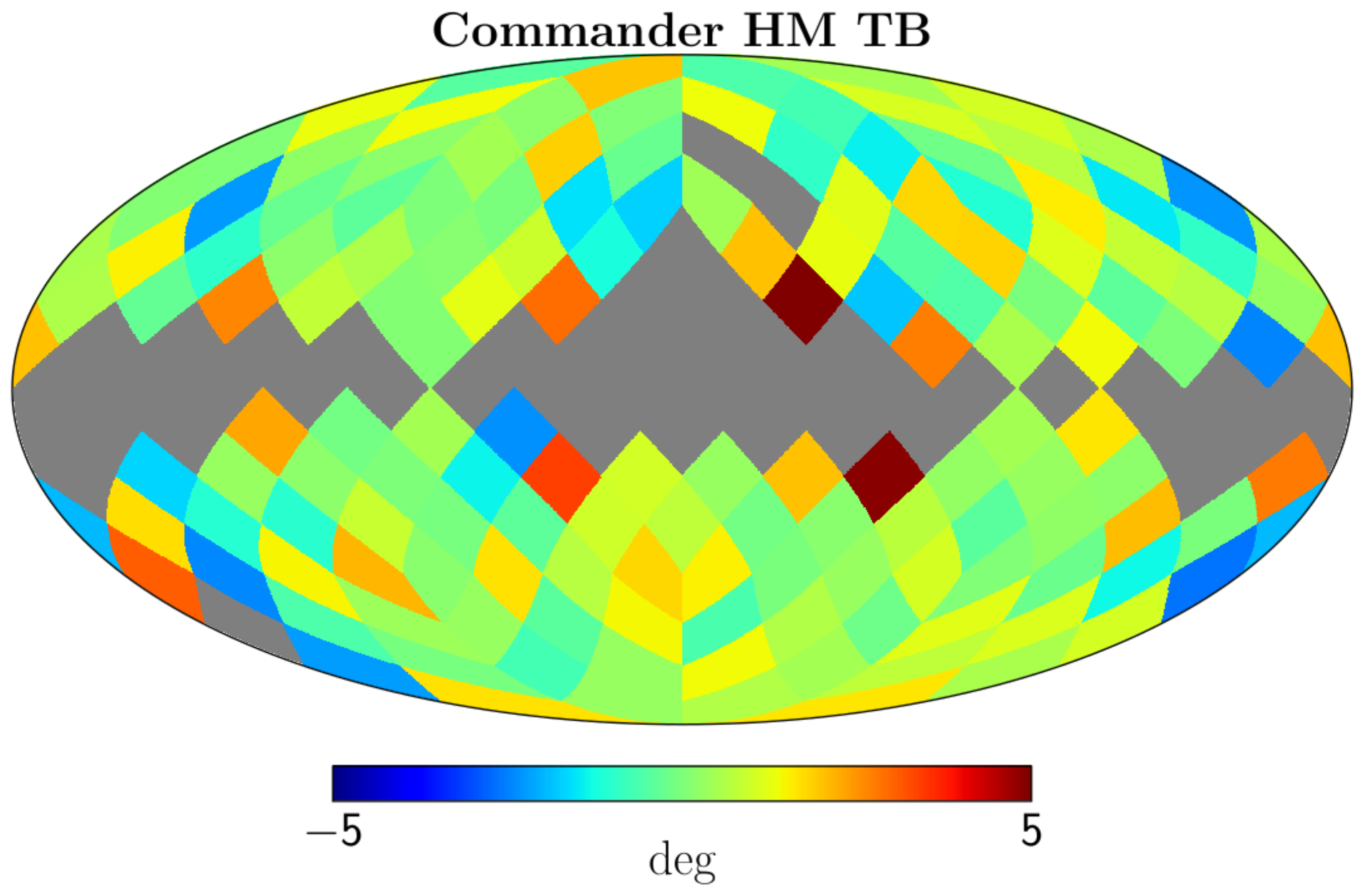}
\includegraphics[width=.45\textwidth]{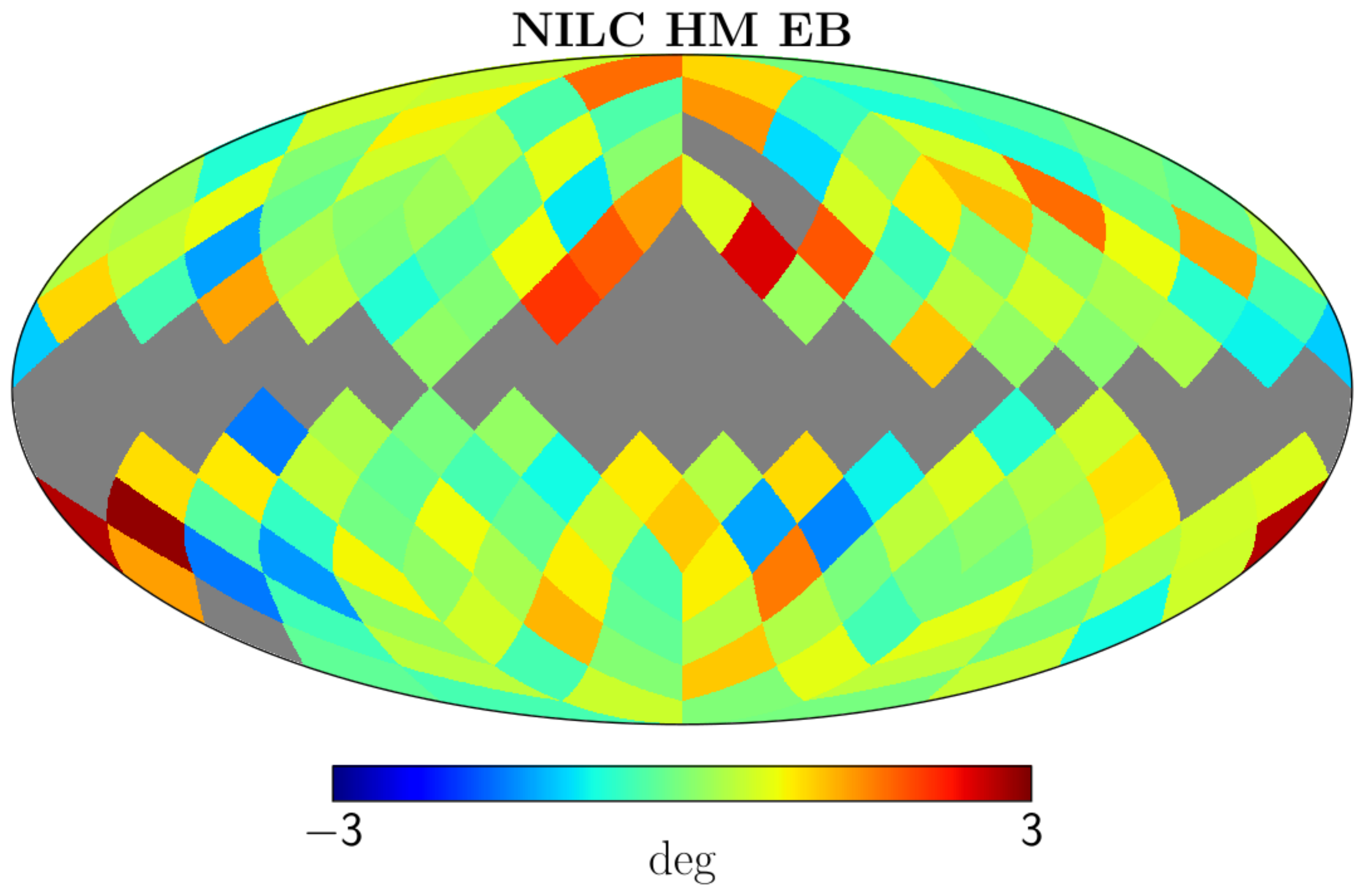}
\includegraphics[width=.45\textwidth]{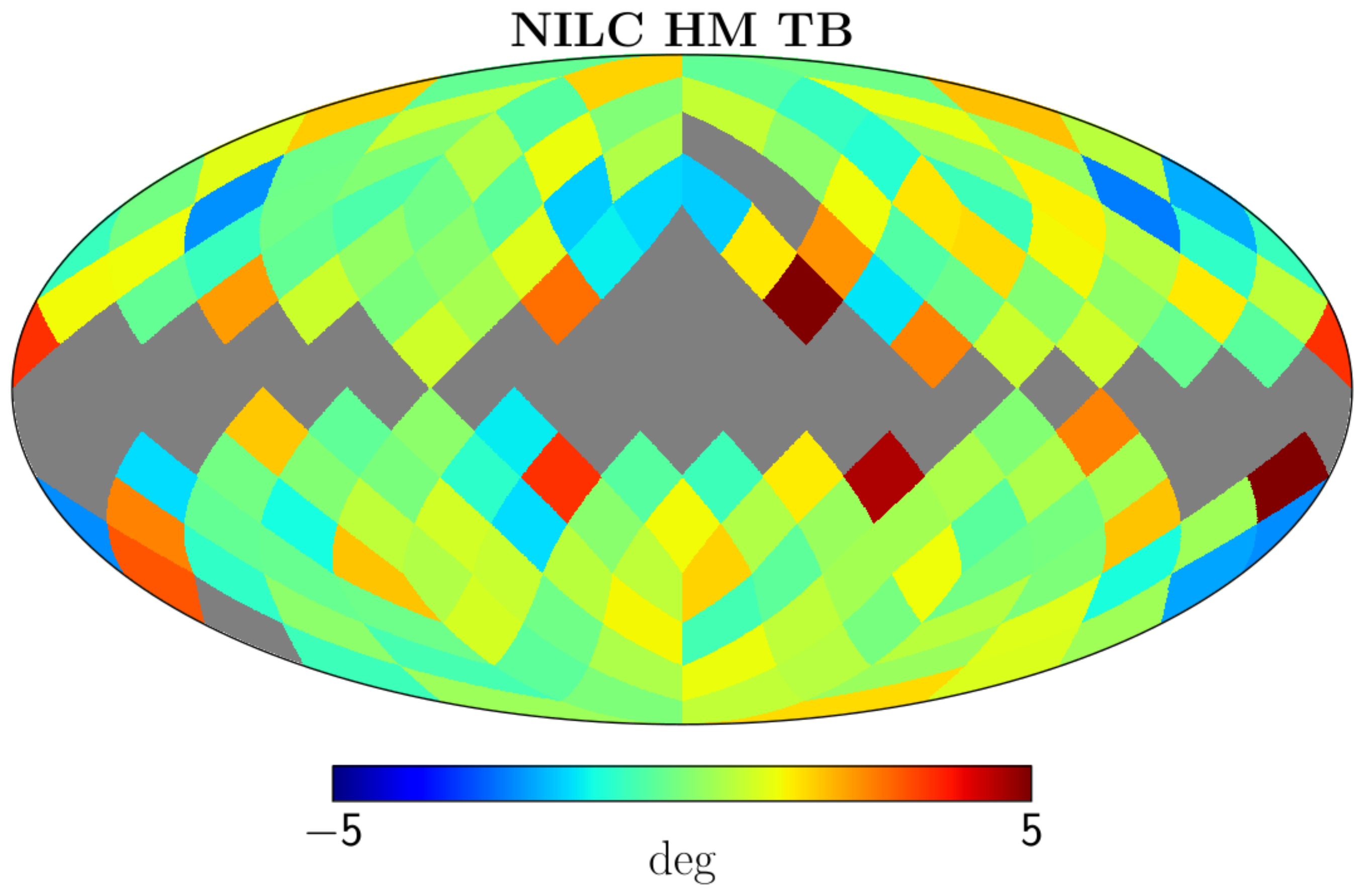}
\includegraphics[width=.45\textwidth]{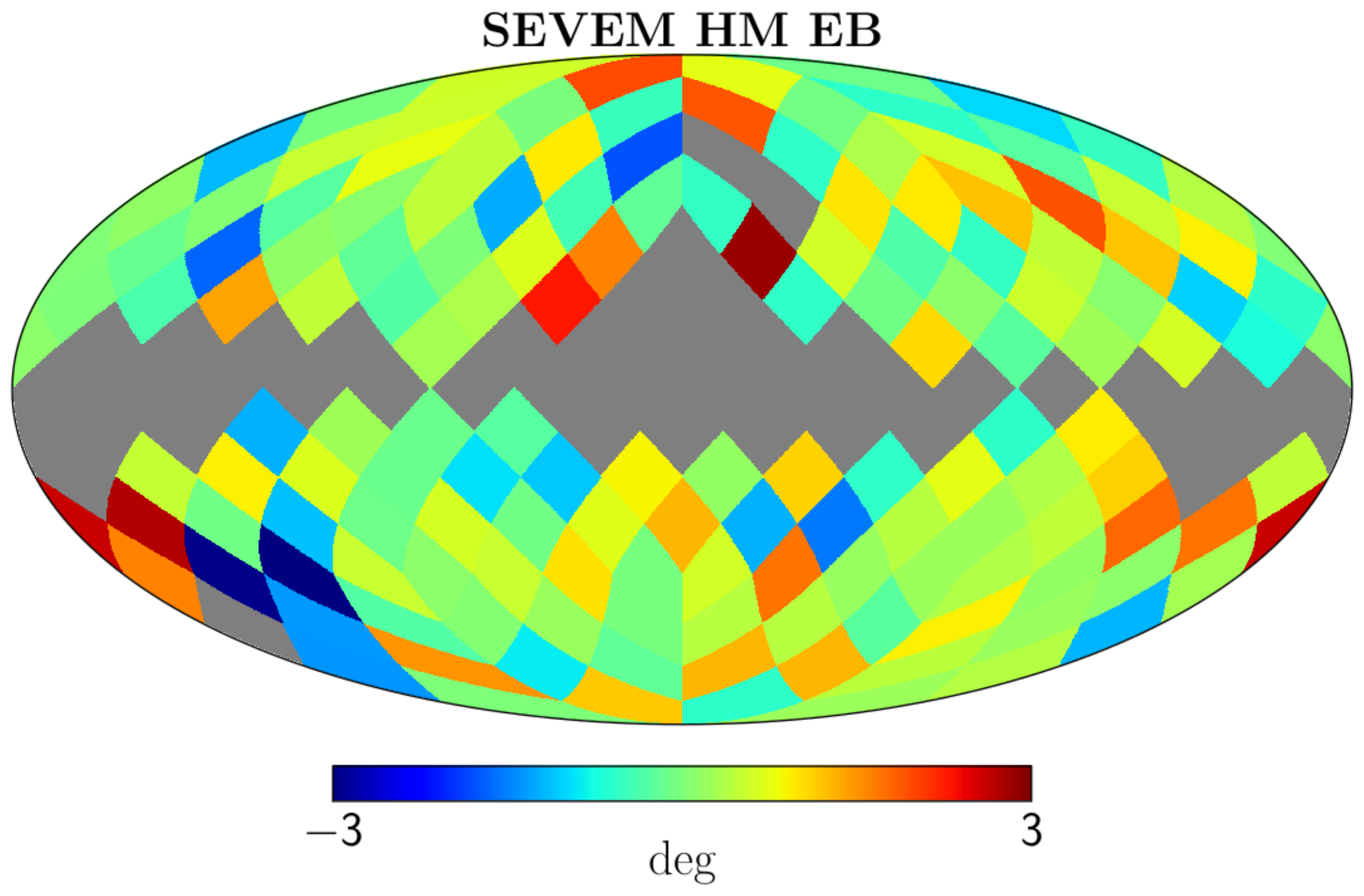}
\includegraphics[width=.45\textwidth]{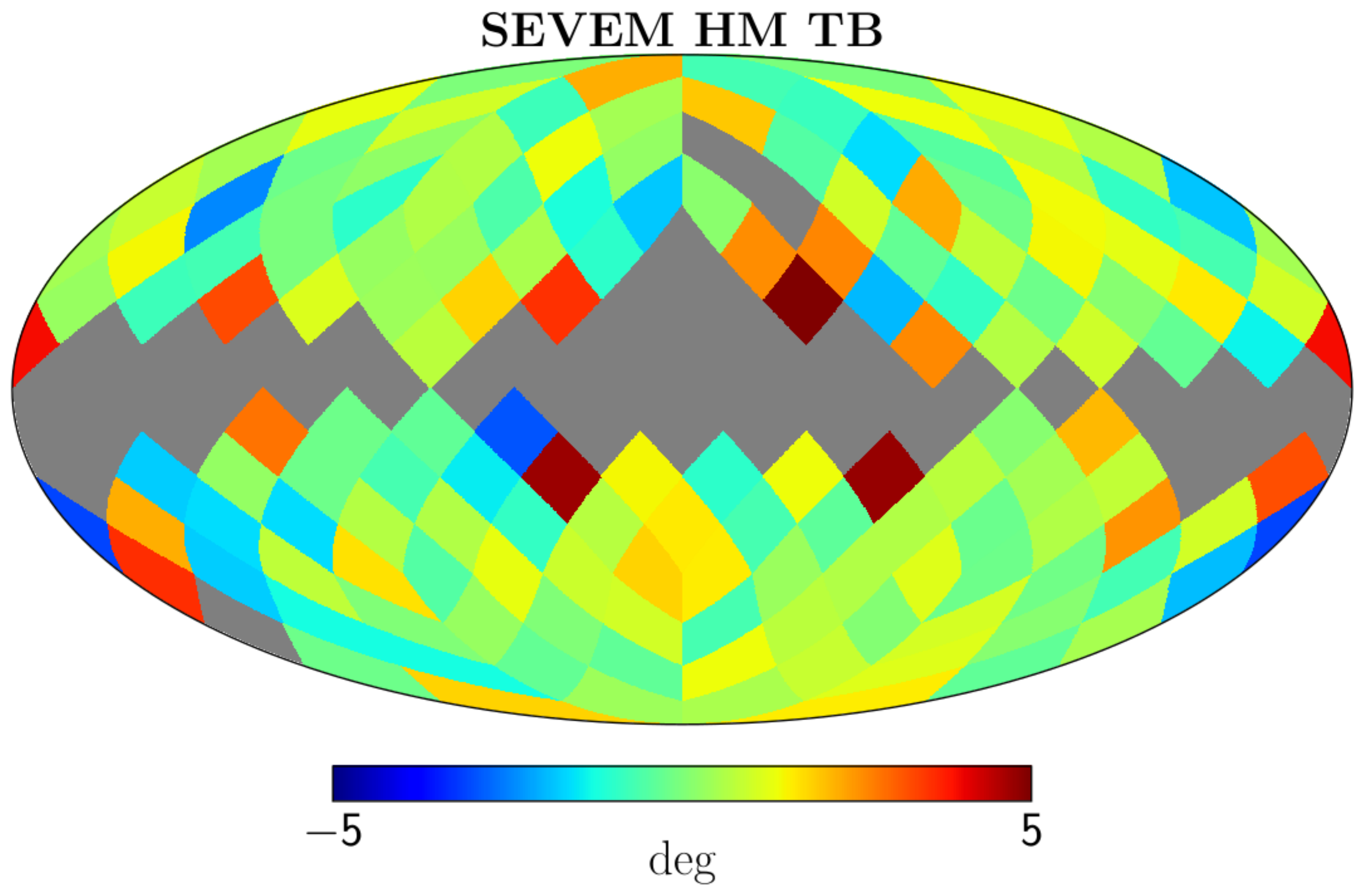}
\includegraphics[width=.45\textwidth]{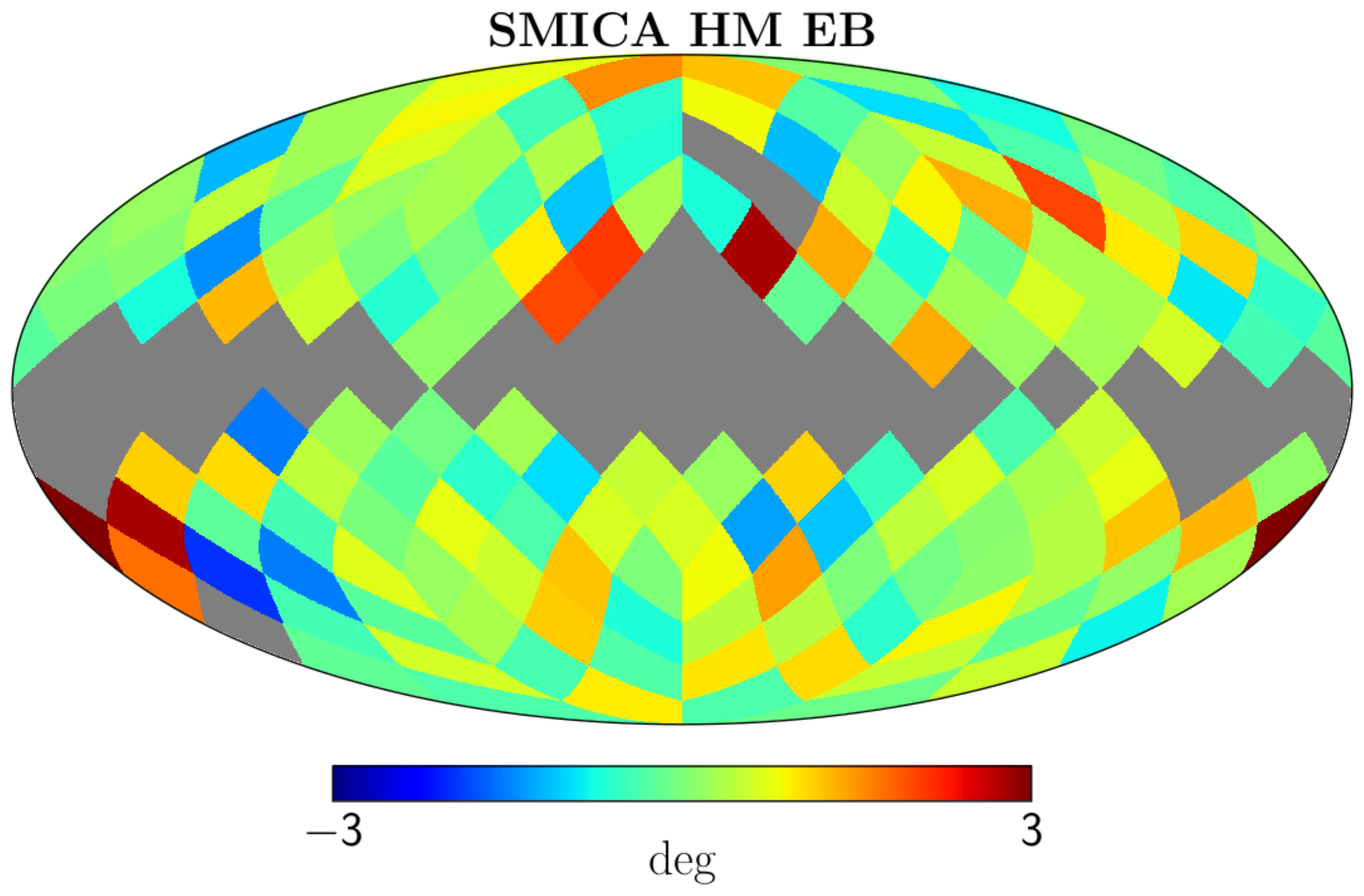}
\includegraphics[width=.45\textwidth]{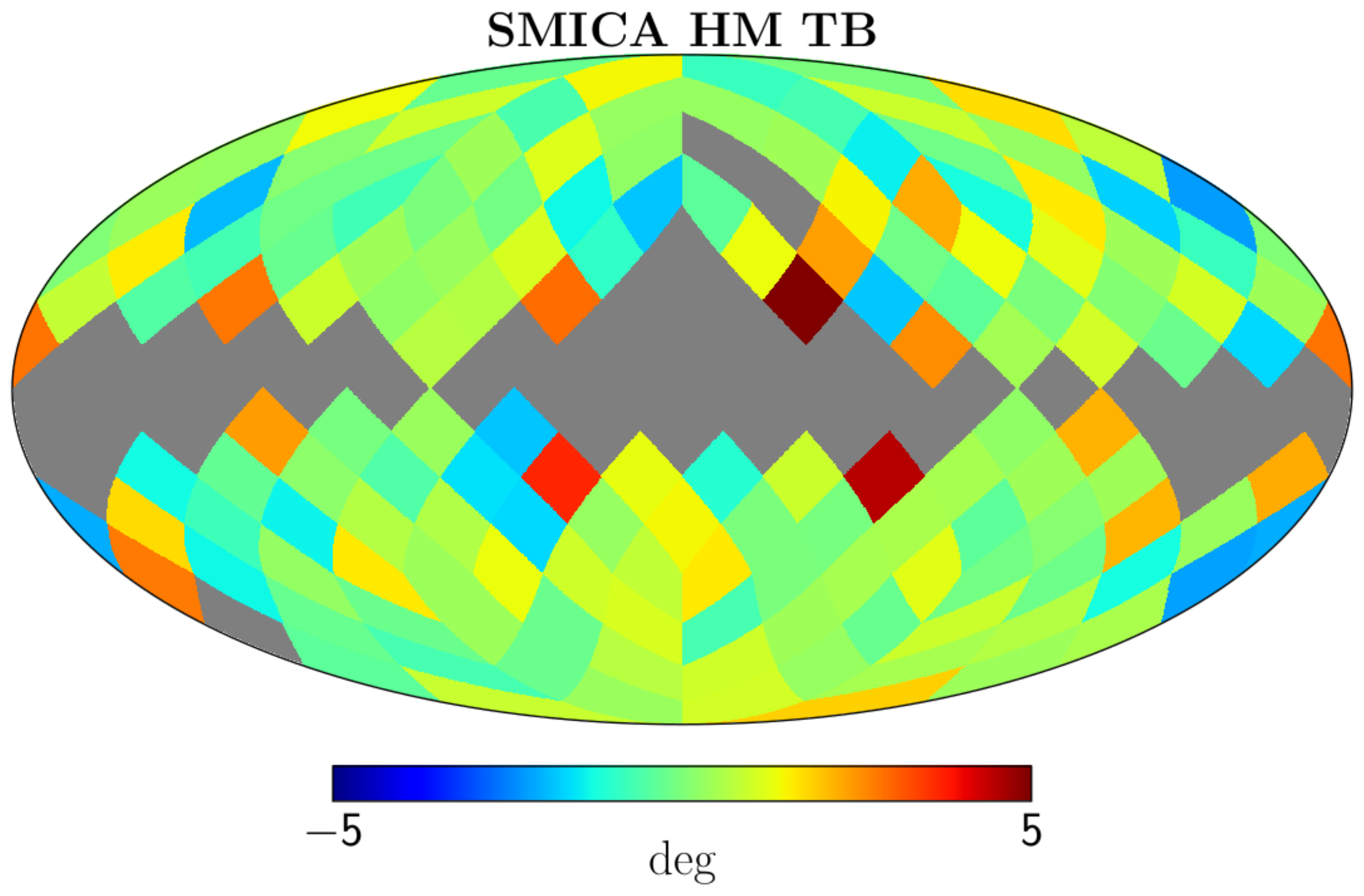}
\caption{Birefringence angle maps at $N_{side}=4$ estimated from {\it Planck} 2018 HM data. Left column is obtained from the minimisation of $\chi_{EB}^2$ and right column from the minimisation of $\chi_{TB}^2$.
From upper to lower panel we show our findings depending on the component separation method, i.e. \commander, \nilc, \sevem\ and \smica. 
Note that the color-scale of the left and right column is chosen to be different because of the intrinsic larger scatter of the TB-based estimator with respect to the companion EB-based.} \label{fig:alpha_maps}
\end{figure}

\begin{figure}[t]
\centering
\includegraphics[width=.45\textwidth]{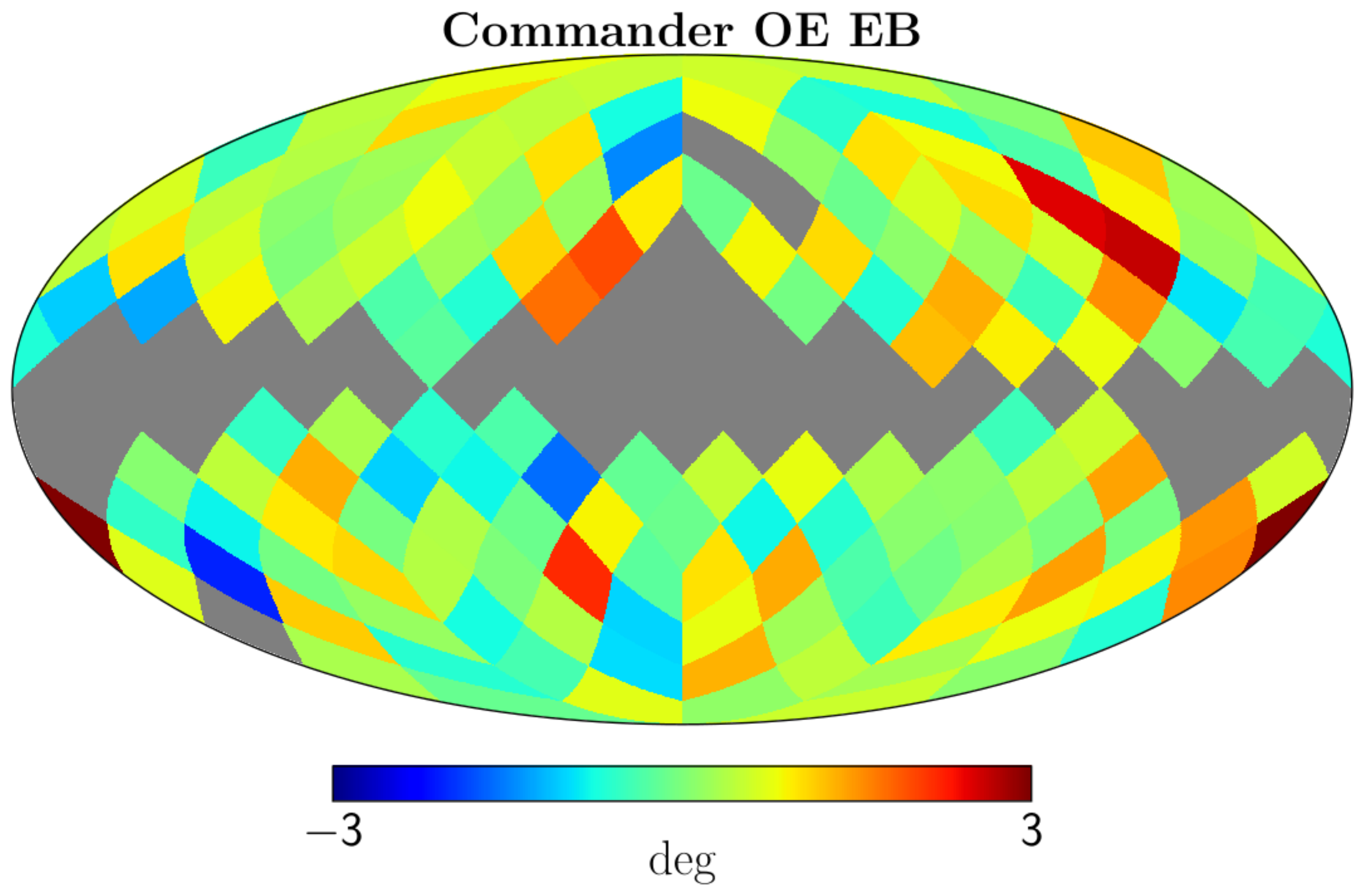}
\includegraphics[width=.45\textwidth]{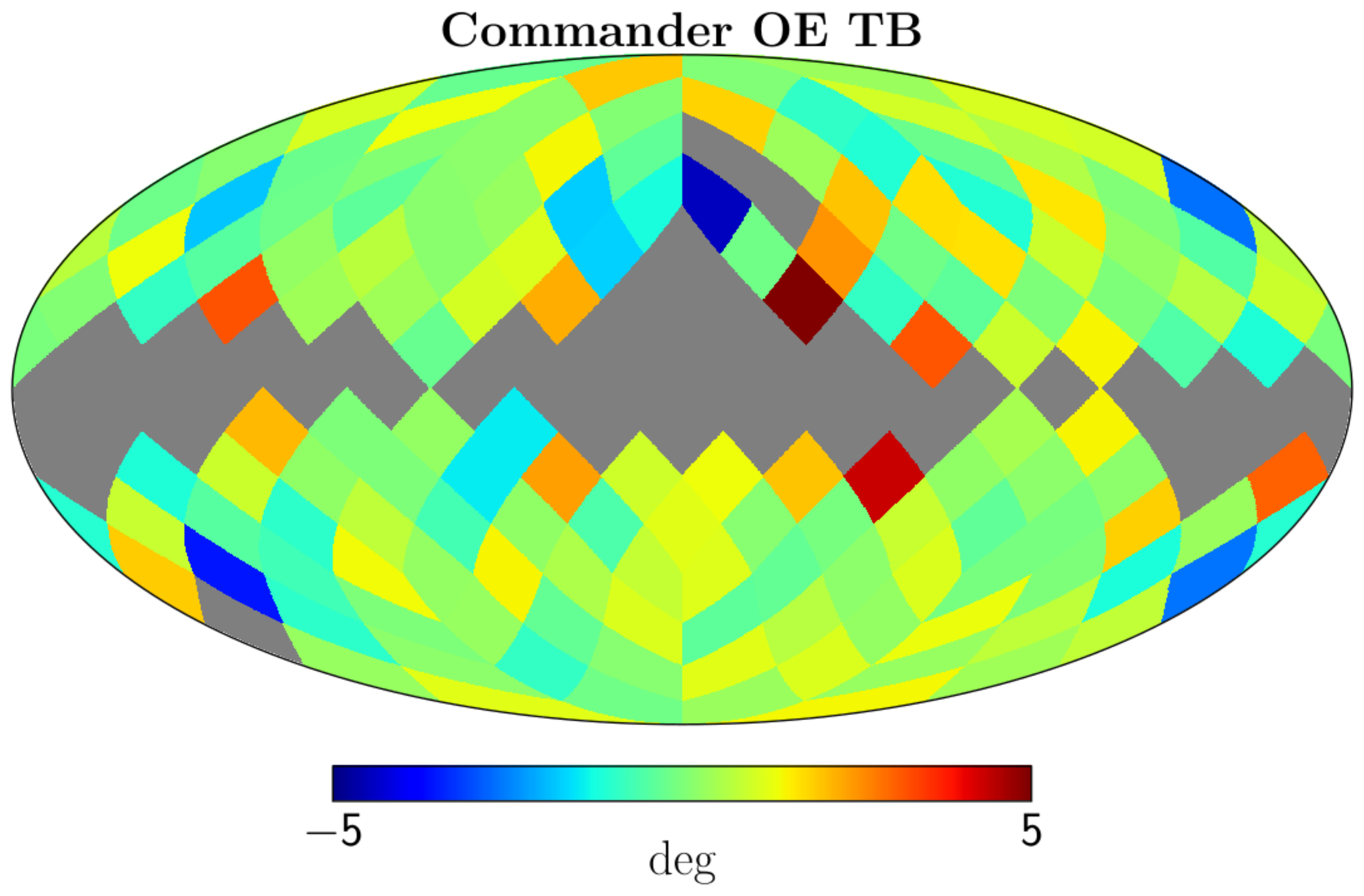}
\includegraphics[width=.45\textwidth]{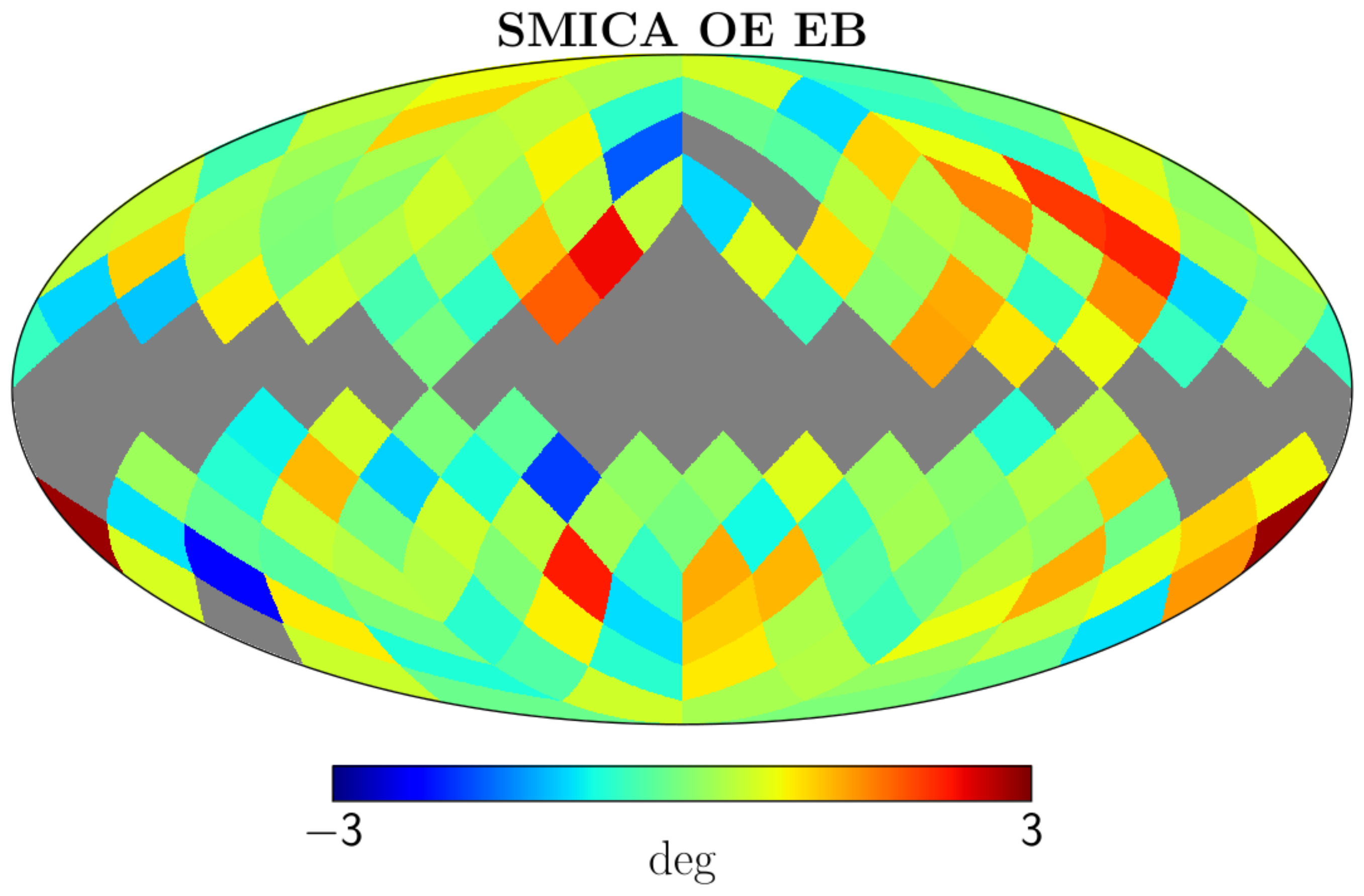}
\includegraphics[width=.45\textwidth]{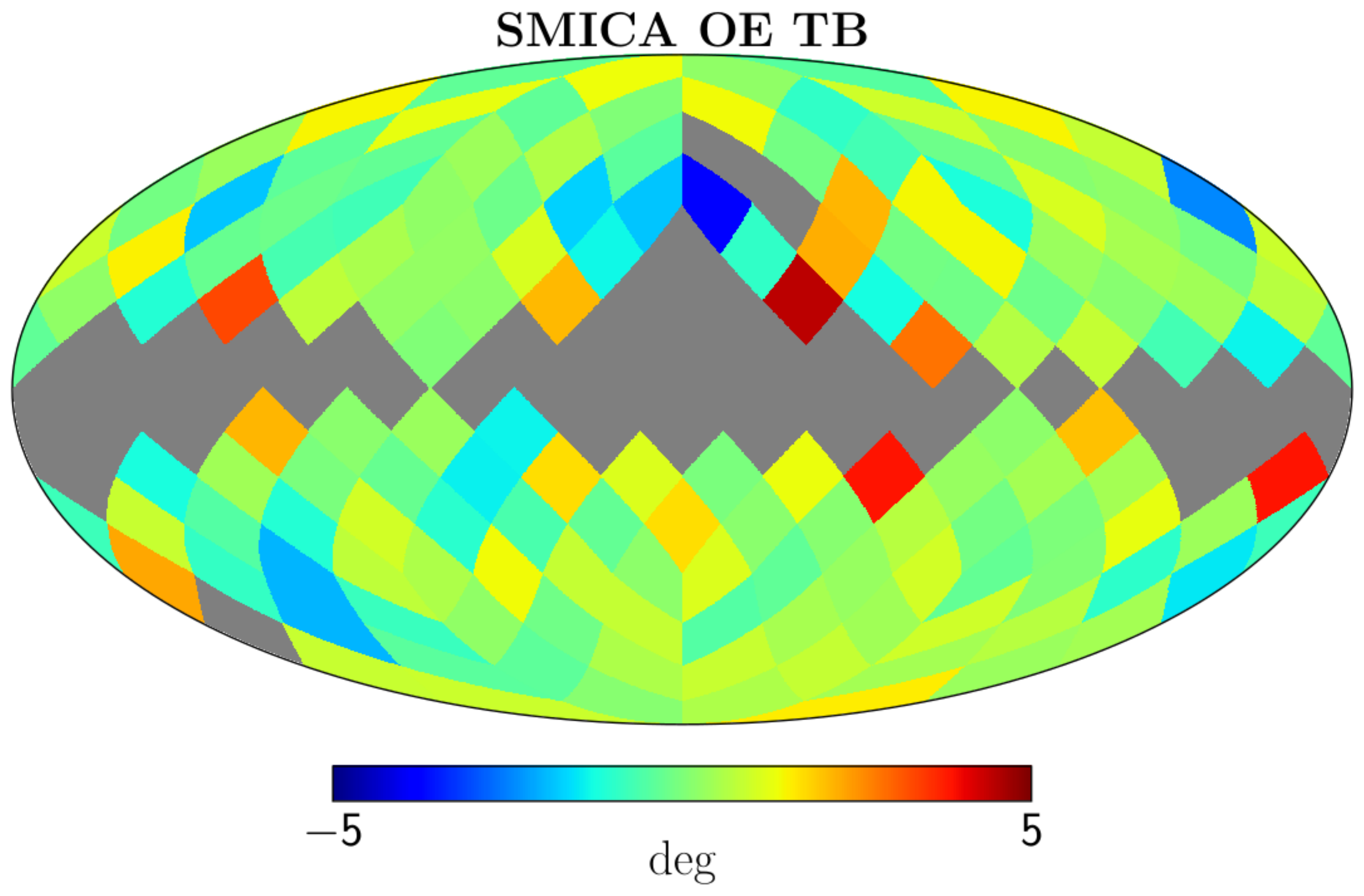}
\caption{Birefringence angle maps at $N_{side}=4$ estimated from {\it Planck} 2018 OE data. Left column is obtained from the minimisation of $\chi_{EB}^2$ and right column from the minimisation of $\chi_{TB}^2$.
From upper to lower panel we show our findings depending on the component separation method, i.e. \commander\ and \smica. Note that the color-scale of the left and right column is chosen to be different because of the intrinsic larger scatter of the TB-based estimator with respect to the companion EB-based.}\label{fig:alpha_maps_OE}
\end{figure}

As performed in \cite{Aghanim:2016fhp} we build $M_{\ell\ell'}^{YY}$ using the FFP10 simulations, which do not contain the birefringence effect.
Therefore, we build such a covariance matrix with $\alpha = 0$ and adopt a simple frequentist approach to test the null hypothesis of no parity violation. 
This minimisation is performed within each large pixel taken under consideration and for both data and FFP10 simulations.
When the latter procedure is applied to HM data we obtain the maps of $\alpha$ at $N_{side}=4$ shown in Figure \ref{fig:alpha_maps}, whereas when applied to OE data we find the maps of $\alpha$ shown in Figure \ref{fig:alpha_maps_OE}.

Note that the variance of the $\alpha$-distribution, obtained with the FFP10 simulations within each large pixel, represents an empirical assessment of the ``noise'' level in the low-resolution map of $\alpha$.
These variances are used to build a pixel-space diagonal anisotropic noise which is essential for the next and last step of the pipeline which consists in applying a quadratic maximum likelihood (QML) estimator in auto-mode \cite{Tegmark:1996qt} to evaluate the APS of the $\alpha$-anisotropies at low resolution. 
More specifically, we use \texttt{BolPol} \cite{Gruppuso:2009ab}, an implementation of the QML method, whose performances with respect to pseudo-$C_{\ell}$ estimators have been evaluated in \cite{Molinari:2014wza}.

\begin{figure}[t]
\centering
\includegraphics[width=.45\textwidth]{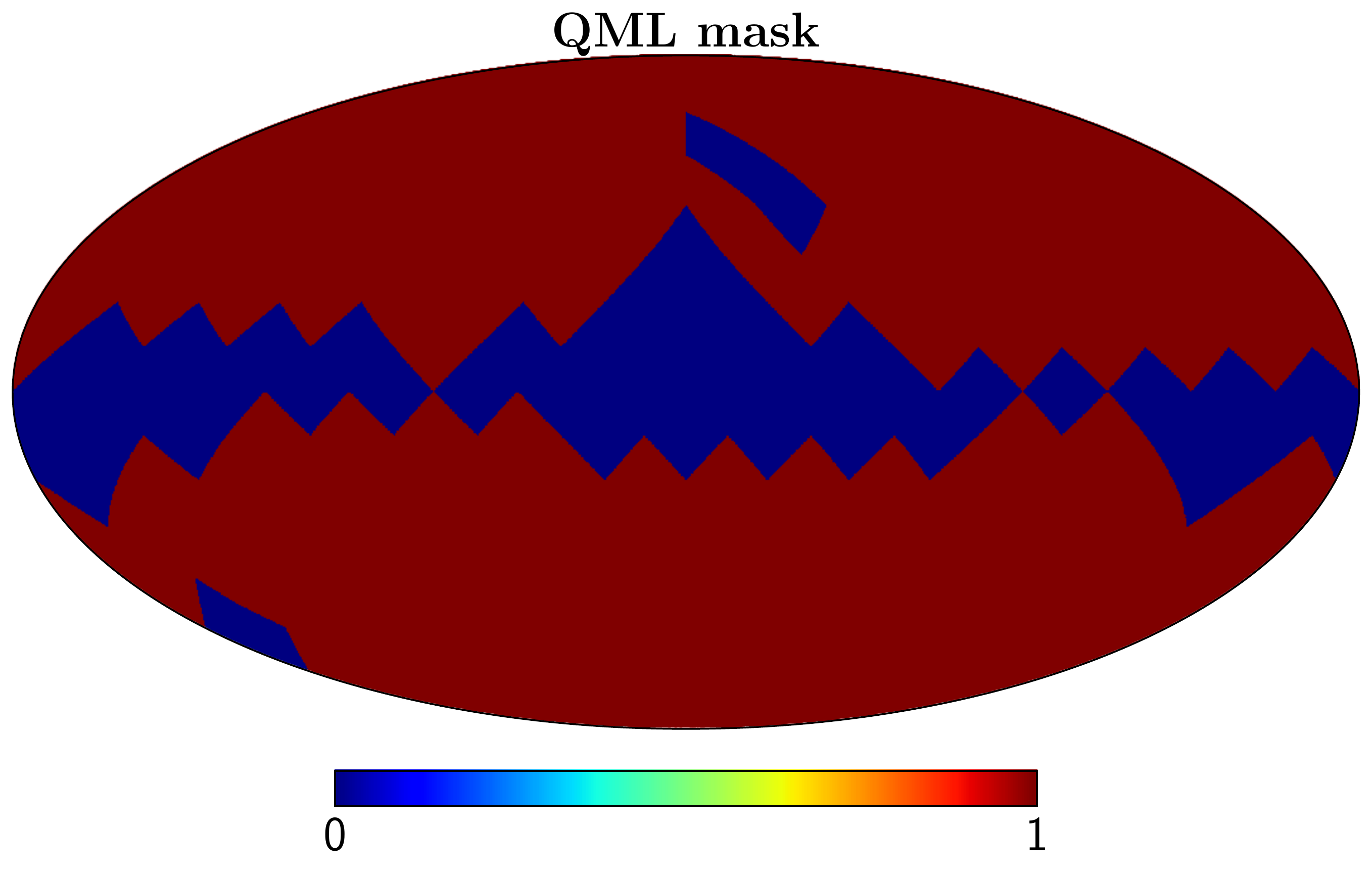}
\includegraphics[width=.45\textwidth]{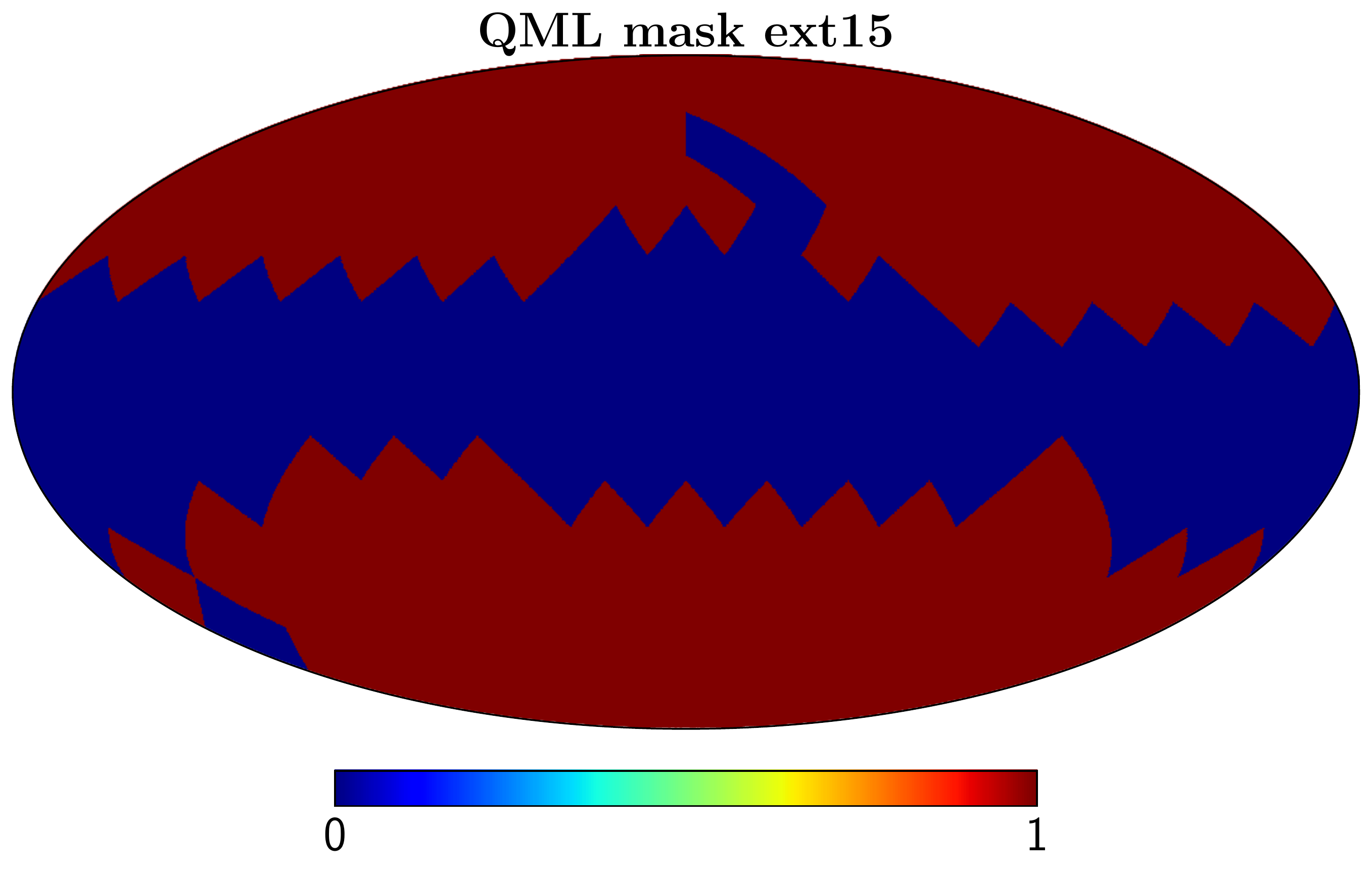}
\caption{Standard (left panel) and extended (right panel) masks used in the extracton of the APS of birefringence map using the QML estimator.}\label{fig:mask_alpha}
\end{figure}

The APS of the maps of $\alpha$ are estimated using the masks shown in Figure \ref{fig:mask_alpha}, referred as standard mask (left panel) and extended mask (right panel).  Our standard mask has been obtained by excising all pixels at $N_{side}=4$ for which more than 50\% of the high resolution ($N_{side}=2048$) pixels they contain, are removed by the \planck\ common mask in temperature or 
polarisation\footnote{In fact, we use as our standard mask the one obtained from the \planck\ high-resolution HM common mask as opposed to OE, since the former is slightly more conservative. Anyway, the difference between the two is small as employing OE would result in incorporating three additional low-resolution pixels.}.
The extended mask
has been obtained by simply enlarging the edges of the standard mask by $\sim$15 degrees, which is the angular scale of the large pixel. 

\begin{figure}[t]
\centering
\includegraphics[width=.45\textwidth]{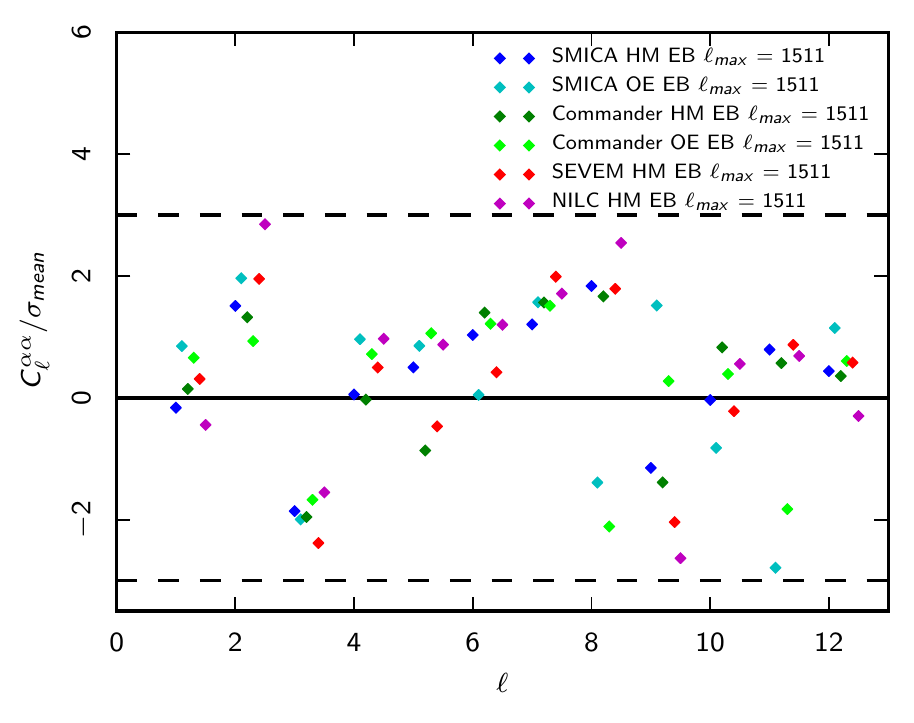}
\includegraphics[width=.45\textwidth]{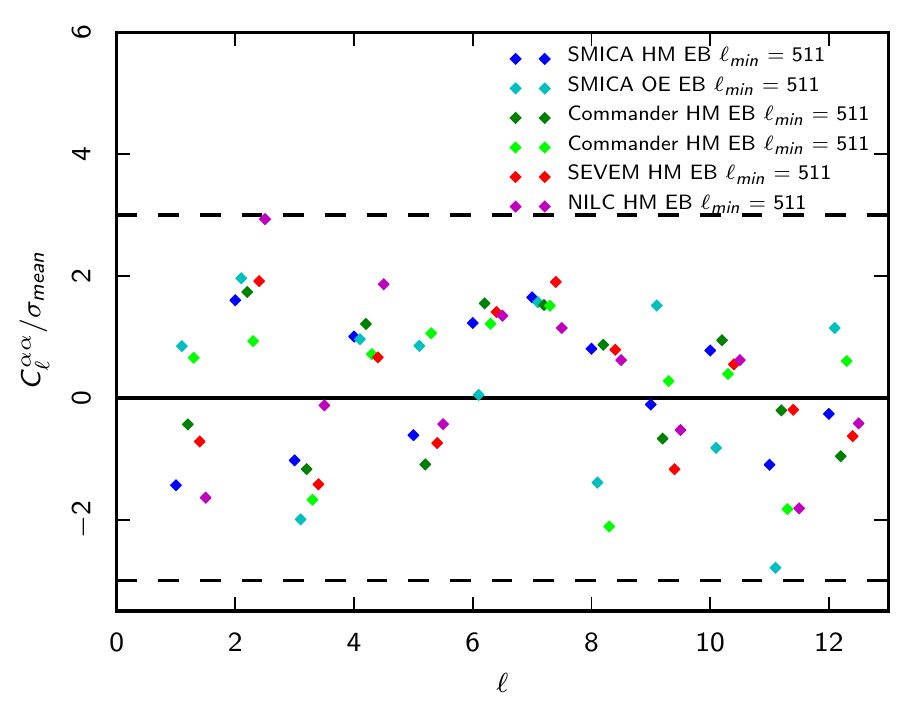}
\caption{MC average of $C_{\ell}^{\alpha \alpha}$ in units of standard deviation of the mean, obtained with FFP10 simulations from $\ell=1$ to $\ell=12$ for the standard mask.
Left panel refers to minimisation of $\chi_{EB}^2$ in the harmonic range $[51-1511]$ whereas right panel is for the $[511-1511]$.
}
\label{fig:plotvalidazione}
\end{figure}

\begin{figure}[t]
\centering
\includegraphics[width=.45\textwidth]{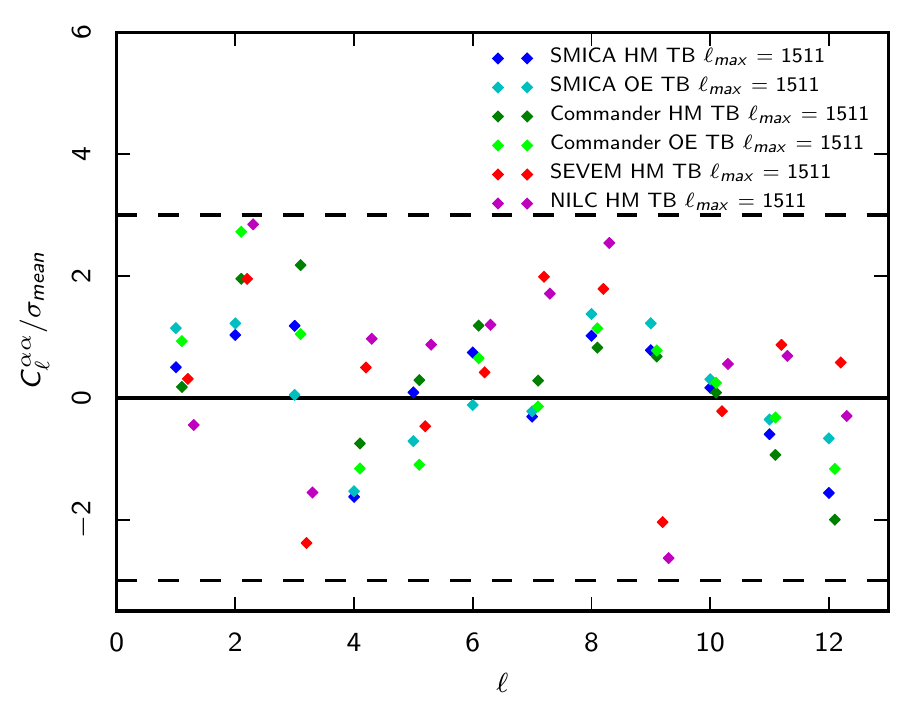}
\caption{As for Figure  \ref{fig:plotvalidazione} but minimising $\chi_{TB}^2$ in  the harmonic range $[51-1511]$.}
\label{fig:plotvalidazioneTB}
\end{figure}

We apply \texttt{BolPol} to the map of $\alpha$ derived from the observed data and each of the 999 maps of $\alpha$ obtained with the FFP10 simulations. The MC averages of the APS in units of standard deviation of the mean, are shown in Figure \ref{fig:plotvalidazione} with $\ell \in [1,12]$ for the standard mask (the monopole, which traces isotropic birefringence is discussed further below). The displayed fluctuations are compatible with no birefringence effect well within $3 \,\sigma_{mean}$ C.L., as expected from the birefringence-free FFP10 simulations. Figure~\ref{fig:plotvalidazione} provides the validation of the whole pipeline for all the considered \planck\ component separation methods with HM and OE splits. In particular, the left panel of Figure \ref{fig:plotvalidazione} is obtained by minimising $\chi_{EB}^2$ in the harmonic range $[51-1511]$ whereas in the right panel such a minimisation is performed in $[511-1511]$. The former range is chosen in analogy to what performed in \cite{Aghanim:2016fhp} and the latter following reference \cite{Akrami:2018mcd}.
In Figure \ref{fig:plotvalidazioneTB} we show the MC averages of $C_{\ell}^{\alpha \alpha}$ (from $\ell=1$ to $\ell=12$ and for the standard mask) given in units of standard deviation of the mean. This was obtained by minimising $\chi_{TB}^2$ in the harmonic range $[51-1511]$. Again, the whole pipeline appears to be validated also through the minimisation of $\chi_{TB}^2$ for all the considered \planck\ component separation methods with HM and OE splits.

Figures \ref{fig:plotvalidazione} and \ref{fig:plotvalidazioneTB} seem to indicate that the residual of the systematic effects present in the FFP10 simulations, as described in Section \ref{dataset}, do not play an important role in the harmonic space of $\alpha$-anisotropies. Further considerations in this respect are reported in the following subsection.

\subsection{Further considerations about residual instrumental FFP10 systematics}
\label{further}

\begin{figure}[t]
\centering
\includegraphics[width=.45\textwidth]{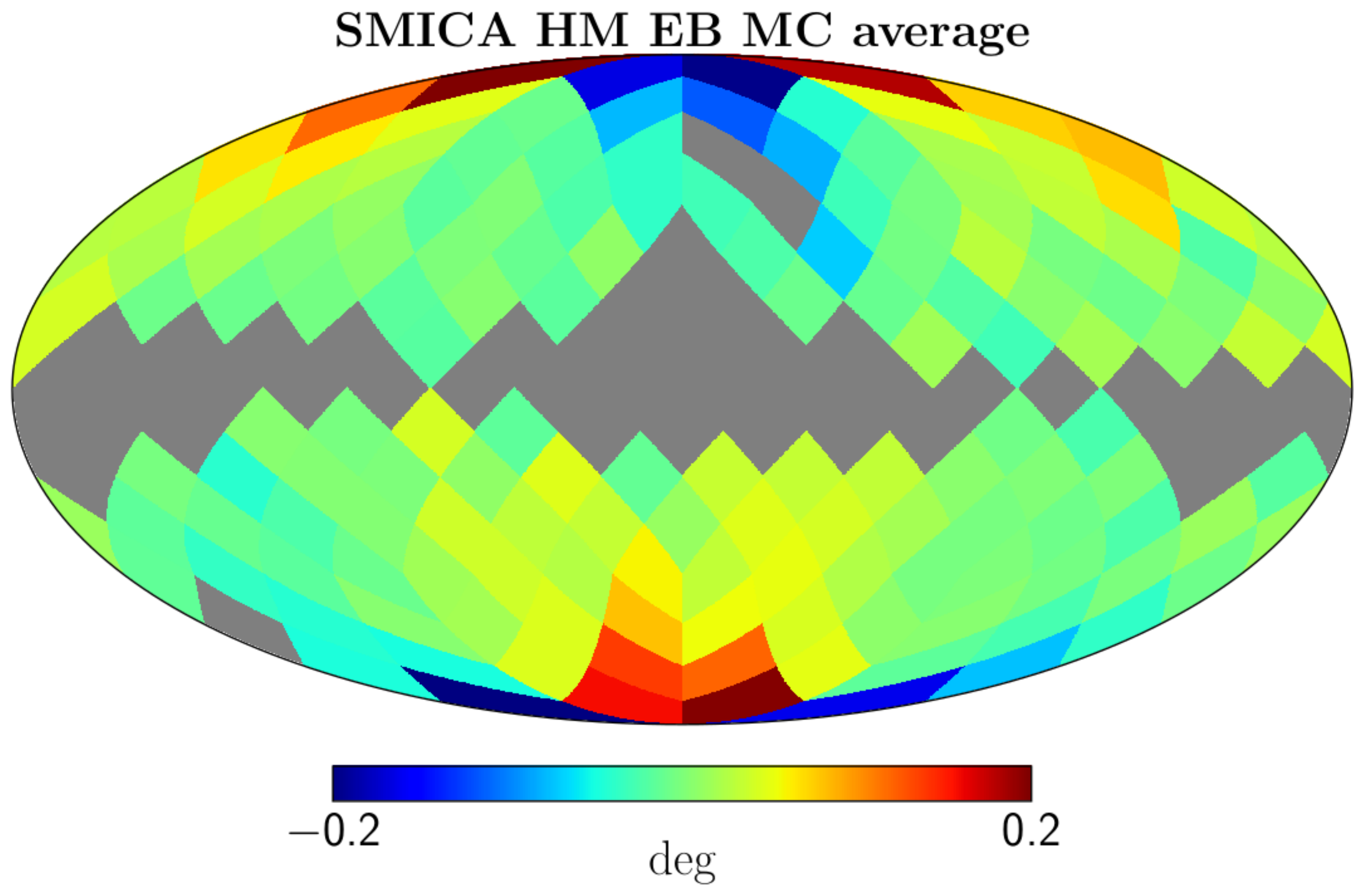}
\includegraphics[width=.45\textwidth]{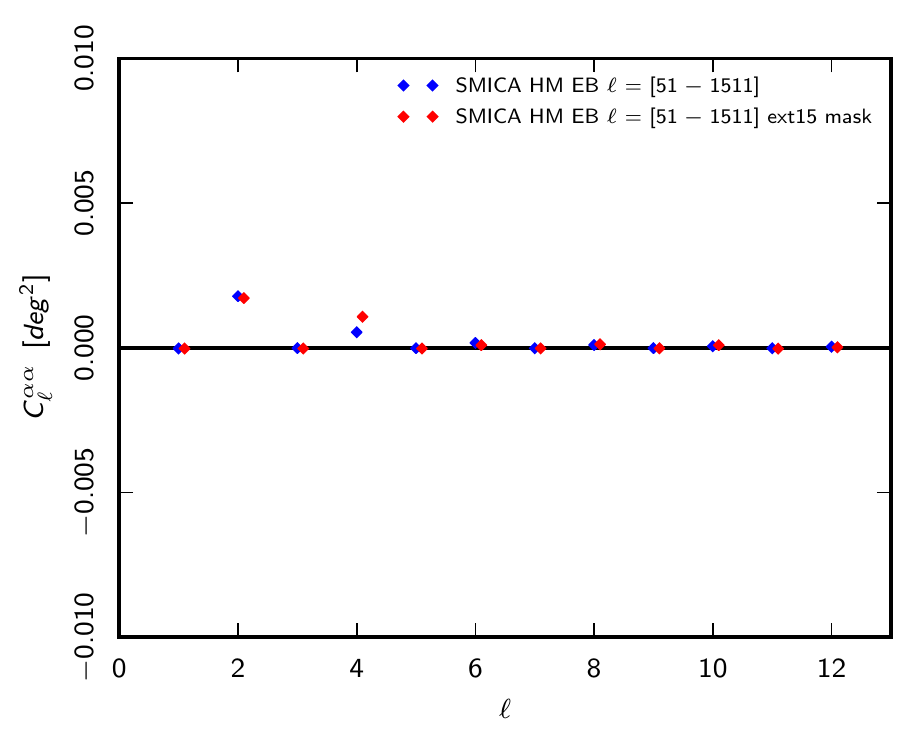}
\caption{Left panel: mean of $\alpha$, obtained from the FFP10  \smica\ HM simulations through the minimisation of $\chi_{EB}^2$ in the harmonic range $[51-1511]$. Right panel: APS of the mean of $\alpha$, see left panel. Blue symbols are for the standard mask, whereas red ones are for the extended mask.  
}
\label{meanFFP10maplevel}
\end{figure}

In the previous section, thanks to Figures \ref{fig:plotvalidazione} and \ref{fig:plotvalidazioneTB}, we have seen that residuals of known systematics, i.e. those that are present in the signal part of the FFP10 simulations, do not have a significant impact on the APS of $\alpha$-anisotropies. However they show up in pixel space, i.e. directly in the map of $\alpha$. 
In the left panel of Figure \ref{meanFFP10maplevel} we display the mean of the MC maps of $\alpha(\hat n)$, obtained from the FFP10 \smica\ HM simulations. As it will be shown below, the structures around the Galactic poles are artefacts due to residuals of systematics present in the FFP10 simulations that do not wash out in cross-spectrum analysis.
It is likely that using a version of the FFP10 simulations whose signal part is computed specifically for HM or OE, not presently available, would lead to 
weaker residuals.
In any case, the residuals shown in Figure~\ref{meanFFP10maplevel}  appear noticeable only because they stand out in the mean field map. 
The same residuals would go substantially unnoticed if a single realization was shown.
The pattern of these systematics residuals is evident: its amplitude is maximum at the Galactic poles and decreases quite symmetrically approaching the Galactic equator.
The  map shown in the left panel of Figure~\ref{meanFFP10maplevel} can be conveniently considered as a template for these systematics: we call it FFP10 template, for sake of brevity. Its APS is shown in the right panel of 
Figure \ref{meanFFP10maplevel}, along with the associated error bars which however are too small to be distinguishable: in fact, the noise level for this template  can be derived by rescaling the single-realisation covariance for $\alpha(\hat n)$ by the number of FFP10 simulations, i.e.\ 999.
The same panel shows that the FFP10 systematic residuals only affect even modes and decrease when increasing $\ell$, so that $\ell=2$ is the multipole mostly contaminated while $\ell=4$ is less polluted. 
All the higher even multipoles are affected in a negligible manner. Instead, the odd multipole are well compatible with zero, a consequence of the symmetry of the FFP10 template around the Galactic equator.
This pattern appears stable with respect to the mask employed.

\begin{figure}[t]
\centering
\includegraphics[width=.45\textwidth]{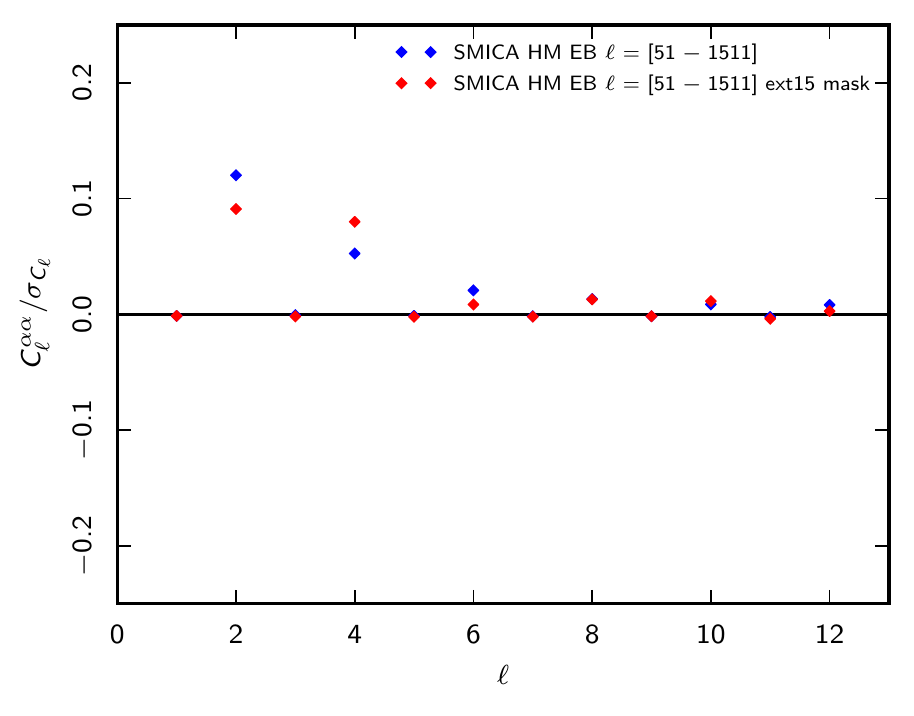}
\includegraphics[width=.45\textwidth]{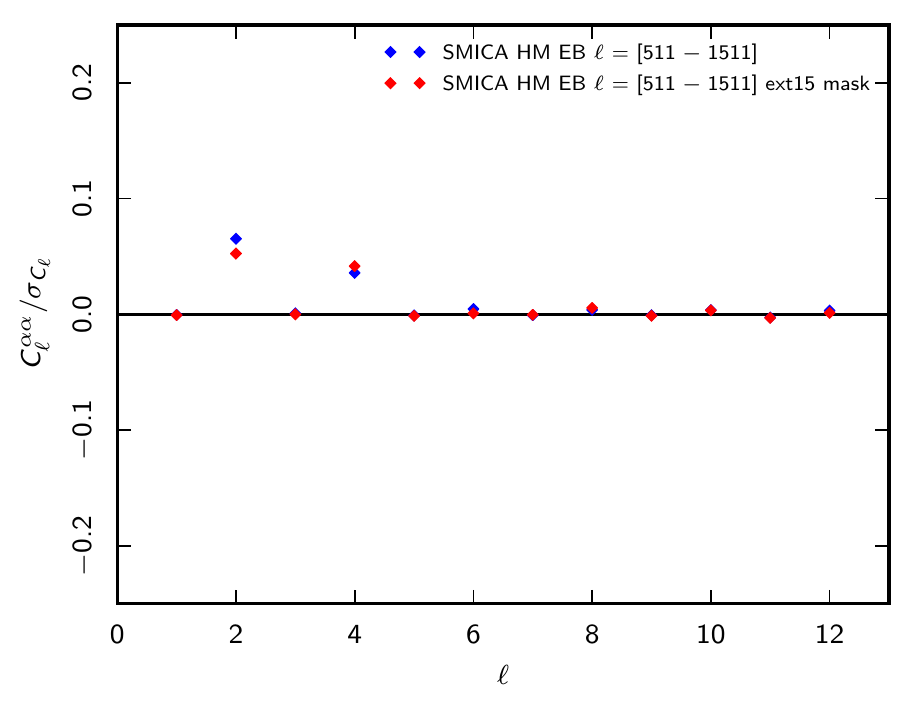}
\caption{APS of the FFP10 template normalised to the statistical uncertainty provided by FFP10 simulations.
Left panel refers to the FFP10 template obtained minimising the $\chi_{EB}^2$ in the harmonic range $[51-1511]$ whereas right panel in $[511-1511]$.}
\label{meanFFP10mapleveloversigma}
\end{figure}

The impact of the FFP10 template on the single realisation of an $\alpha$-map can be quantified by plotting its APS in units of the uncertainty of the single realisation, as shown in Figure \ref{meanFFP10mapleveloversigma}: the template's amplitude is around at most $\sim$10$\%$ of the single realisation error, both for the template derived from EB in the harmonic range $[51-1511]$ (left panel) and for the range $[511-1511]$ (right panel). In short, we are able to detect the FFP10 template with high confidence thanks to the high number of simulations involved, but its impact at the level of the single realisation is very small. We nonetheless account for this template in the subsequent analysis when needed.
As mentioned above, the results shown here regarding the FFP10-template are obtained from the FFP10 \smica\ simulations in the HM splits. However, very similar results are also found for the other component separation methods and for the OE data splits.

\begin{figure}[t]
\centering
\includegraphics[width=.45\textwidth]{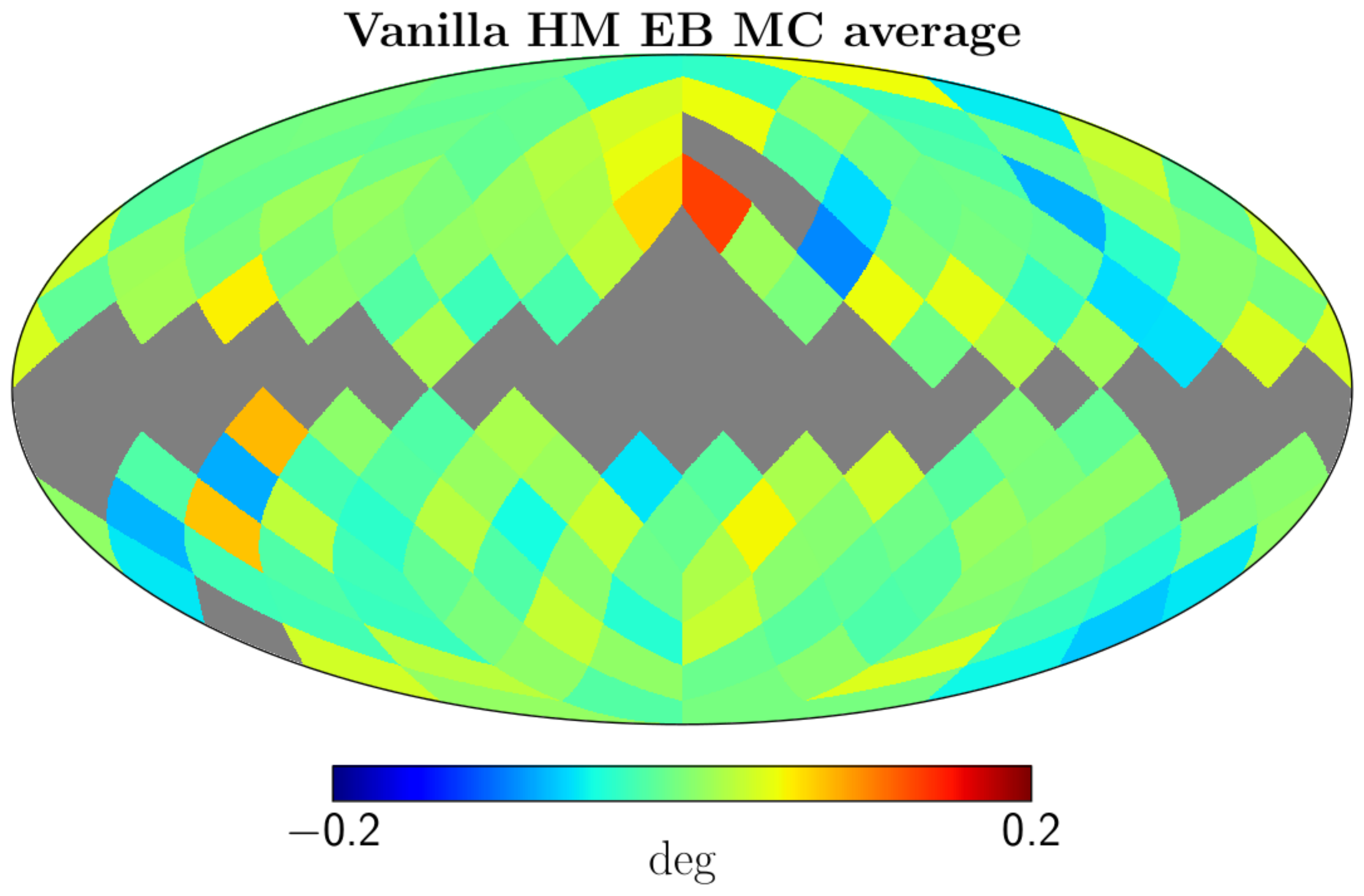}
\includegraphics[width=.45\textwidth]{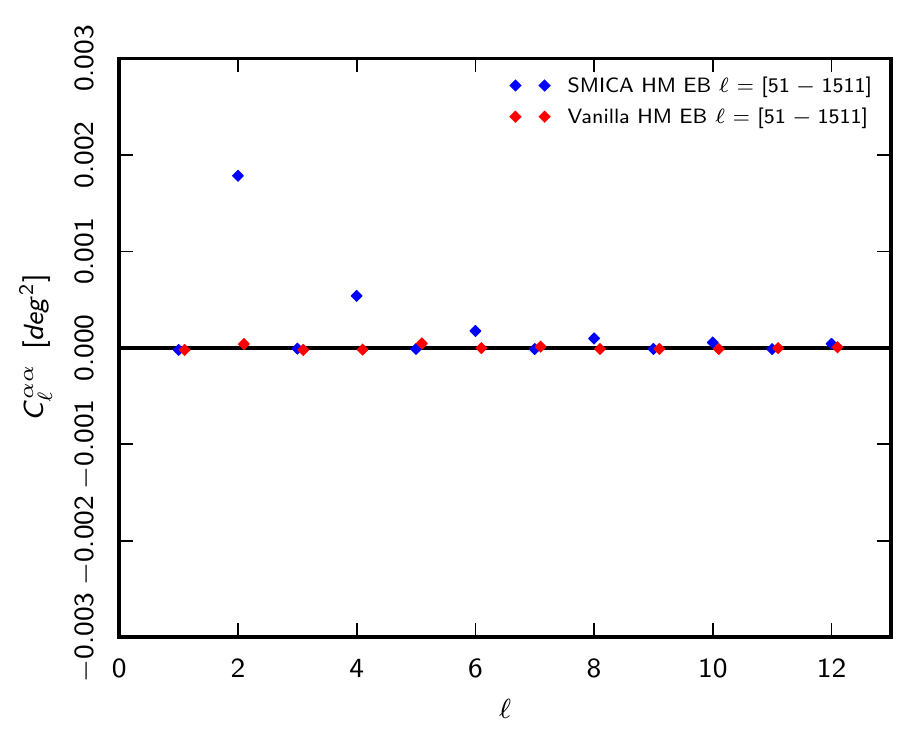}
\caption{Left panel: mean of $\alpha$, obtained from ``vanilla simulations'' (systematics-free $\Lambda$CDM signal plus FFP10 noise for the $143$ GHz channel) through the minimisation of $\chi_{EB}^2$. Right panel: APS of the map mean $\alpha$ shown in the left panel, computed considering the standard mask, (red symbols). The blue symbols refer to the case of Figure~\ref{meanFFP10maplevel}, i.e.\  \smica\ full-mission signal estimates plus the corresponding FFP10 HM noise}
\label{meanVanillamaplevel}
\end{figure}

In order to prove that the pattern found in Figure \ref{meanFFP10maplevel} really arises due to the systematics modeled in the FFP10 simulations, we replace the signal part of the latter with a ``vanilla'', systematics-free, set of simulations extracted from a $\Lambda$CDM model, while leaving the FFP10 noise part unaltered. For this case no residuals are present at the level of the $\alpha$-map, 
as shown in the left panel of Figure \ref{meanVanillamaplevel}.
The corresponding APS is well compatible with zero as shown in the right panel of Figure \ref{meanVanillamaplevel}.
This demonstrates that the known non-idealities present in the signal part of the FFP10 simulations are responsible for the pattern shown in Figure \ref{meanFFP10maplevel}. As mentioned above, this template is derived from the signal part of FM simulations, the only configuration available for the \planck\ 2018 release. Strictly speaking, this is not the proper configuration to use since our analysis is based on HM and OE data splits. 
Had we employed simulations generated for the latter splits, currently not available, the resulting systematic residuals would have likely been even lower. 

The outcome of the analysis carried out in this section is that the systematics modelled in the FFP10 simulations are detectable at the map level but are negligible or weak in harmonic space. 
In the spirit of being cautious we have anyway successfully repeated the validation procedure removing the FFP10 template from each of the simulations, for all the cases 
considered\footnote{We do not report here all the validations for sake of brevity and also because they are similar to what already shown in Section  \ref{Description}.}.

To sum up, in the remainder of this work, we always specify when we do include or not the de-biasing FFP10 template. 
The effect is weak and at most representing a small fraction of the statistical uncertainty. 
In addition, since it is not known how much similar is the FM set of the signal part of the FFP10 simulations to the corresponding HM or OE splits, it is not clear to what extent we are allowed to use 
the FFP10 template on the data. As it will be clear in the following, we typically do not de-bias the data with the FFP10 template. For some cases, we check the impact of this operation, which is again found to be 
small.

\section{Settings of the baseline configuration}
\label{analysis}

The pipeline for all the considered cases is {\it independently} validated on FFP10 simulations in Section \ref{validation}. 
We now perform null-tests comparing data and simulations and use them to select the settings for what we call ``baseline'' configuration.

\subsection{Null tests and systematics assessment}
\label{systematics}

\begin{figure}[t]
\centering
\includegraphics[width=.25\textwidth]{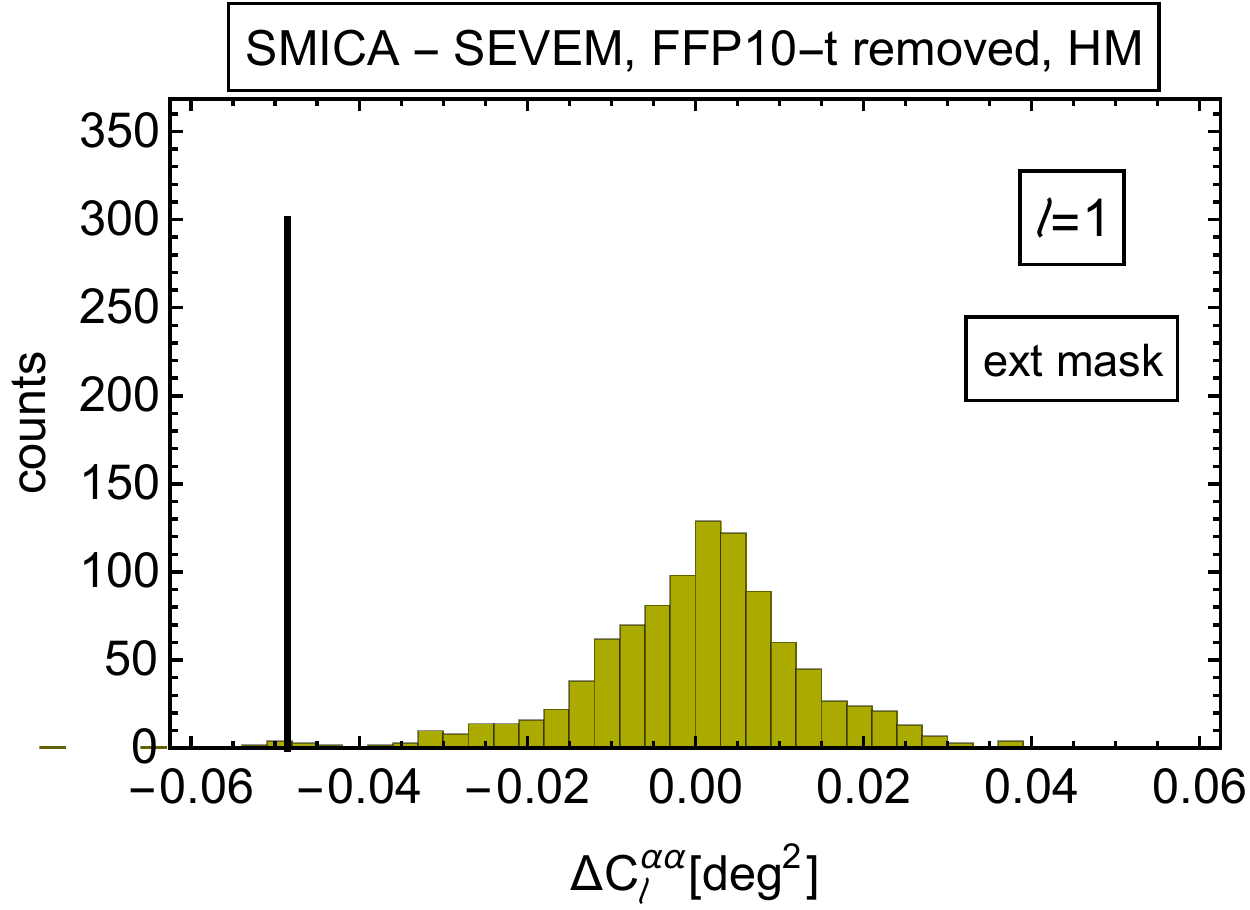}
\includegraphics[width=.25\textwidth]{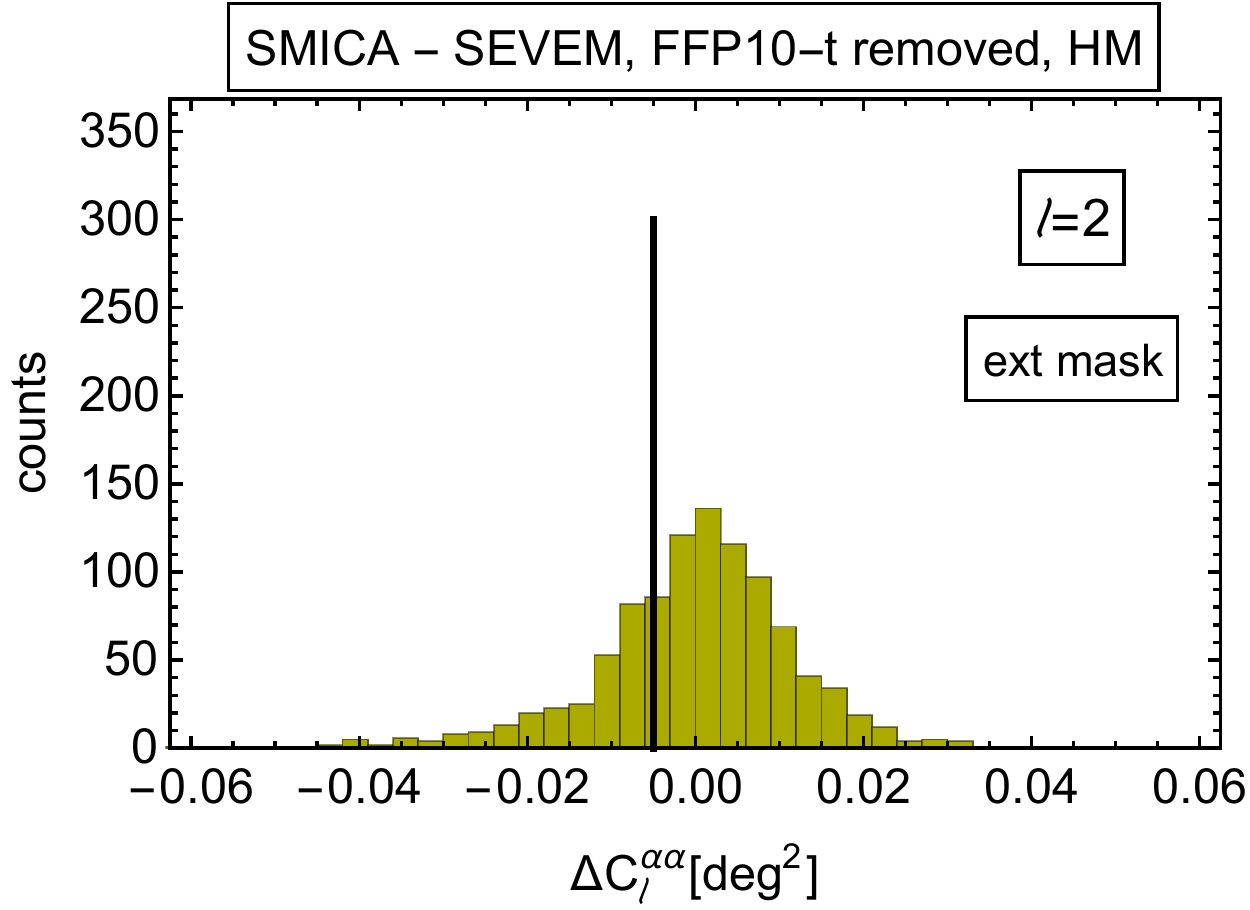}
\includegraphics[width=.25\textwidth]{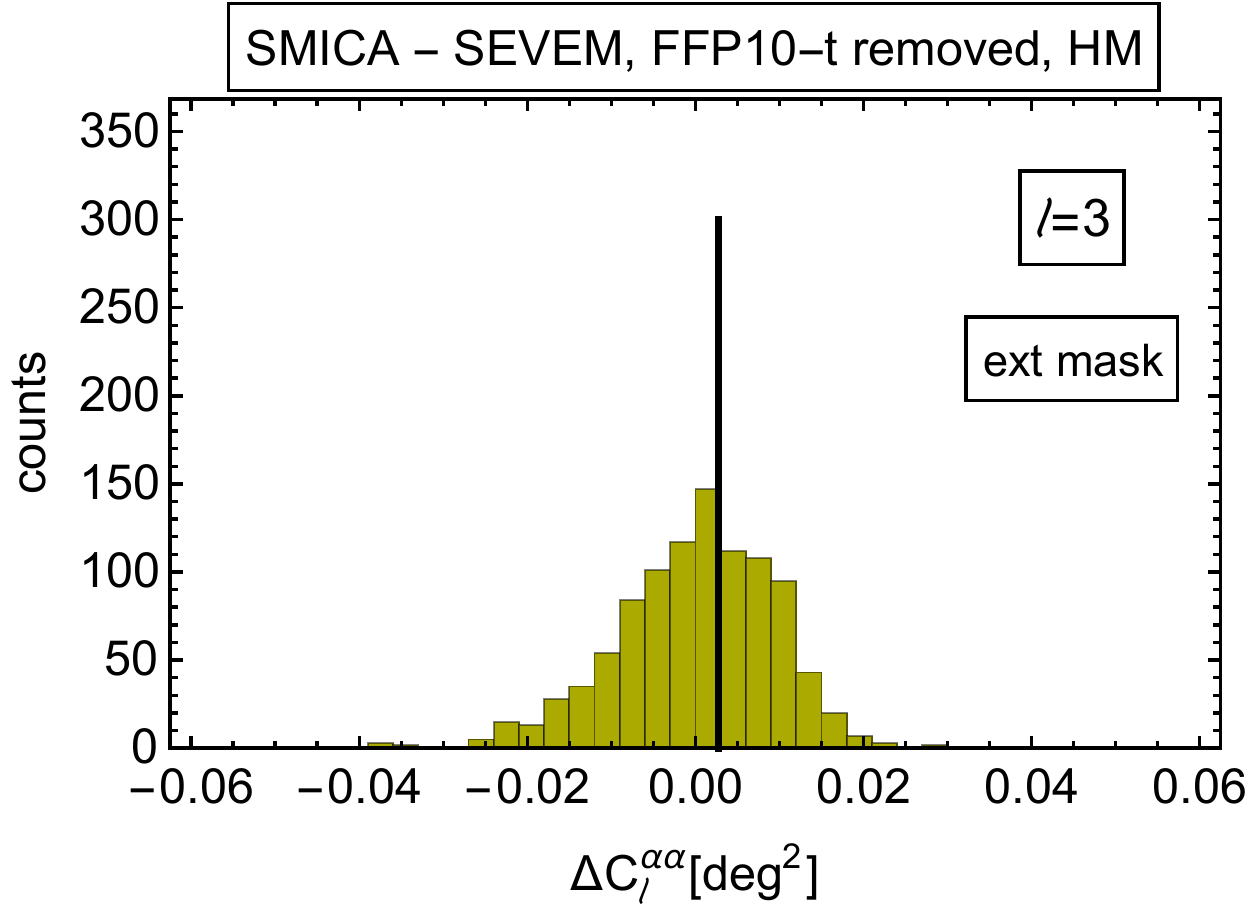}
\includegraphics[width=.25\textwidth]{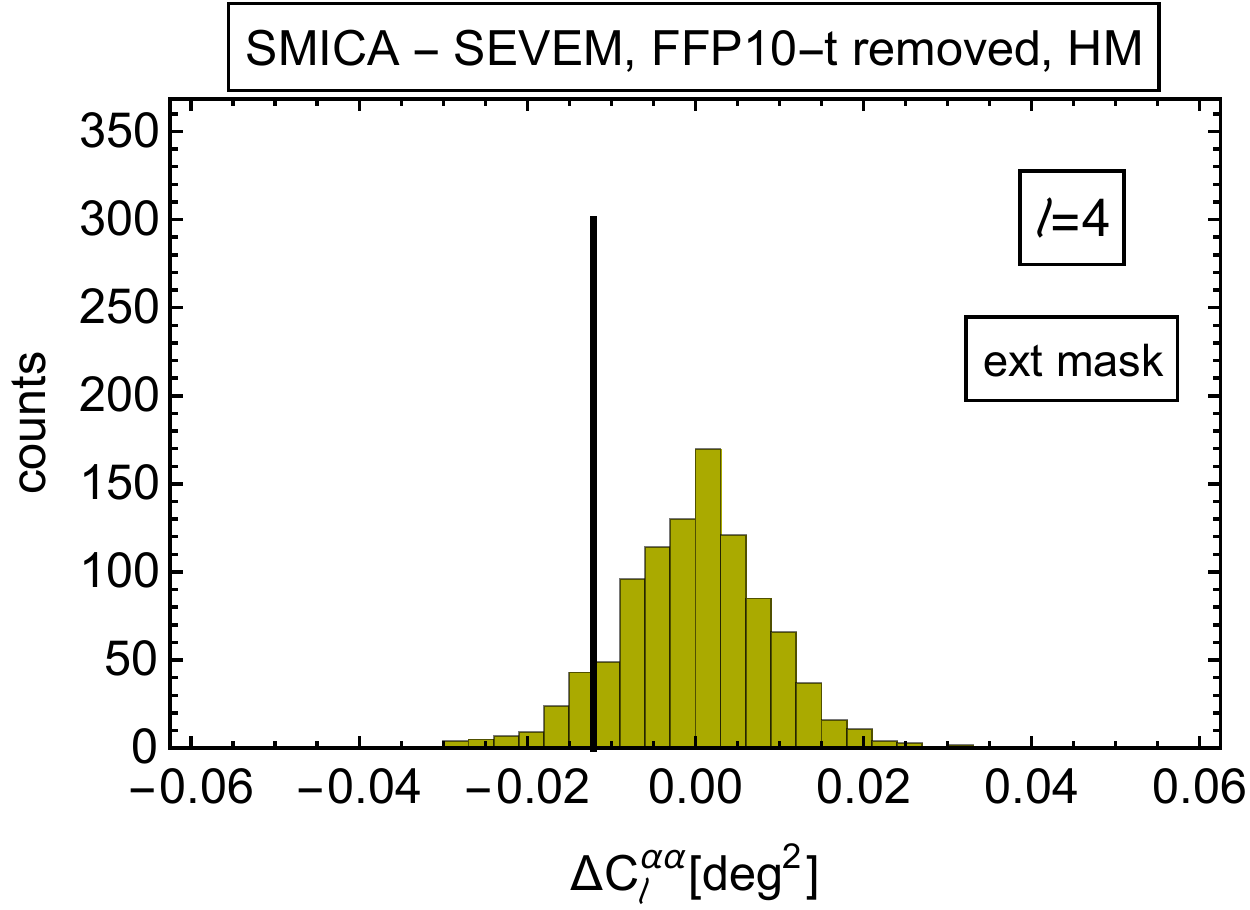}
\includegraphics[width=.25\textwidth]{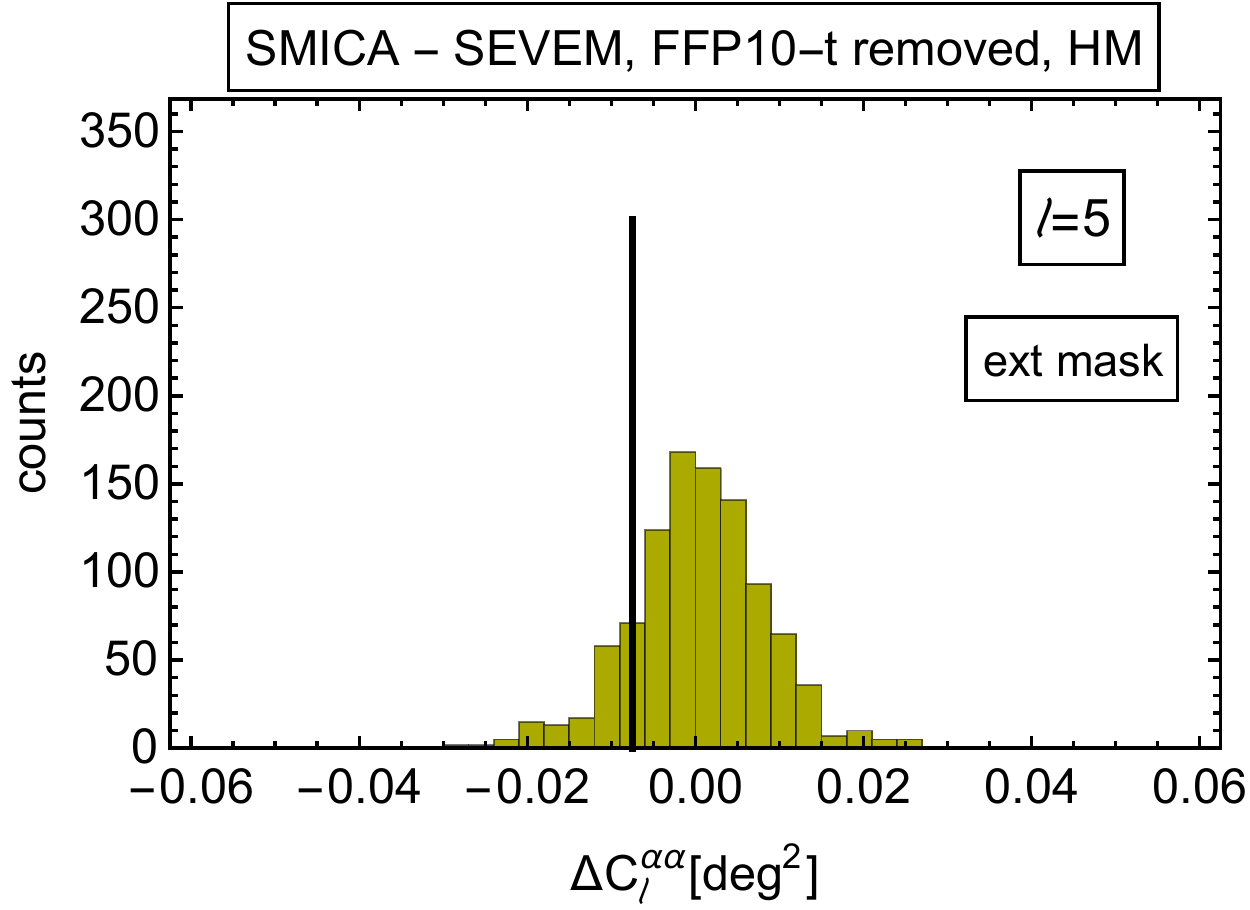}
\includegraphics[width=.25\textwidth]{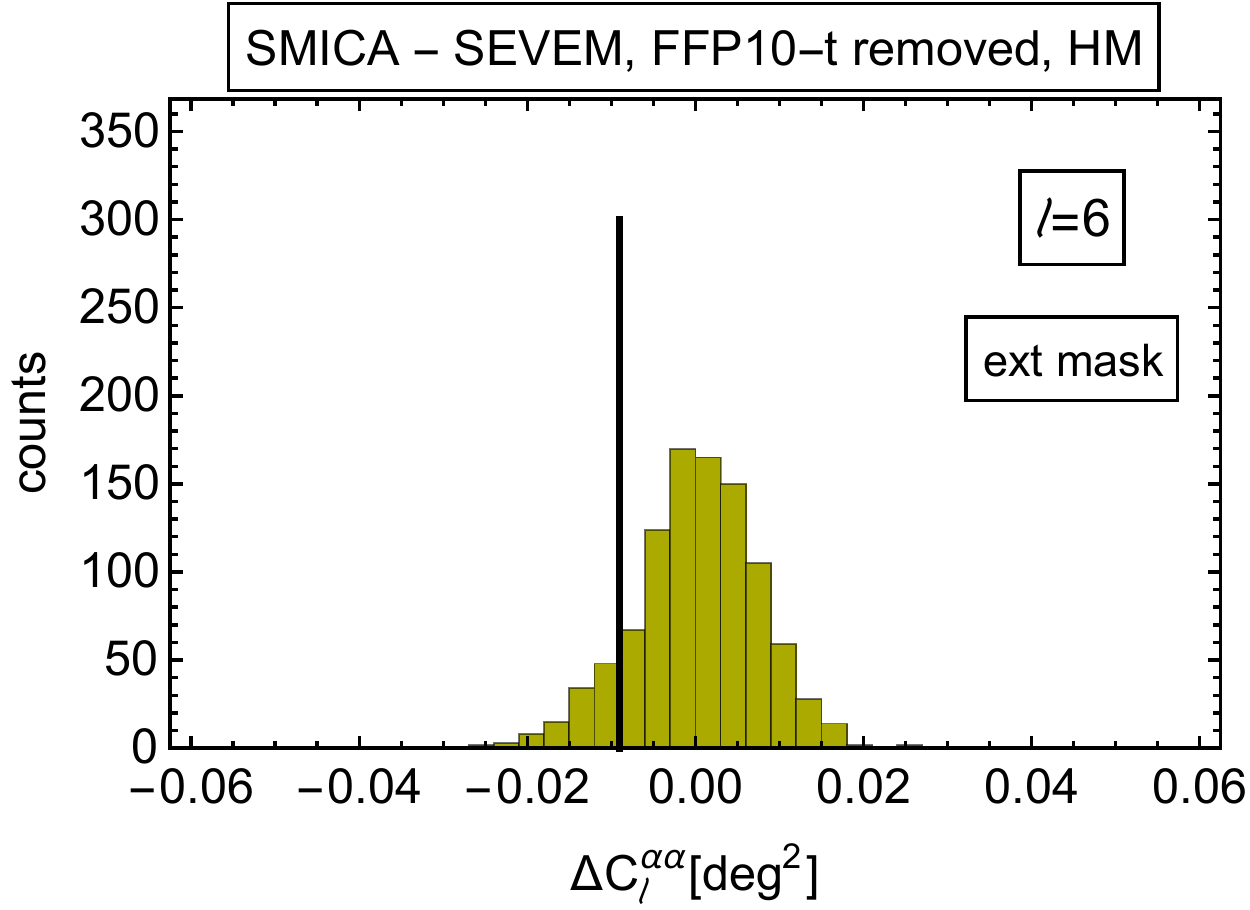}
\includegraphics[width=.25\textwidth]{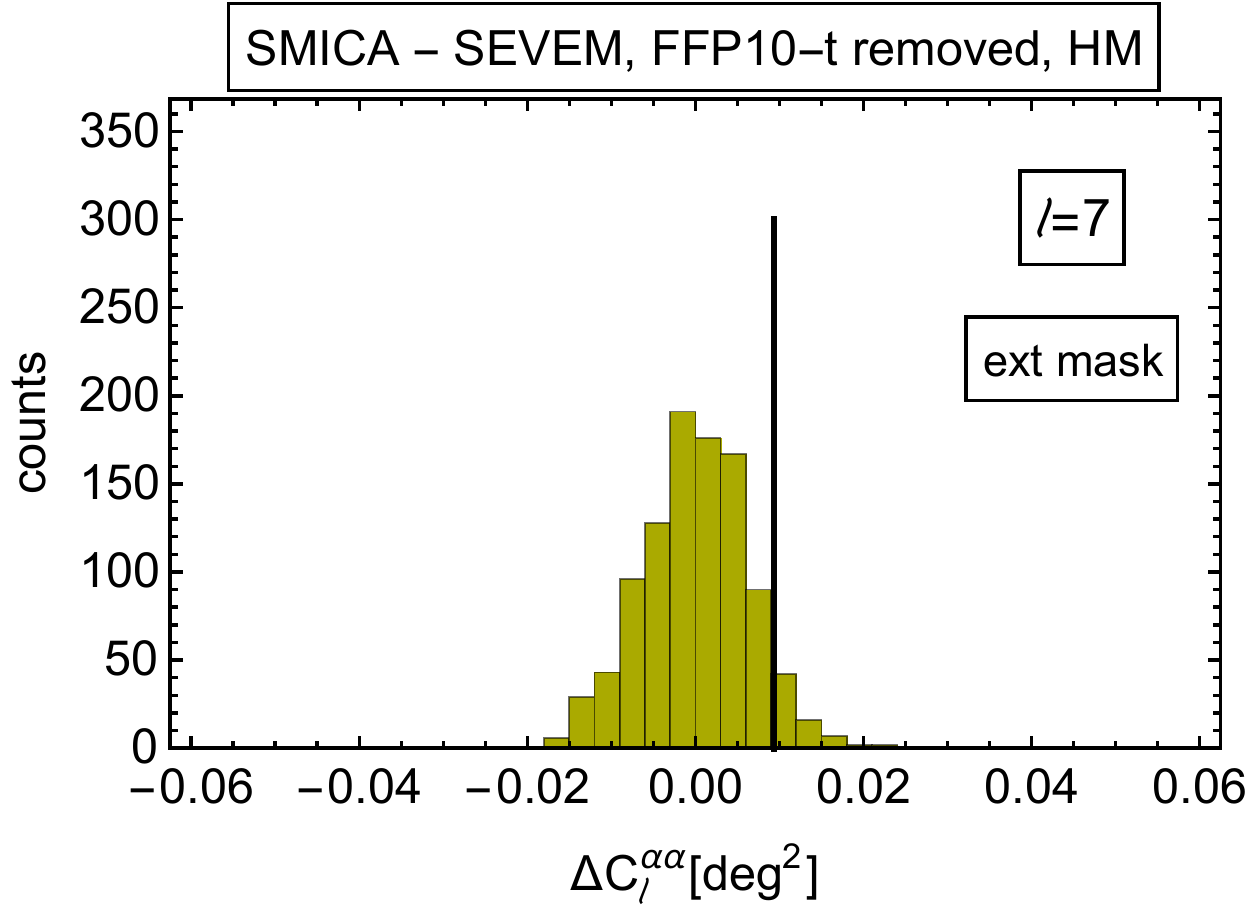}
\includegraphics[width=.25\textwidth]{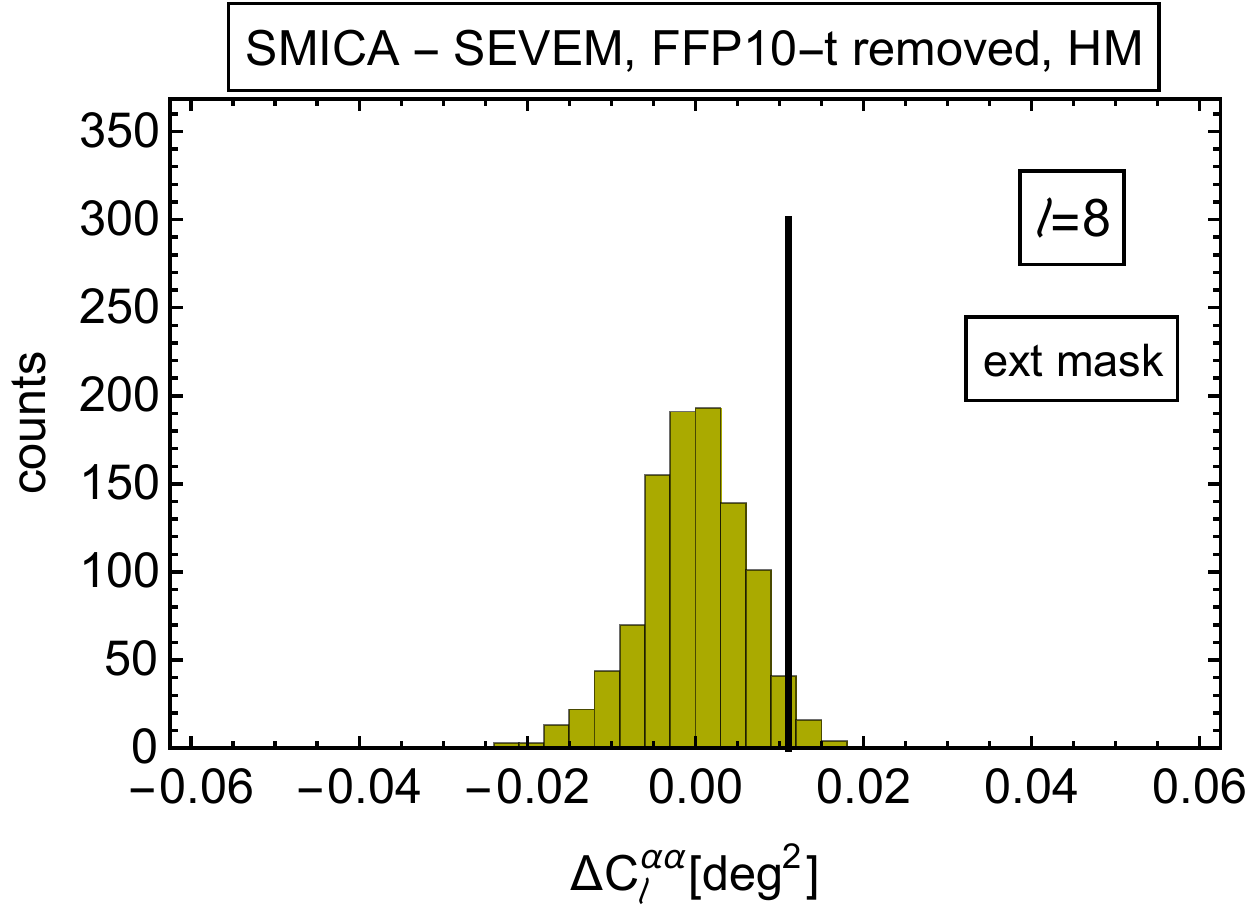}
\includegraphics[width=.25\textwidth]{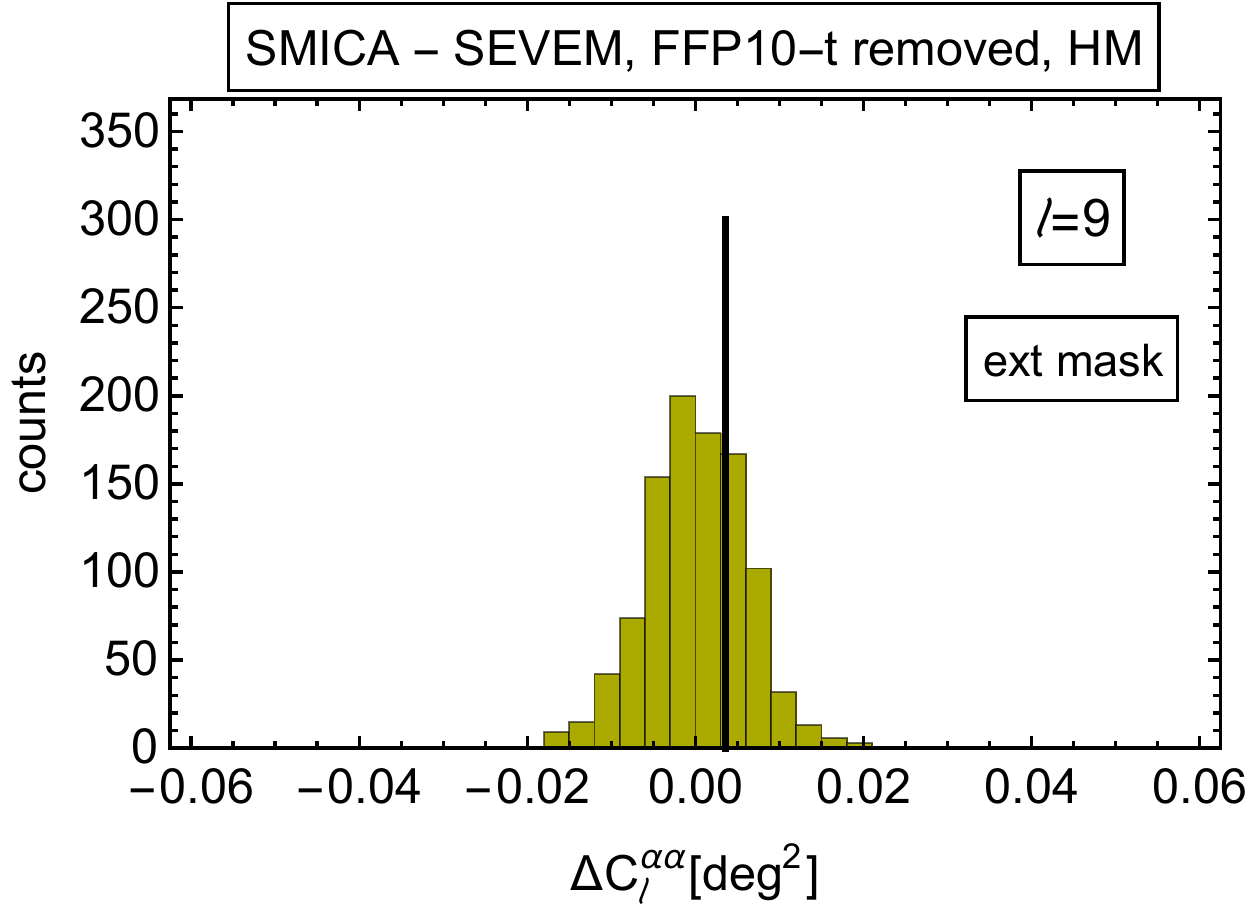}
\includegraphics[width=.25\textwidth]{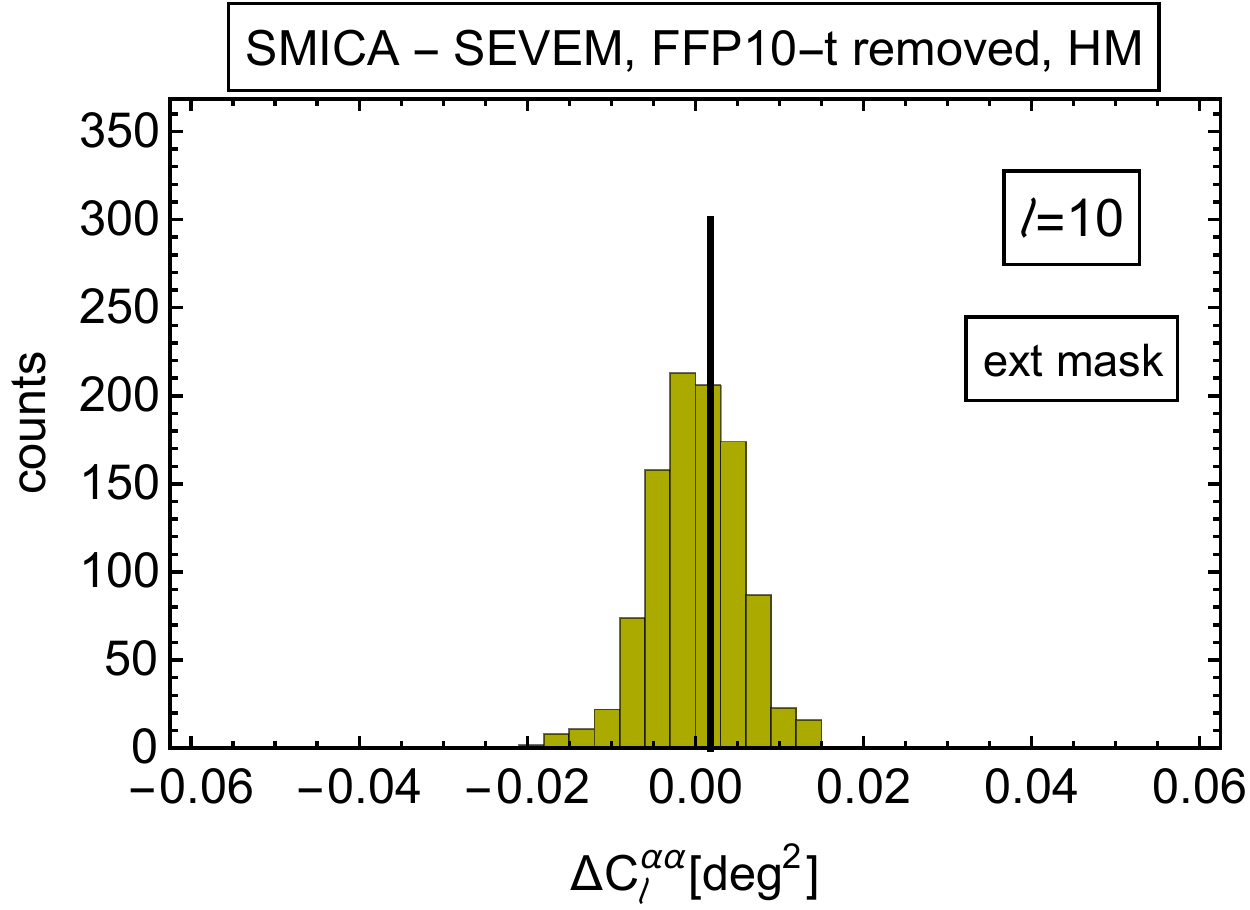}
\includegraphics[width=.25\textwidth]{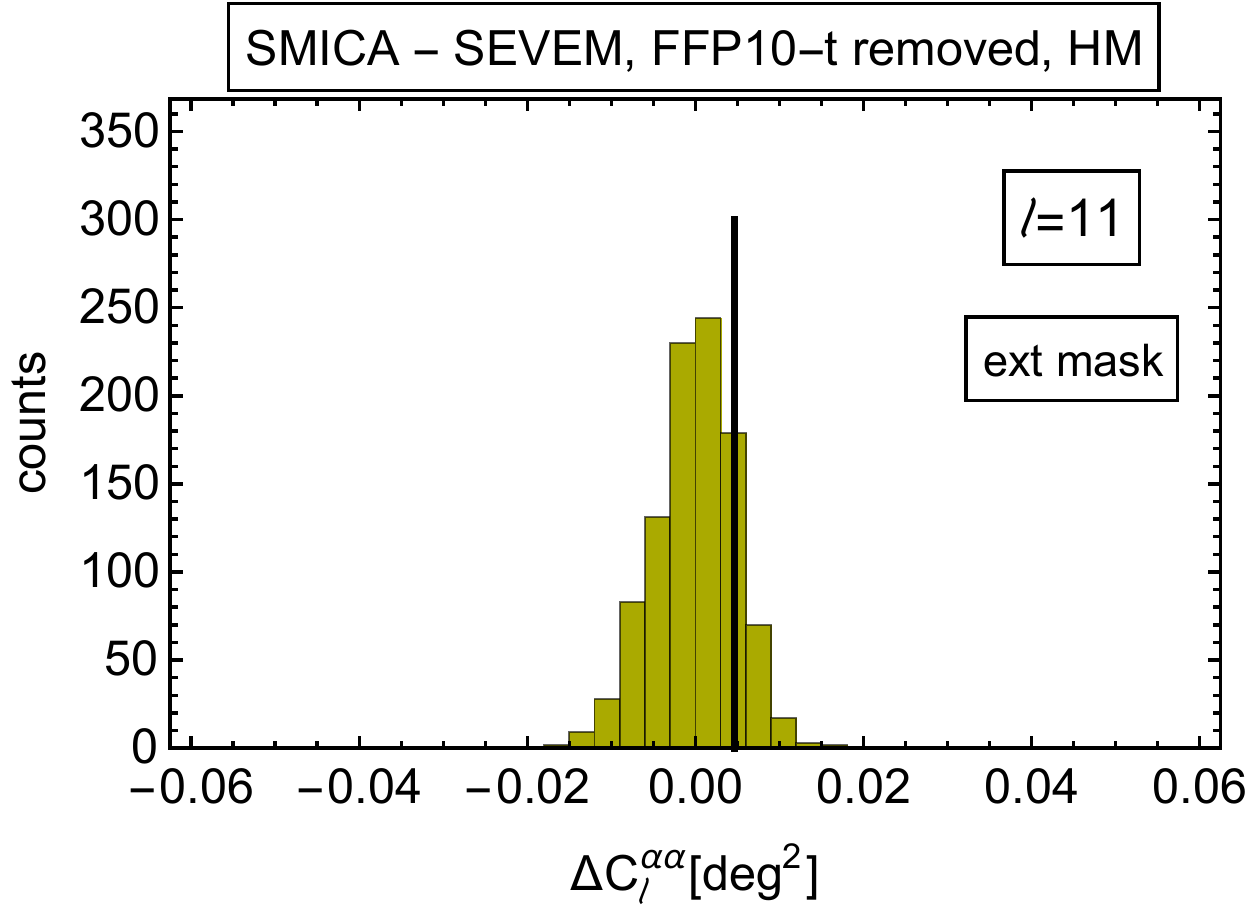}
\includegraphics[width=.25\textwidth]{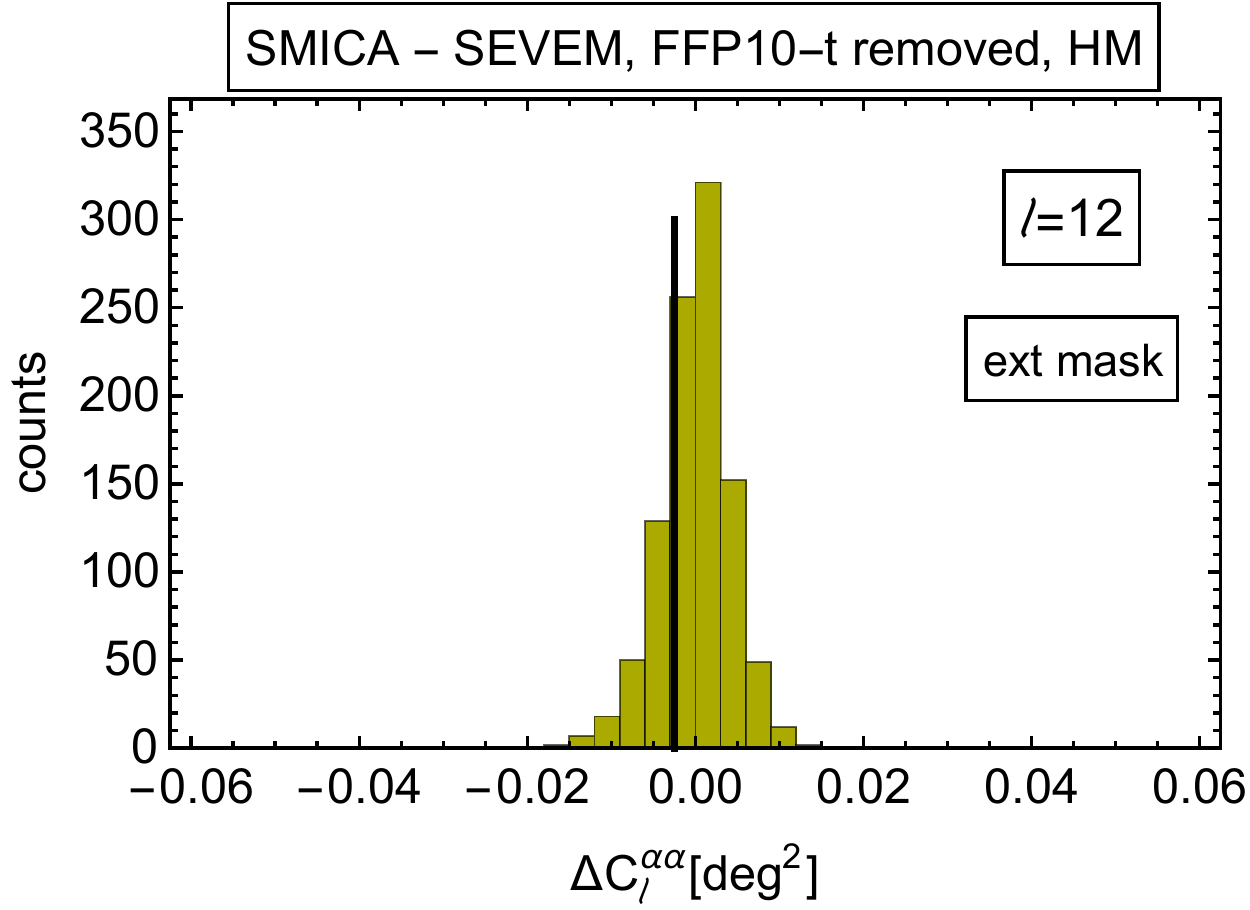}
\caption{Difference of $C^{\alpha \alpha}_{\ell}$ as computed from \smica\ and \sevem\ for each realisation of the FFP10 simulations (histograms) and for the data (vertical bars), see Eq.~(\ref{nulltest1}). 
Each panel refers to a single multipole from $\ell=1$ to $\ell=12$. In this case the FFP10 template has been taken into account and removed. The employed mask is the extended one and the data split is HM.}
\label{fig:nulltestextmasksmicasevem}
\end{figure}

First of all, the uncertainties obtained by the TB pipeline are much larger, as already mentioned in Section \ref{Description}, because of the cosmic variance induced by the CMB temperature fluctuations. 
Therefore we focus here on the EB-based pipeline. 
We start fixing the data split and evaluate all the possibile differences, $\Delta^{X}_{Y} C^{\alpha \alpha}_{\ell}$, defined as
\be
\Delta^{X}_{Y} C^{\alpha \alpha}_{\ell} = C^{\alpha \alpha, (X)}_{\ell} - C^{\alpha \alpha, (Y)}_{\ell} 
\, ,
\label{nulltest1}
\ee
for $\ell=1,...,12$, where $X$ and $Y$ refer to the component separation methods \commander, \nilc, \sevem\ and \smica\ for the HM split and \commander\ and \smica\ for the OE split.
Eq.~(\ref{nulltest1}) is computed 
for each realisation\footnote{For each realisation (and for the data) this difference cancels common fluctuations that, correctly, are not counted in the statistical budget.}, and the set obtained for each given $\ell$ is employed to build the histogram representing the empirical probability distribution function at that multipole. Such distribution is then compared to the same quantity computed on data.
Considering the standard mask, it is interesting to note that for the HM split, we obtain a very good consistency among \commander, \nilc\ and \smica\ for all multipoles, while when we introduce \sevem\ we begin experiencing anomalous percentages or outliers. In particular, \smica\ and \sevem\ turn out to have 5 estimates out of 12, which are at 
2.4$\sigma$ or more.
Under the assumption that each of the $\Delta^{X}_{Y} C^{\alpha \alpha}_{\ell}$ can be treated as an independent Gaussian realisation\footnote{We checked that the inverse of the Fisher matrix of the $C^{\alpha \alpha}_{\ell}$ is dominated by its diagonal part.} we have computed that for 12 independent Gaussian realisations having five of them at least at 2.4$\sigma$ is a realisation which happens very rarely (with a probabilty around $0.0003\%$). Similarly \commander\ and \sevem\ have 4 estimates out of 12 which are at 2.26$\sigma$ or more: this means that they are consistent at the level of $0.015\%$. Finally for \nilc\ and \sevem\ we find 4 estimates out of 12 at 2.51$\sigma$ or more and this translates into a probability of the order of $0.0011\%$. These percentages are very stable with respect to the removal of the FFP10 template. As for the HM split, we find that also for the OE split \commander\ and \smica\ are very well consistent with each other.
When we take into account the extended mask, we find that \sevem\ turns out to be in agreement with all the other component separation methods and that among all the multipoles, the dipole is the one less compatible but still with acceptable percentages. Specifically, comparing \smica\ and \sevem, see Figure \ref{fig:nulltestextmasksmicasevem}, we have only one multipole (the dipole) out of the 12 which is at more than 2.41$\sigma$: this is a very common situation which happens $16.02\%$ of realisations. In addition we compute that the dipole is at 2.46$\sigma$ and the probability to have one multipole out of 12 at that distance or more is $14.4\%$.
When we compare \commander\ and \sevem\ now we have only one multipole (again the dipole) at more than 2.26$\sigma$: as before, this is a likely situation corresponding to $22.1\%$ of realisations. In this case the dipole is computed to be at $3.1\sigma$ and the probability to find one multipole out of 12 at that fluctuation is still large as $2.3\%$.
Finally, when we compare \nilc\ and \sevem\ we do not have multipoles at more than $2.51\sigma$ which corresponds to $86.4\%$. Here, the dipole (which is the less compatible multipole) is at $2.12\sigma$ and having one multipole out of 12 at that distance or more, corresponds to $27.9\%$ of time.
All the other combinations are still well compatible for each multipole and independently of the kind of data split or whether we remove or not the FFP10 template.
In conclusion the null test on different component separation methods for fixed data split, clearly suggests the use of the extended mask.

We perform now a second null test which is formally similar to that given in Eq.~(\ref{nulltest1}) but now the difference is taken for a fixed method of component separation with different data split.
Specifically in this case we have considered
\be
\Delta_ {DS} C^{\alpha \alpha}_{\ell} = \frac{1}{2}\left( C^{\alpha \alpha, (X,HM)}_{\ell} - C^{\alpha \alpha, (X,OE)}_{\ell} \right)
\, ,
\label{nulltest2}
\ee
where $X$ refer to \commander\ and \smica. As before, also for this null test, Eq.~(\ref{nulltest2}) is computed for the same realisation of the FFP10 simulations and the set of the obtained $\Delta_ {DS} C^{\alpha \alpha}_{\ell}$ are used to build the histogram representing the expected probability distribution function, which is then confronted with the data.
We start taking into account the estimates obtained with the standard mask. We notice that both \commander\ and \smica\ have the estimates for $\ell=4, 6, 8$ and $10$ falling in the tails of the corresponding histograms with very similar pattern. We evaluate the total probability to exceed (PTE) considering the sum over all the considered multipoles of the absolute value of Eq.~(\ref{nulltest2}). 
We find a total PTE of 
$0.6\%$ for \commander\ and 
$0.5\%$ \smica. These percentages are obtained removing the FFP10 template, but even without removing it, their PTEs are very stable.
When we consider the extended mask, we have a total PTE of 
$6.6\%$ for \commander\ and 
$14.9\%$ for \smica.
We note that the multipole which provides the lowest compatibility is $\ell=8$ both for \commander\ and \smica. 

Based on the above considerations, we conservatively choose for our final constraints the case of the EB pipeline for the harmonic range $[51-1511]$ and the extended mask.
This is what we call baseline configuration.
Moreover, because of a better performance of \smica\ in the second of the aforementioned null tests, we decide to use the results from this method as representative for the present paper.
 
\section{Results}
\label{results}

In this Section we provide the constraints on the birefringence effect and its cross-correlation with the CMB temperature field from CMB 2018 \planck\ data for all the component separation methods in the HM data split mode.
In addition, we provide the results for the OE split of \commander\ and \smica. As already stated in Section \ref{dataset}, the latter are delivered only for consistency check: 
they are not used for our final results because of a better performance of the HM data split in terms of null tests, see \cite{Aghanim:2019ame}.
Moreover, our results are given considering the baseline configuration, defined in Section \ref{analysis} above. However, for stability tests, we also provide for alternative settings.

\subsection{APS of $\alpha$-anisotropies at large scales}

\begin{figure}[t]
\centering
\includegraphics[width=.49\textwidth]{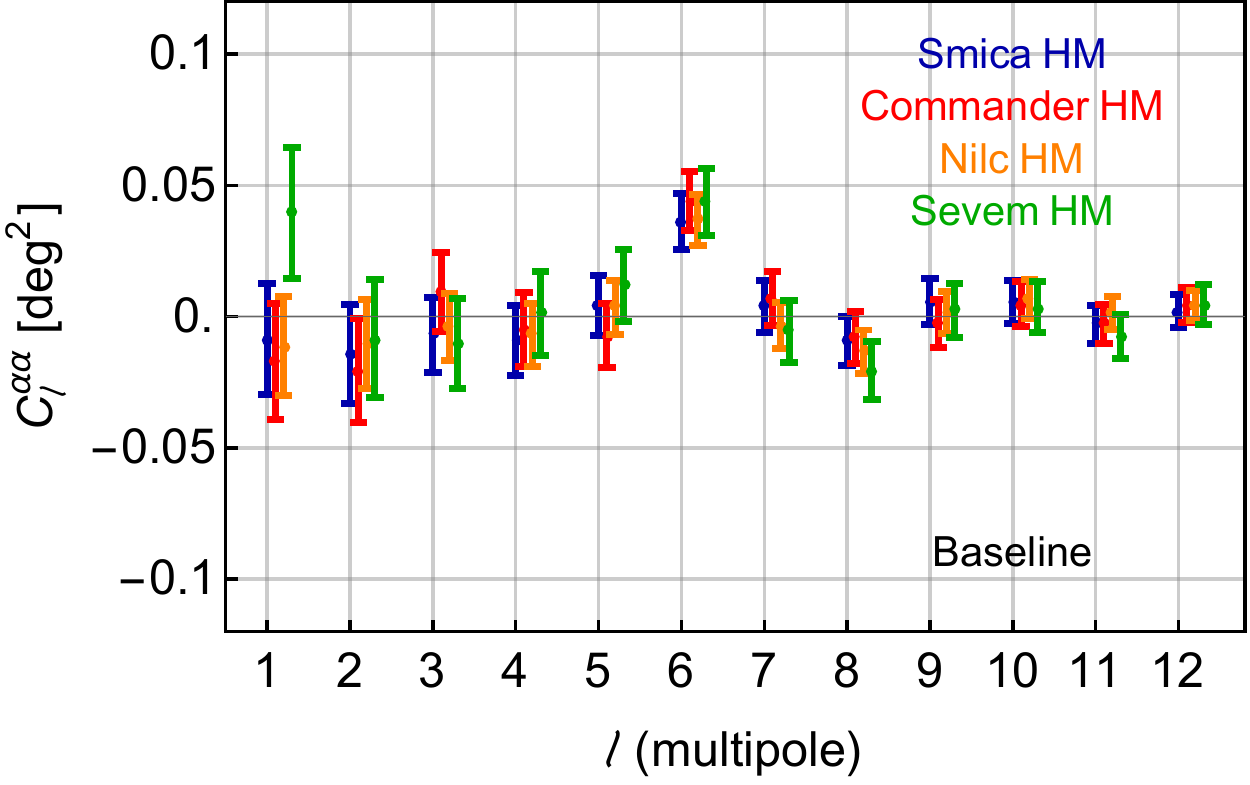}
\includegraphics[width=.49\textwidth]{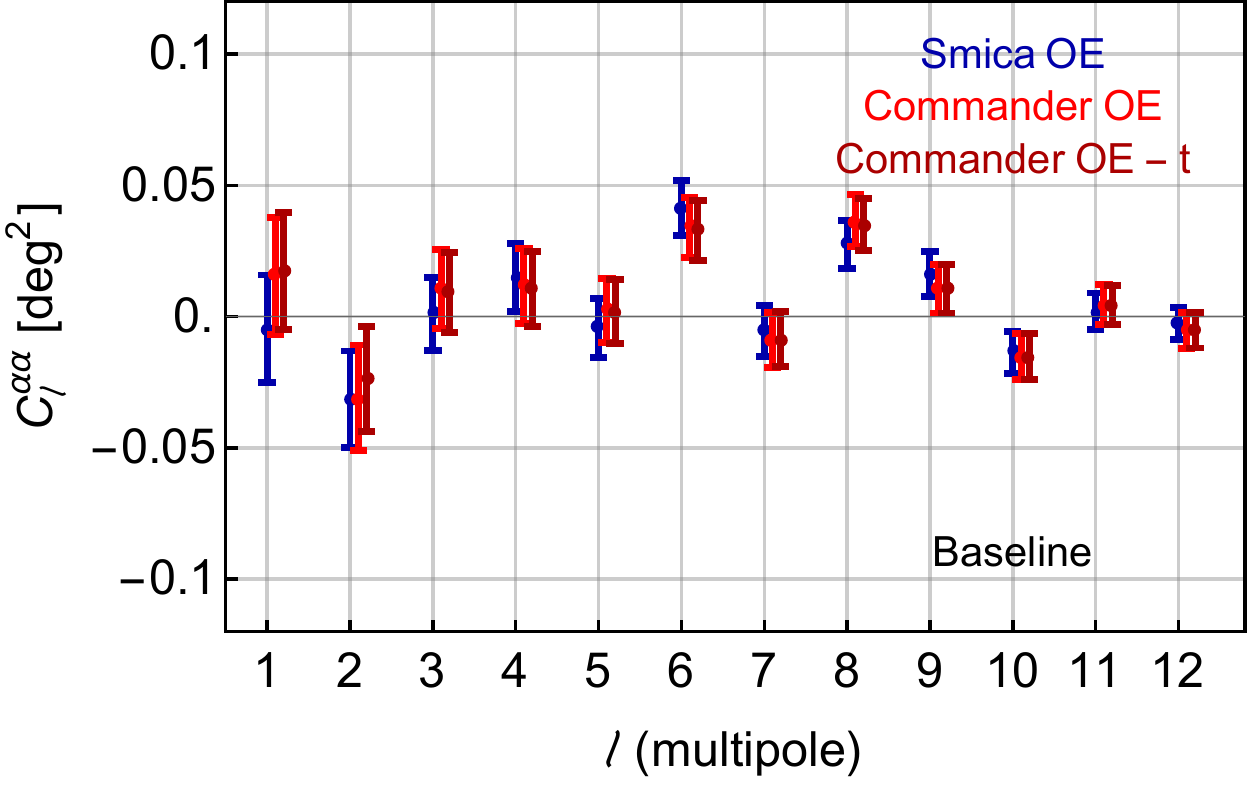}
\caption{APS of the birefringence angle at low multipoles. These estimates have been obtained through the EB pipeline for $\ell \in [51-1511]$ and the extended mask, i.e. the baseline configuration. 
Left panel is for the HM split and right panel for the OE split.
In the right panel, the label ``-t'' stands for the removal of the FFP10 template.
}
\label{fig:spectrabirefringencewithtoterror}
\end{figure}

The APS of the birefringence angle at low multipoles is shown in Figure \ref{fig:spectrabirefringencewithtoterror} for the baseline configuration. 
The left panel of Figure \ref{fig:spectrabirefringencewithtoterror} shows the spectra for the HM split of \commander, \nilc, \sevem\ and \smica\ whereas the right panel displays the spectra for the OE split of \commander\ and \smica. 
For comparison we show also \commander\ OE with the removal of the FFP10 template: the impact of this template is a small fraction of the statistical uncertainty for $\ell=2$ and negligible in practice for all the other multipoles.
The error bars shown in Figure \ref{fig:spectrabirefringencewithtoterror} are those given by the Fisher matrix of the QML estimator and represent the statistical fluctuations of the FFP10 MC set.

\begin{figure}[t]
\centering
\includegraphics[width=.49\textwidth]{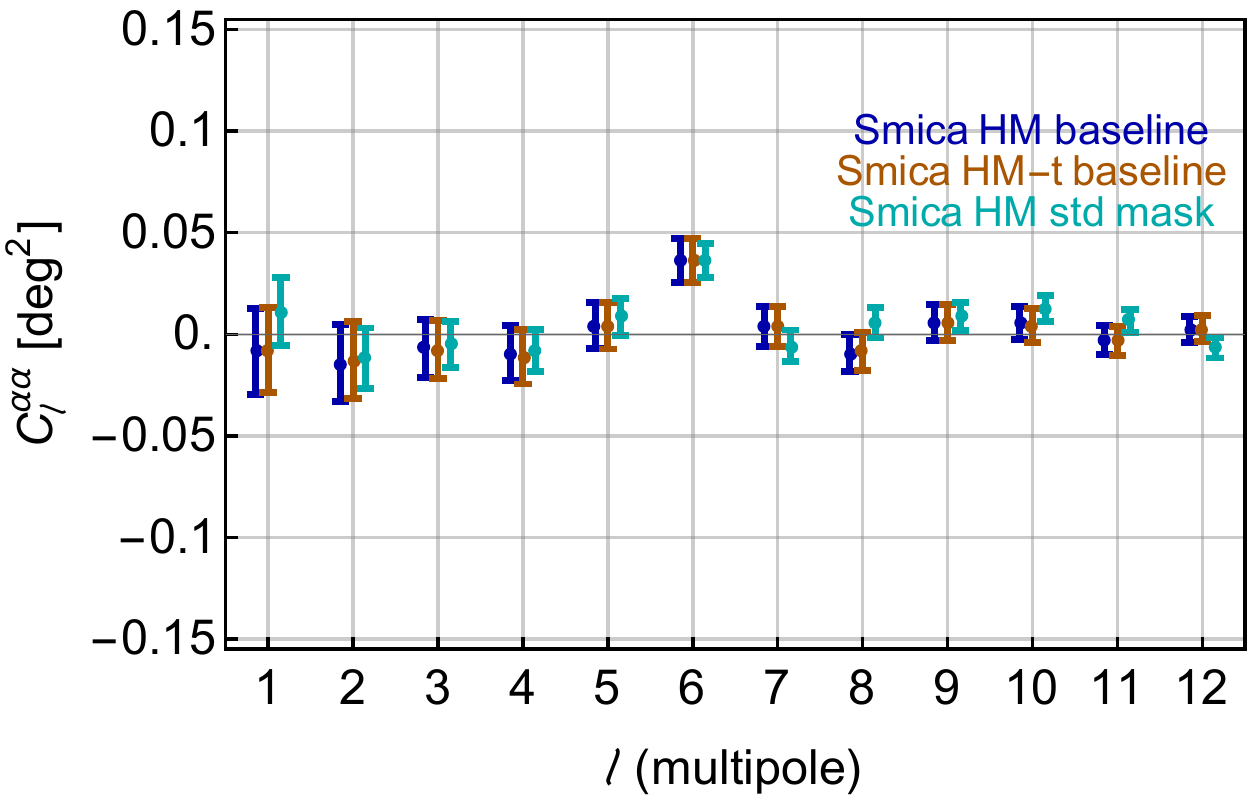}
\includegraphics[width=.49\textwidth]{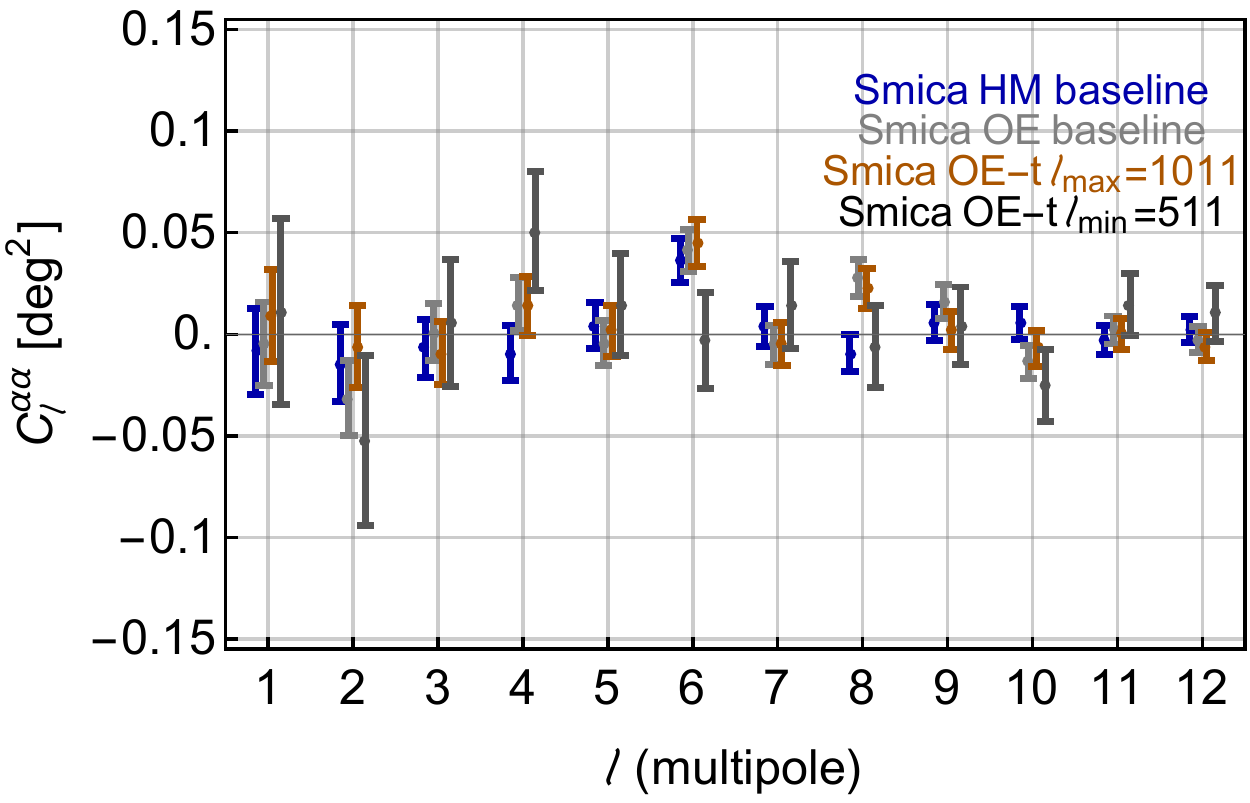}
\includegraphics[width=.49\textwidth]{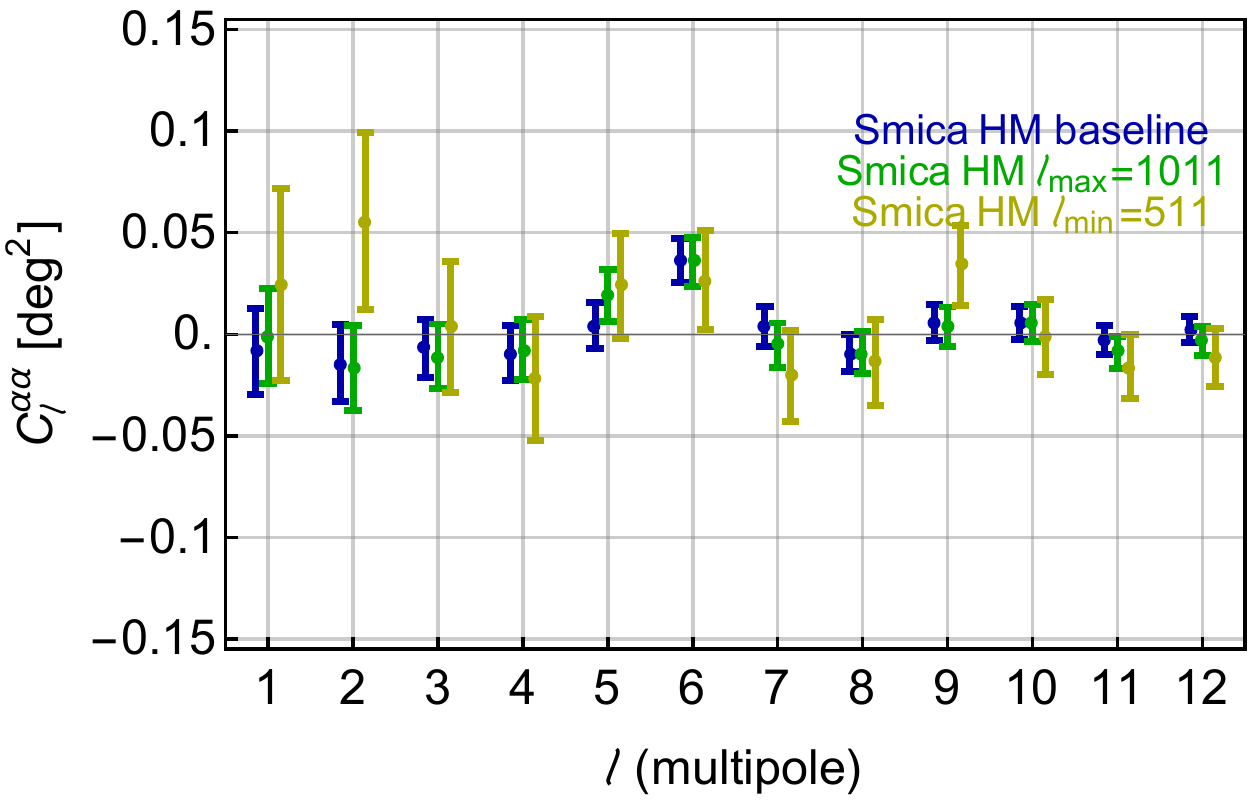}
\includegraphics[width=.49\textwidth]{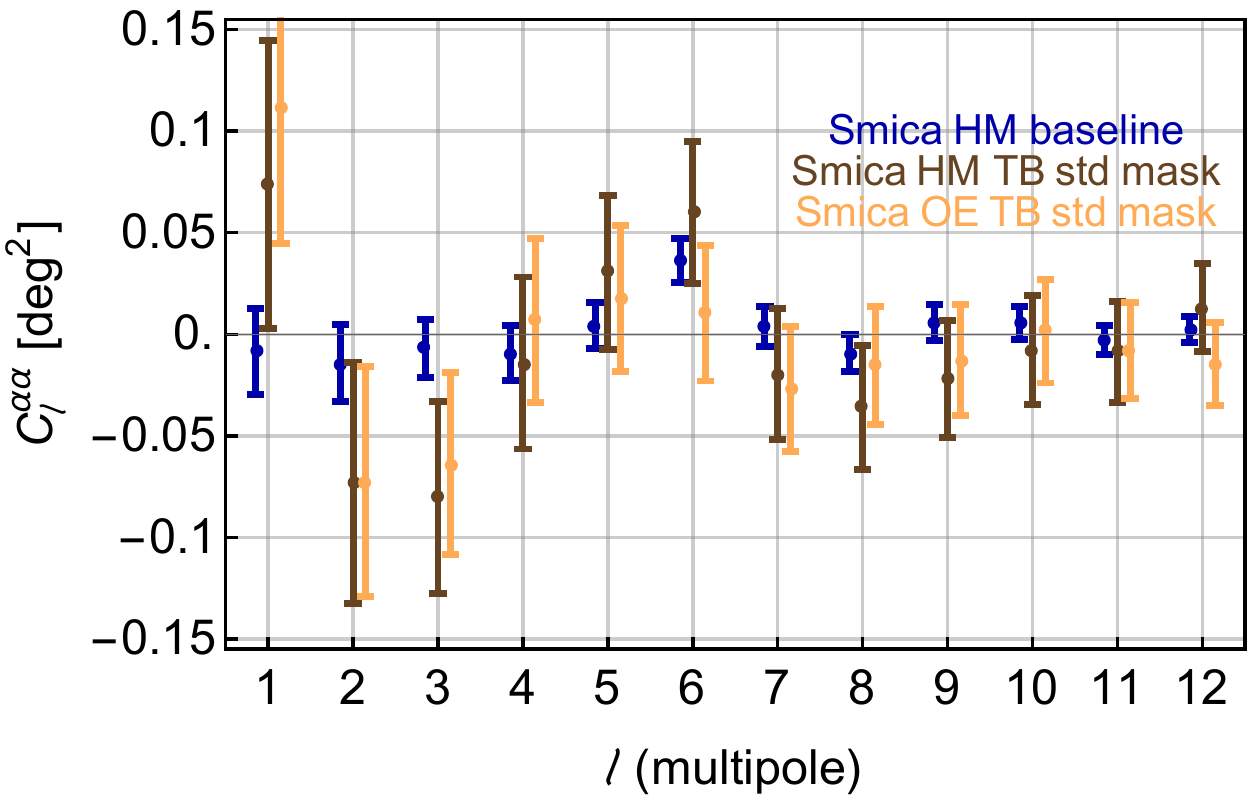}
\caption{Robustness of the APS of the birefringence angle at low multipoles as obtained from \smica\ HM in the baseline configuration.
The stability is studied against different sky fractions and the FFP10 template removal (upper left panel), against the OE data split in different multipole ranges (upper right panel), 
against different multipole ranges (lower left panel) and against the TB-based pipeline (lower right panel).}
\label{fig:spectrabirefringencewithtoterrortest}
\end{figure}

The robustness of our results for \smica\ HM baseline are provided in Figure \ref{fig:spectrabirefringencewithtoterrortest}. 
In the upper left panel of Figure \ref{fig:spectrabirefringencewithtoterrortest} we show the APS of the anisotropic birefringence for different sky fractions and considering the removal of the FFP10 template (labeled with ``-t'' in the plot): again, as before in the right panel of Figure \ref{fig:spectrabirefringencewithtoterror} for \commander\ OE, also here the impact of this template turns out to be in practice negligible. 
The stability of the \smica\ HM baseline versus the multipoles ranges considered for the CMB is given in lower left panel of Figure \ref{fig:spectrabirefringencewithtoterrortest}: the consistency appears again very good.
The stability of the \smica\ HM baseline appears good for all multipoles when compared to \smica\ OE (again in baseline configuration). However, for $\ell=8$ the \smica\ HM and OE baselines appear at $\sim$3$\sigma$, see upper right panel of Figure \ref{fig:spectrabirefringencewithtoterrortest}.
This discrepancy at $\ell=8$ seems to be due to the first 500 multipoles of the OE split: when we remove them, the estimate is again compatible with the corresponding HM split and with zero.
In the lower right panel of Figure \ref{fig:spectrabirefringencewithtoterrortest} we compare the \smica\ baseline versus the pipeline based on the TB spectrum:
this quantifies the different statistical efficiency of the TB and EB based pipelines. 

\begin{figure}[t]
\centering
\includegraphics[width=.49\textwidth]{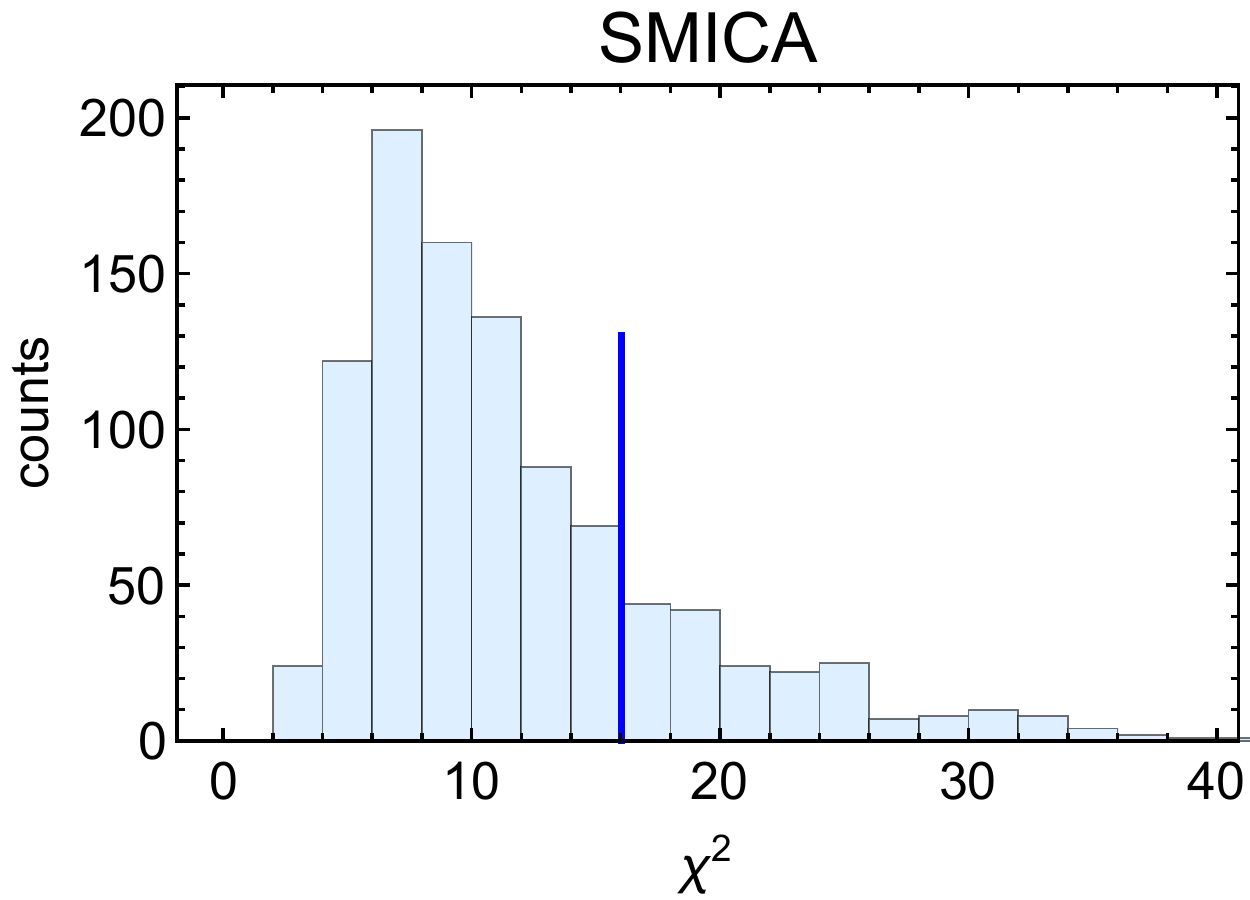}
\includegraphics[width=.49\textwidth]{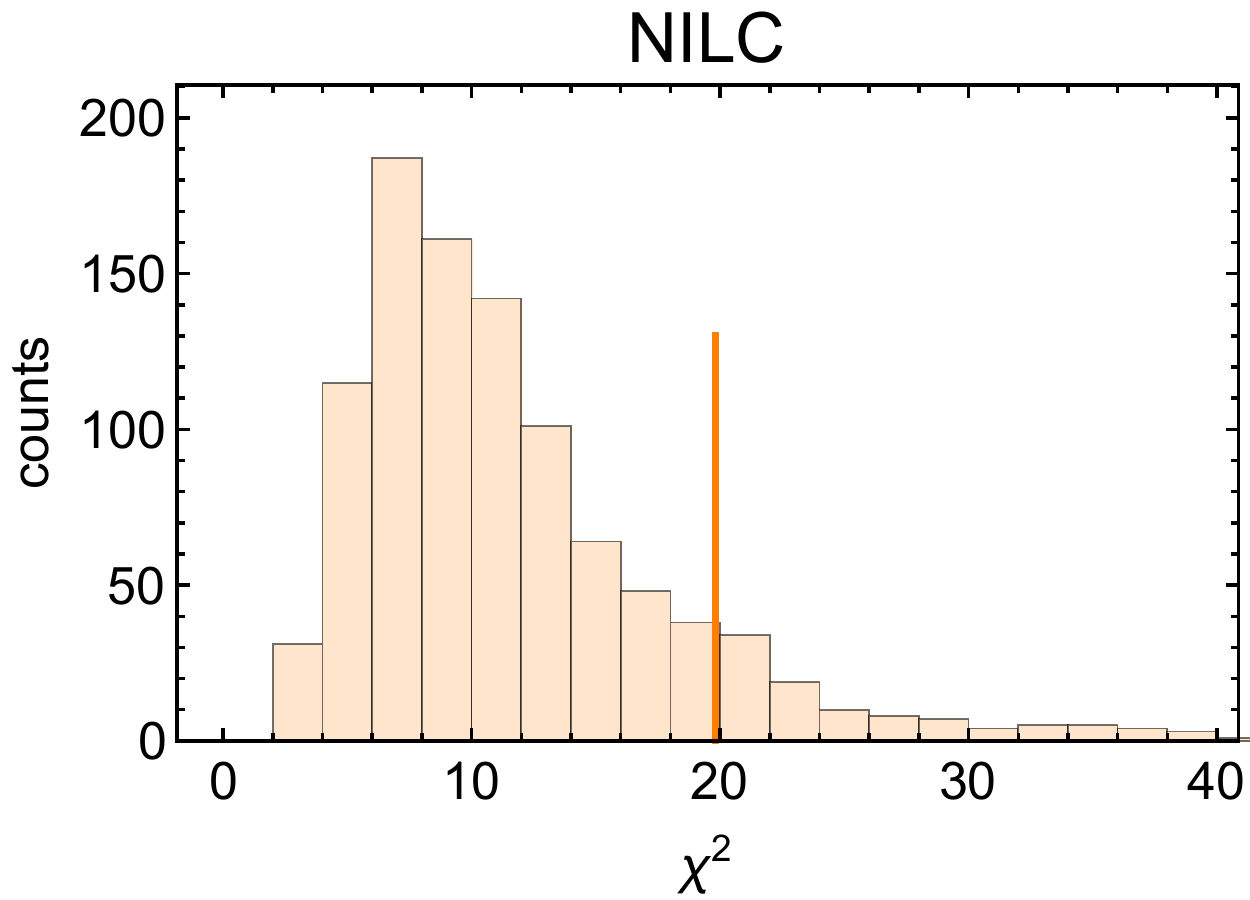}
\includegraphics[width=.49\textwidth]{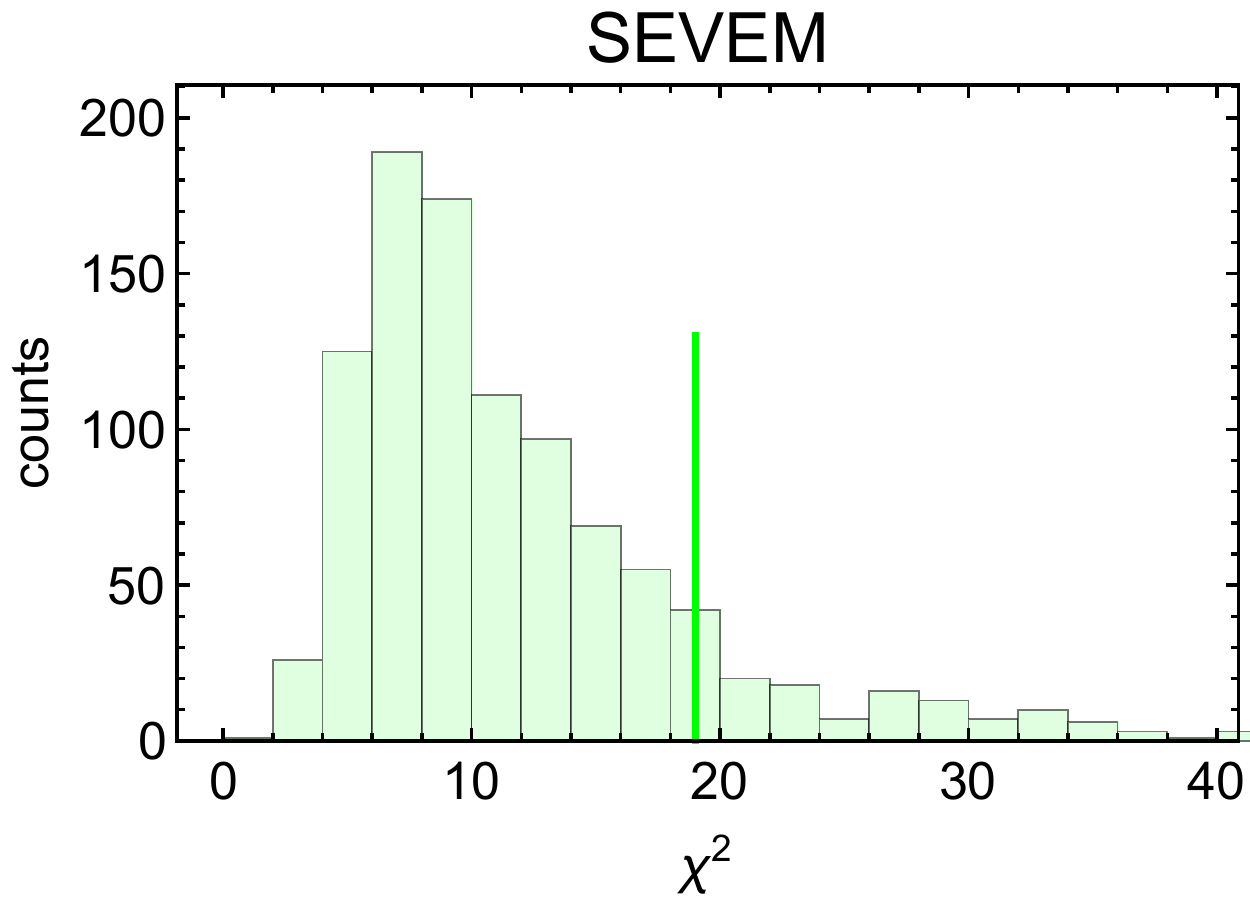}
\includegraphics[width=.49\textwidth]{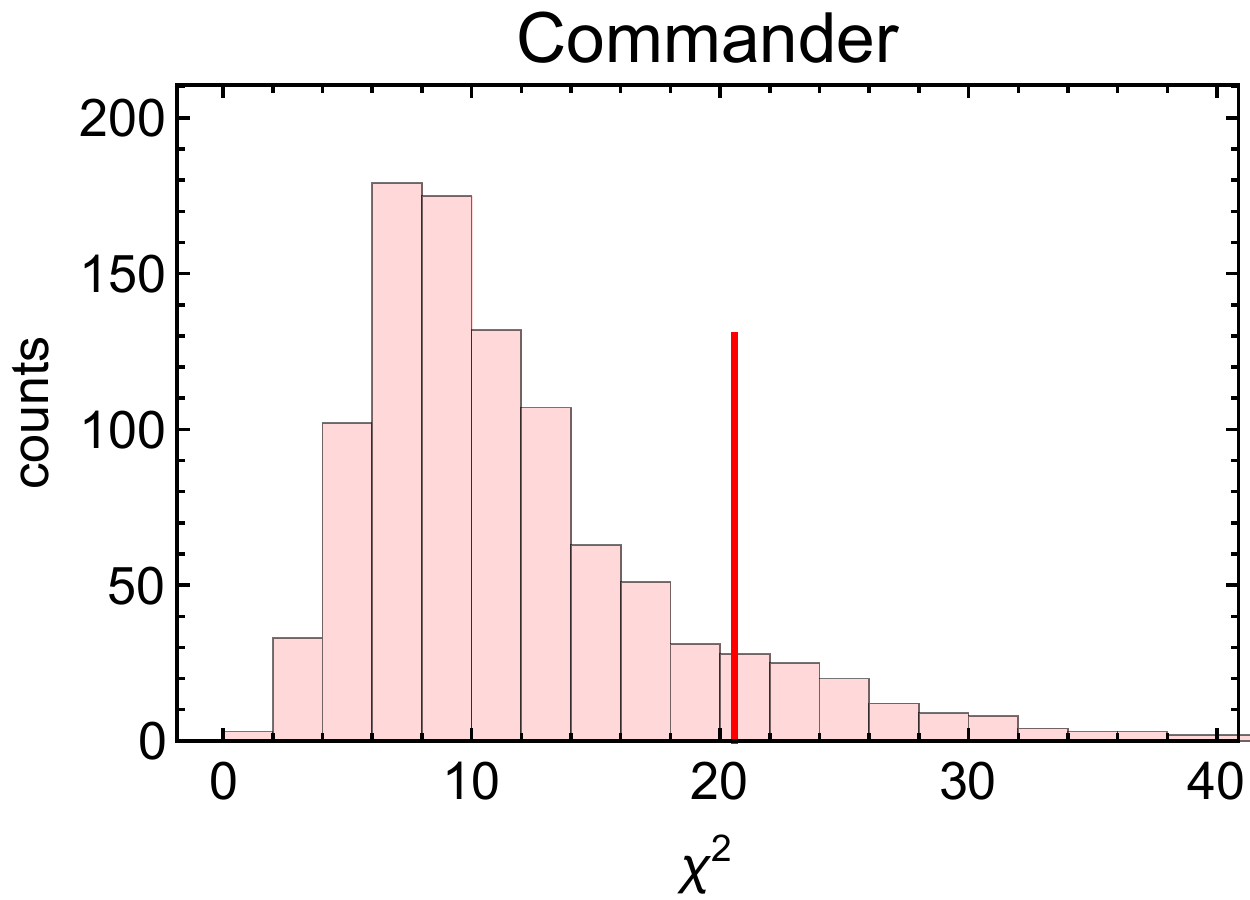}
\caption{Histogram of the harmonic $\chi^2$, defined in Eq.~(\ref{chiquadroarmonico}), obtained from \smica\ (upper left), \nilc\ (upper right), \sevem\ (lower left), \commander\ (lower right) HM FFP10 simulations for the baseline configuration and considering the removal of the FFP10 template. In each panel the vertical bar stands for the corresponding value computed from HM observation.}
\label{fig:chiquadroarmonico}
\end{figure}

In Figure \ref{fig:chiquadroarmonico} we plot the histograms of the harmonic $\chi^2$ defined as
\begin{equation}
\chi^2 = \sum_{\ell,\ell^{\prime}=1}^{12} C_{\ell}^{\alpha} \langle C_{\ell}^{\alpha} C_{\ell^{\prime}}^{\alpha} \rangle^{-1} C_{\ell^{\prime}}^{\alpha} \, ,
\label{chiquadroarmonico}
\end{equation}
and obtained for our baseline with \smica, \nilc, \sevem\, \commander\ HM FFP10 simulations. In each panel of Figure \ref{fig:chiquadroarmonico} the vertical bars represent the corresponding values computed with the 
HM data.
The probability to exceed (PTE) is $20.1 \%$, $11.5 \%$, $12.8 \%$ and $11.5 \%$ for \smica, \nilc, \sevem\, \commander\ respectively. Therefore, Figure \ref{fig:chiquadroarmonico} already shows a good compatibility of the \planck\ 
2018 observations with null anisotropic birefringence at low multipoles, within the error budget traced by de-biased FFP10 simulations. Similar results are obtained without the use of the FFP10 template.
When we consider the baseline for the OE split we obtain smaller PTEs. More precisely we obtain $0.7\%$ for \commander\  and $0.8\%$ for \smica.
These percentages are mainly driven by $\ell=8$ for \commander\ and $\ell=6$ for \smica, as can be seen from the right panel of Figure \ref{fig:spectrabirefringencewithtoterror}. 
The PTEs become $6.9\%$ for \commander\ and $7.7\%$ for \smica\ when we remove from the analysis $\ell=8$ and $\ell=6$, respectively. 
 
\subsection{APS of the cross-correlation of $\alpha$-anisotropies and CMB temperature map}

In this section we focus on the the cross-correlation between CMB temperature and $\alpha$-anisotropy maps. 
This represents a further element of novelty of the present paper since to our knowledge this is the first evaluation of this cross-correlation at angular scales larger than $\sim 15$ degrees.
The maps of $\alpha(\hat n)$ have been obtained through the minimisation of $\chi_{EB}^2$ in the harmonic range $[51-1511]$ on HM FFP10 simulations as validated in Section \ref{validationcross}.
The $C_{\ell}^{\alpha T}$ are estimated using pixels out of the extended mask and considering the CMB anisotropy temperature map used in the \planck\ likelihood code at low multipoles \cite{Aghanim:2019ame} which is provided by the \commander\ algorithm once downsampled at $N_{side}=4$.

\subsubsection{$C^{\alpha T}_{\ell}$ on simulations}
\label{validationcross}

\begin{figure}[t]
\centering
\includegraphics[width=.49 \textwidth]{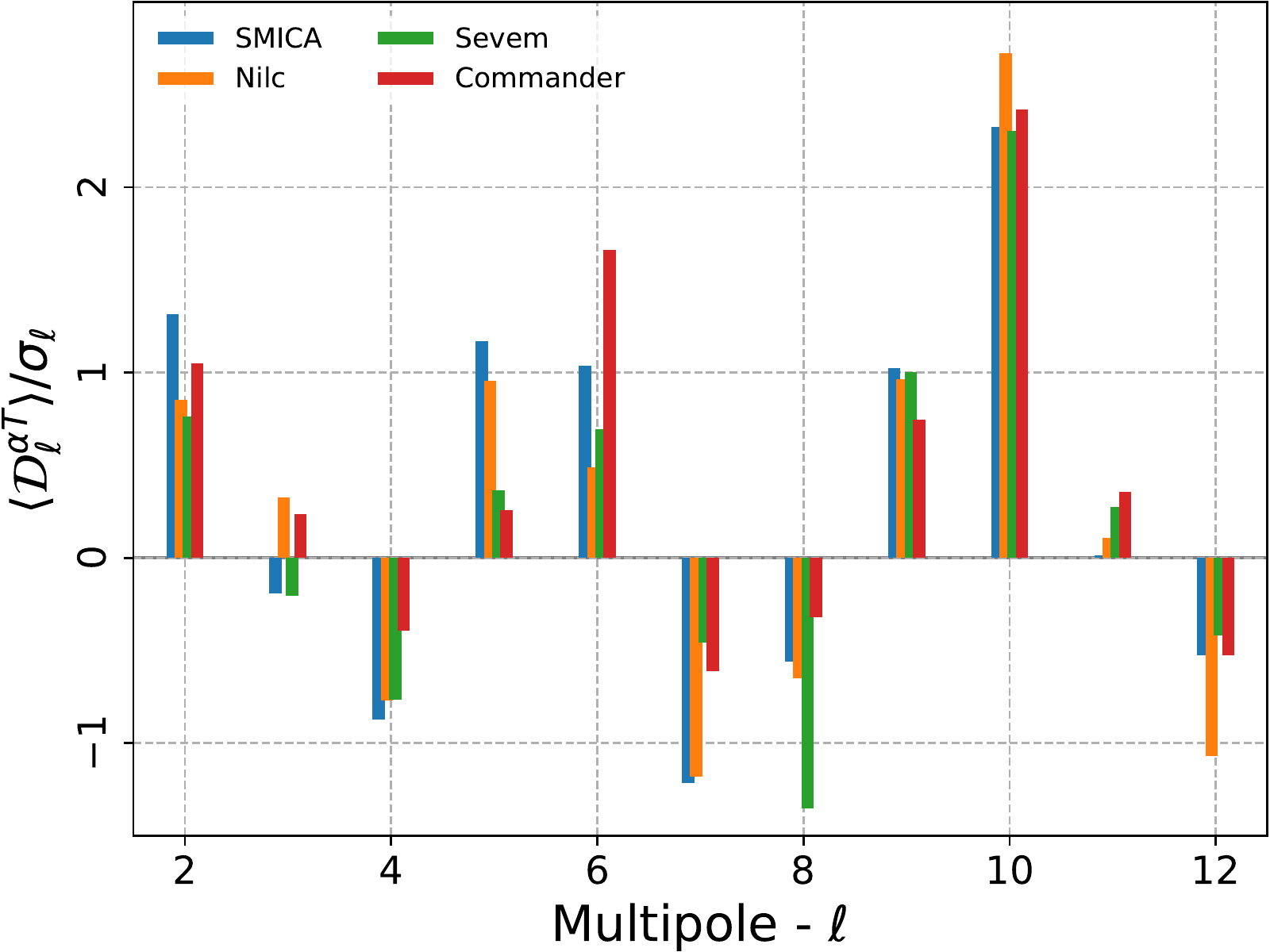}
\includegraphics[width=.49 \textwidth]{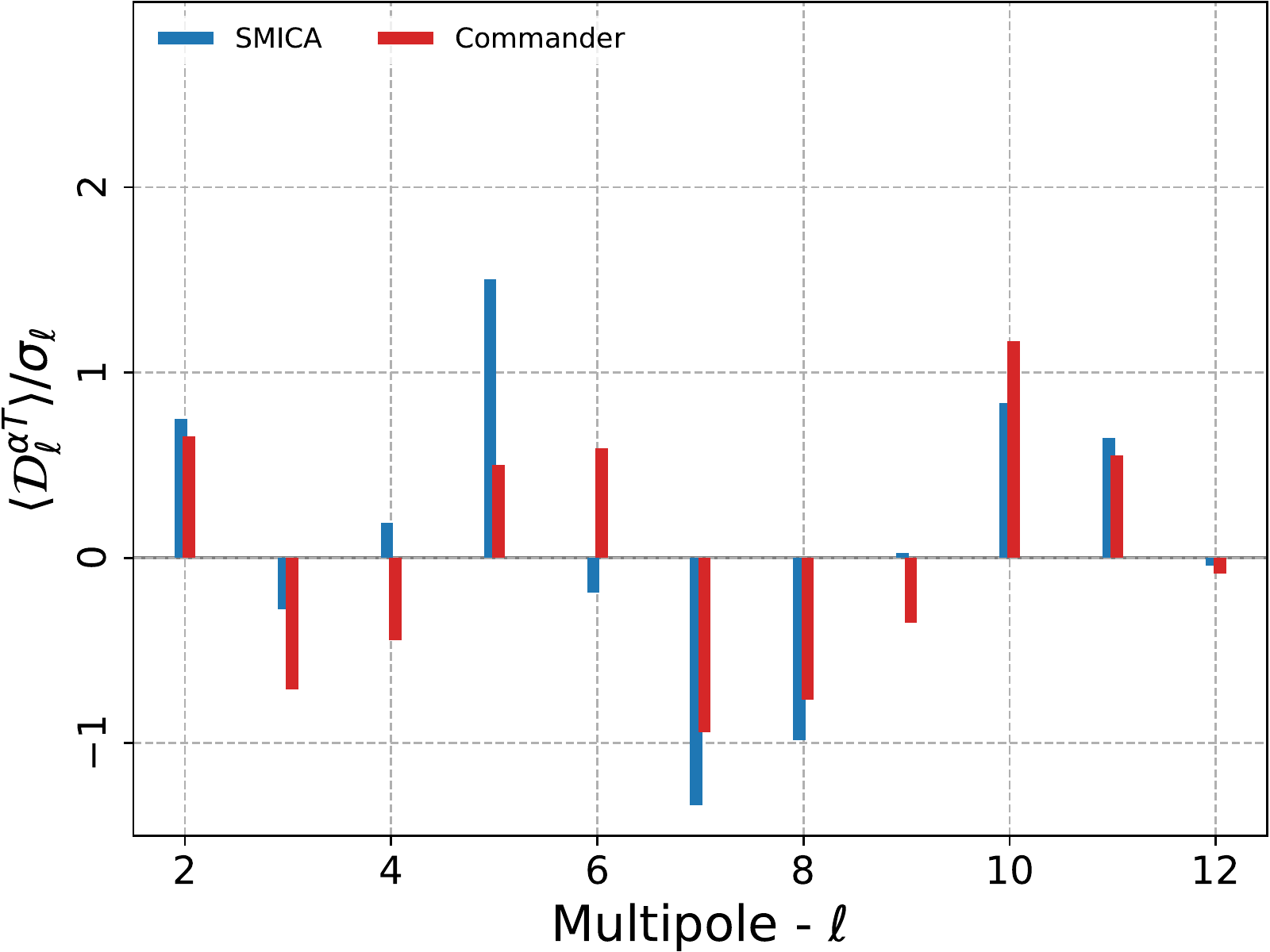}
\caption{Validation of the cross-correlation $C_{\ell}^{\alpha T}$ spectra. Deviation of the mean of $D_{\ell}^{\alpha T} \equiv \ell (\ell +1) C_{\ell}^{\alpha T}/(2 \pi)$ from null in units of standard deviation of the mean itself, $\sigma_{\ell}$ (y-axis), versus the multipole $\ell$ (x-axis). The maps $\alpha(\hat n)$ are built through the minimisation of $\chi_{EB}^2$ in the harmonic range $[51-1511]$ on the HM (left panel) and OE (right panel) FFP10 simulations. The $C_{\ell}^{\alpha T}$ are estimated using pixels out of the extended mask.
}
\label{fig:valisationCelltTalpha}
\end{figure}

We start considering simulations and take into account the maps of $\alpha(\hat n)$ built with the HM and OE splits of the FFP10 simulations through the minimisation of $\chi_{EB}^2$ in the harmonic range $[51-1511]$ without any de-bias of the FFP10 template. The CMB temperature maps have been randomly generated at low resolution using the cosmological parameters found by {\it Planck} \cite{Aghanim:2018eyx}.
We estimate the angular cross spectrum $C^{\alpha T}_{\ell}$ from $\ell=2$ to $\ell=12$ for all the component separation methods 
using the implementation of the cross-QML method already adopted in \cite{Pagano:2019tci}.  
In Figure \ref{fig:valisationCelltTalpha} we show the deviation of the mean of $D_{\ell}^{\alpha T} \equiv \ell (\ell +1) C_{\ell}^{\alpha T}/(2 \pi)$ from null in units of standard deviation of the mean itself, $\sigma_{\ell}$.
The fluctuations of the mean in the considered harmonic range, show amplitudes compatible with no effect as expected since the simulations do not contain the birefringence effect. 
This validates the procedure for the cross-angular power spectrum, $C^{\alpha T}_{\ell}$.

\subsubsection{$C^{\alpha T}_{\ell}$ on data}

\begin{figure}[t]
\centering
\includegraphics[width=.49 \textwidth]{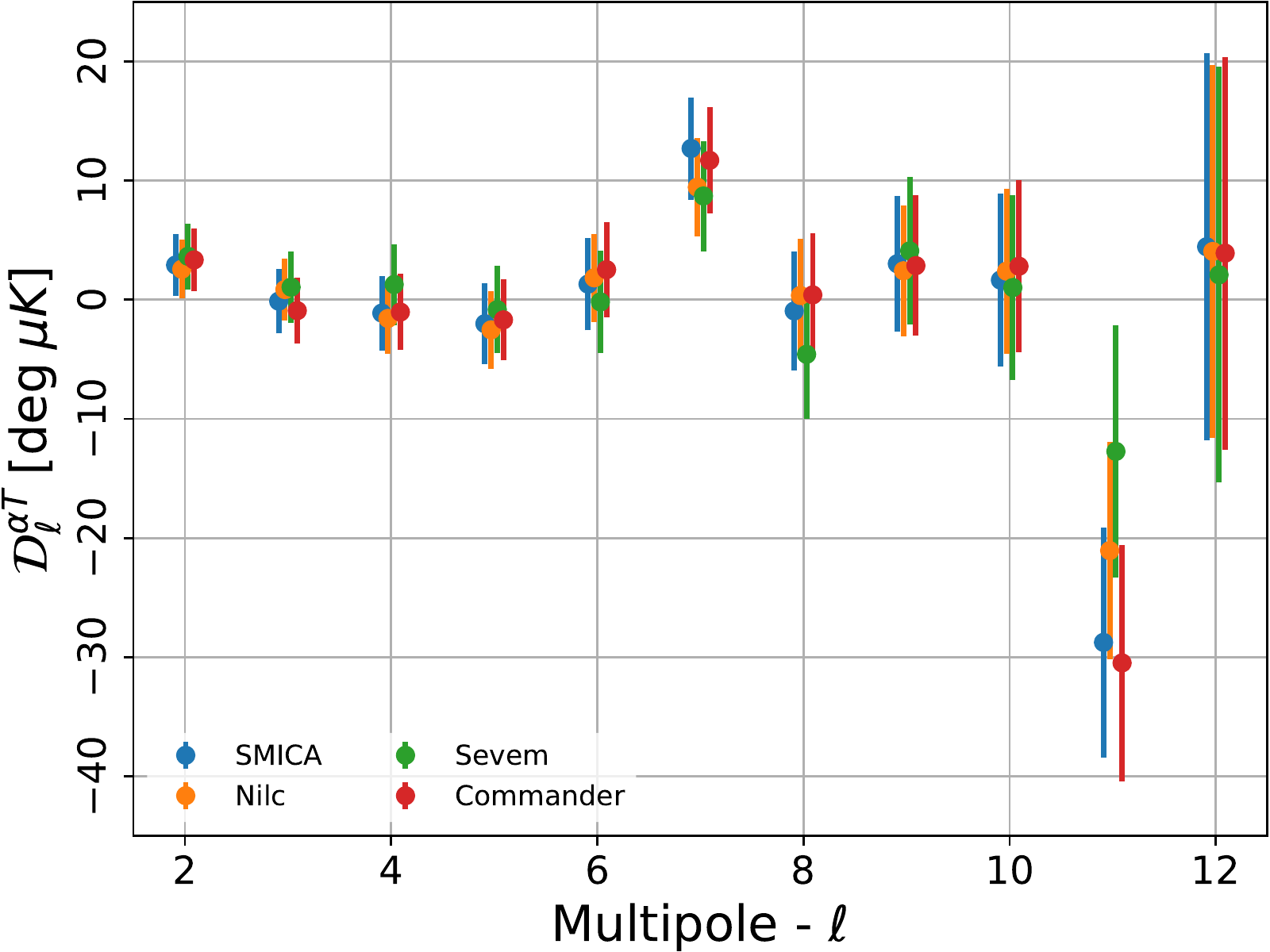}
\includegraphics[width=.49 \textwidth]{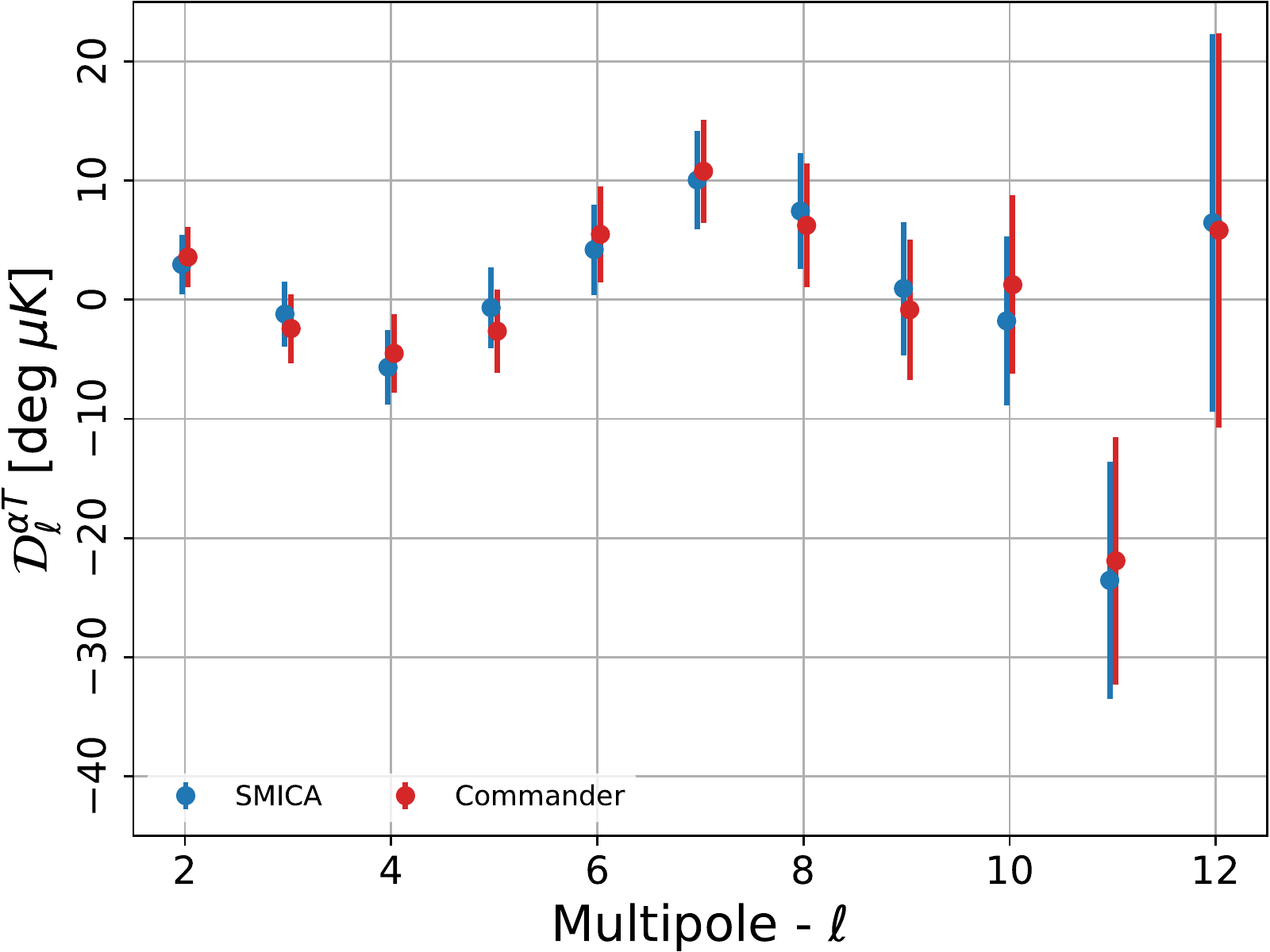}
\caption{Cross-correlation $C_{\ell}^{\alpha T}$ spectra for all the considered component separation methods as measured by {\it Planck} 2018 data expressed as $D_{\ell}^{\alpha T}$ (y-axis) versus the multipole $\ell$ (x-axis). 
Left panel for the HM split and right panel for the OE split.
}
\label{fig:CelltTalpha}
\end{figure}

\begin{figure}[t]
\centering
\includegraphics[width=.49 \textwidth]{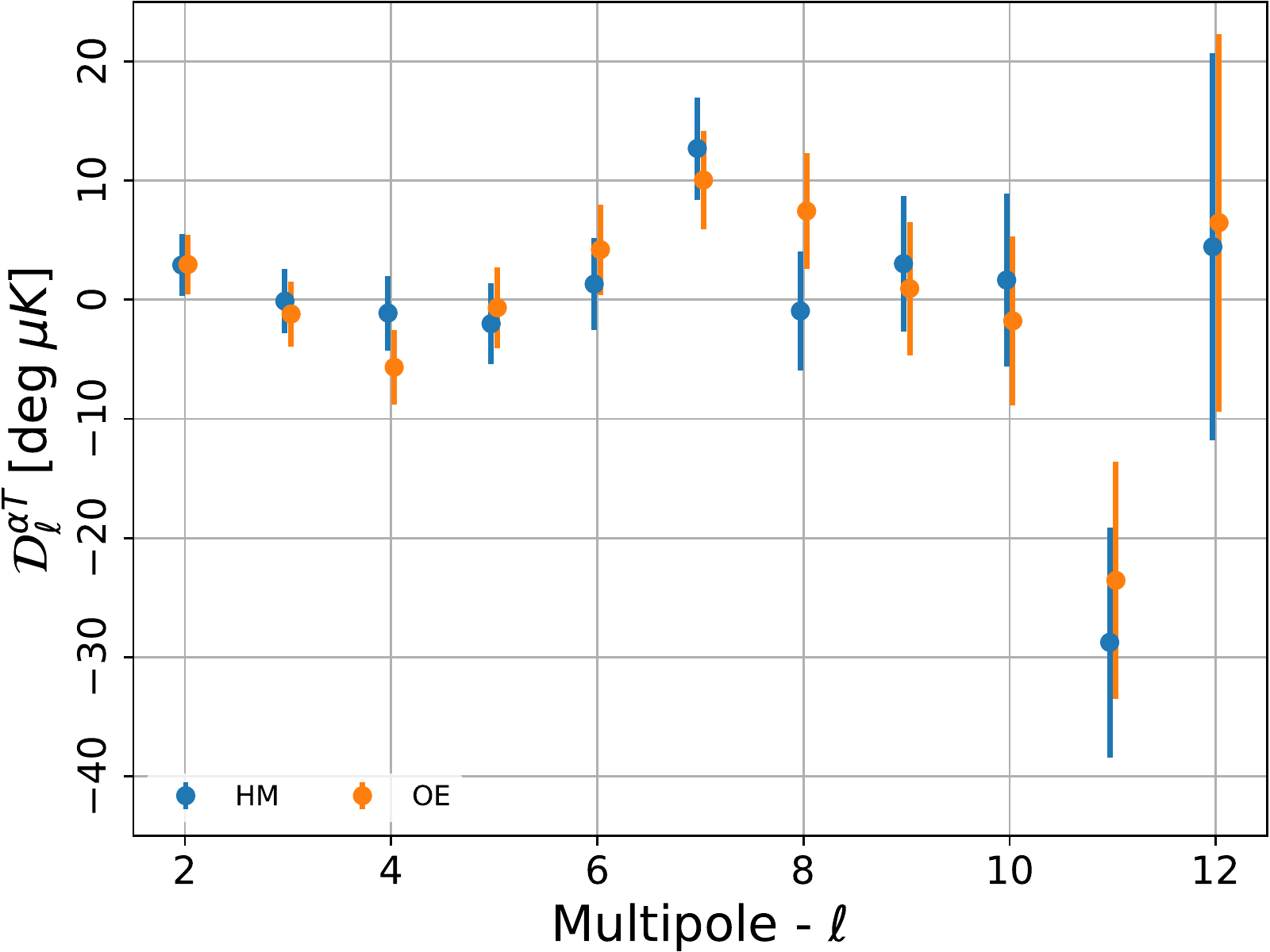}
\caption{Comparison of $C_{\ell}^{\alpha T}$ obtained from \smica\ HM and \smica\ OE data split, expressed as $D_{\ell}^{\alpha T}$ (y-axis) versus the multipole $\ell$ (x-axis).
}
\label{fig:CelltTalpharobustness_SMICA_HM_OE}
\end{figure}

In Figure \ref{fig:CelltTalpha} we show the $C^{\alpha T}_{\ell}$ for the HM pipeline (left panel) and for the OE pipeline (right panel). In each panel all the component separation methods available are displayed.  
Figure \ref{fig:CelltTalpha} shows that all the HM estimates are consistent almost always within 1$\sigma$ C.L. and the same can be said for the estimates based on OE pipeline.
Figure \ref{fig:CelltTalpharobustness_SMICA_HM_OE} provides a direct comparison of the \smica\ estimates obtained through the HM and the OE pipeline: the agreement between the two sets of estimates is within 1$\sigma$ C.L.
but for $\ell=4$ and $\ell=8$ where it is within 2$\sigma$ C.L.. Similar agreement is found for the \commander\ HM and OE estimates.

\begin{figure}[t]
\centering
\includegraphics[width=.86 \textwidth]{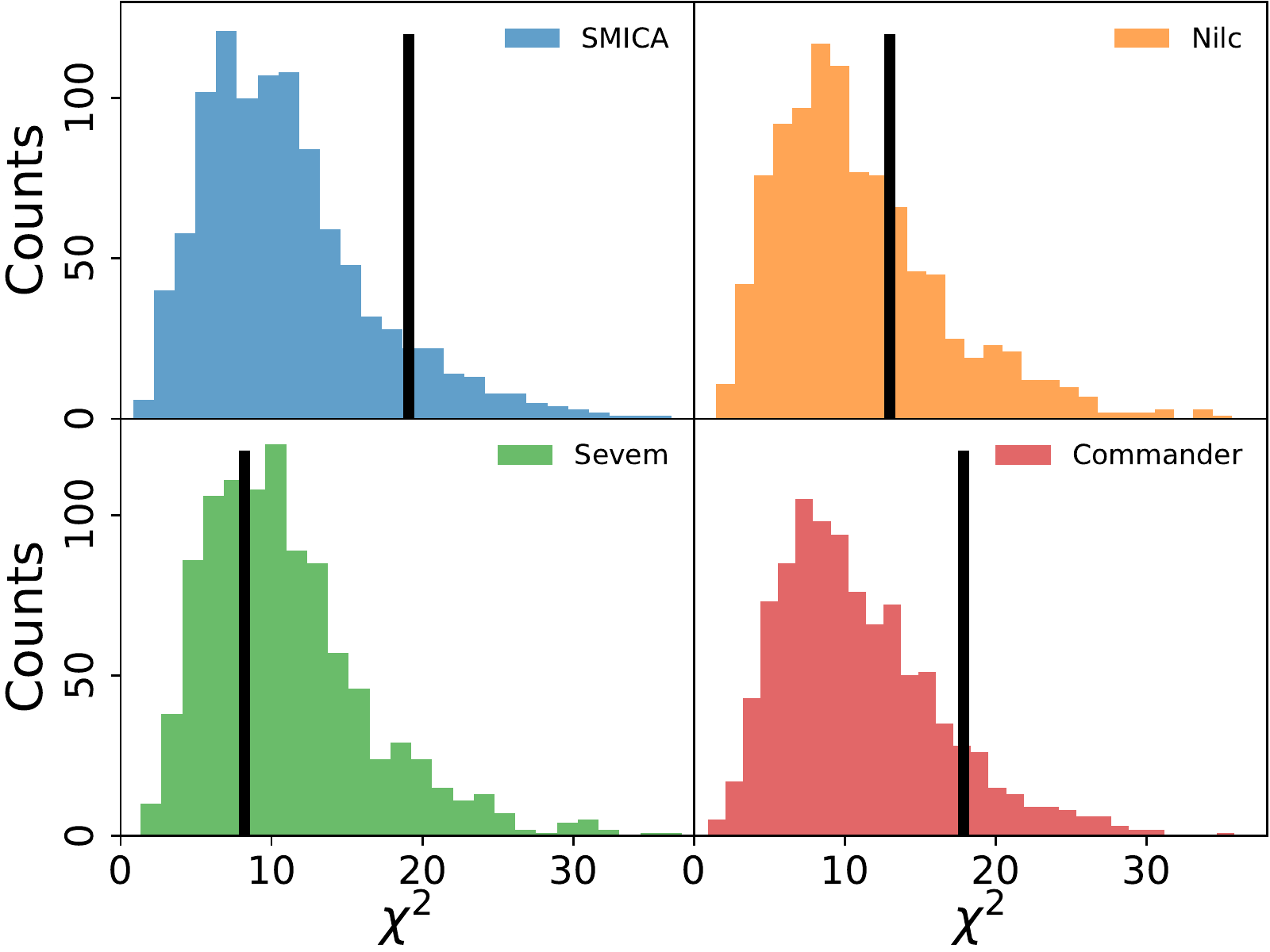}
\caption{$\chi^2$-analysis for the null effect using $C_{\ell}^{\alpha T}$ spectra for the HM pipeline.
}
\label{fig:CelltTalphanull}
\end{figure}

\begin{figure}[t]
\centering
\includegraphics[width=.86 \textwidth]{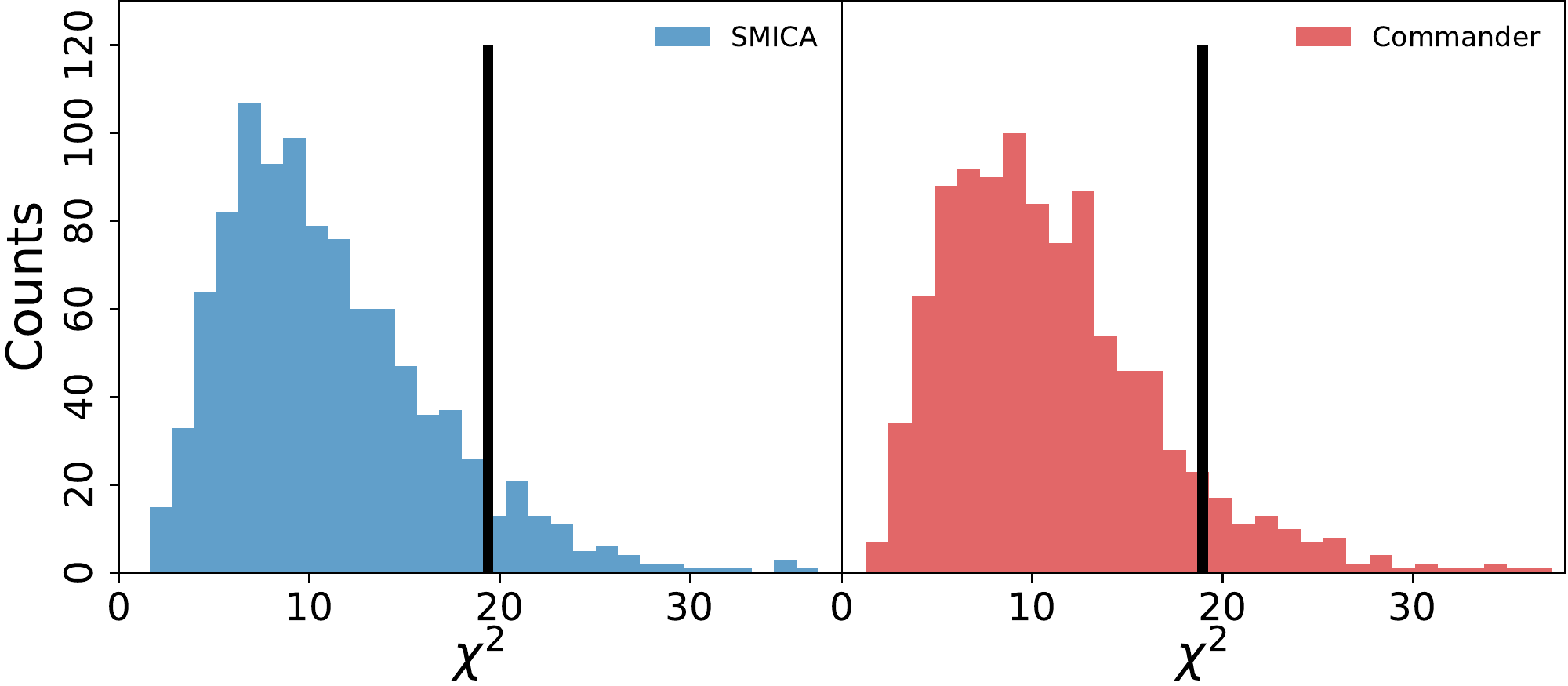}
\caption{$\chi^2$-analysis for the null effect using $C_{\ell}^{\alpha T}$ spectra for the OE pipeline.
}
\label{fig:CelltTalphanull_OE}
\end{figure}

All the $C^{\alpha T}_{\ell}$ are well compatible with null. This is quantified by the $\chi^2$-analysis given in Figures \ref{fig:CelltTalphanull} and \ref{fig:CelltTalphanull_OE} for HM and OE respectively. 
In particular, the probabilities to exceed, PTE, we evaluate from simulations and data for the HM pipeline are $9.6 \%$ for \smica, $29.7 \%$ for \nilc, $64.8 \%$ for \sevem\ and $10.8 \%$ for \commander.
The corresponding PTE for the OE pipeline is  $8.0 \%$ for \smica\ and $8.6 \%$ for \commander.

\begin{figure}[t]
\centering
\includegraphics[width=.49 \textwidth]{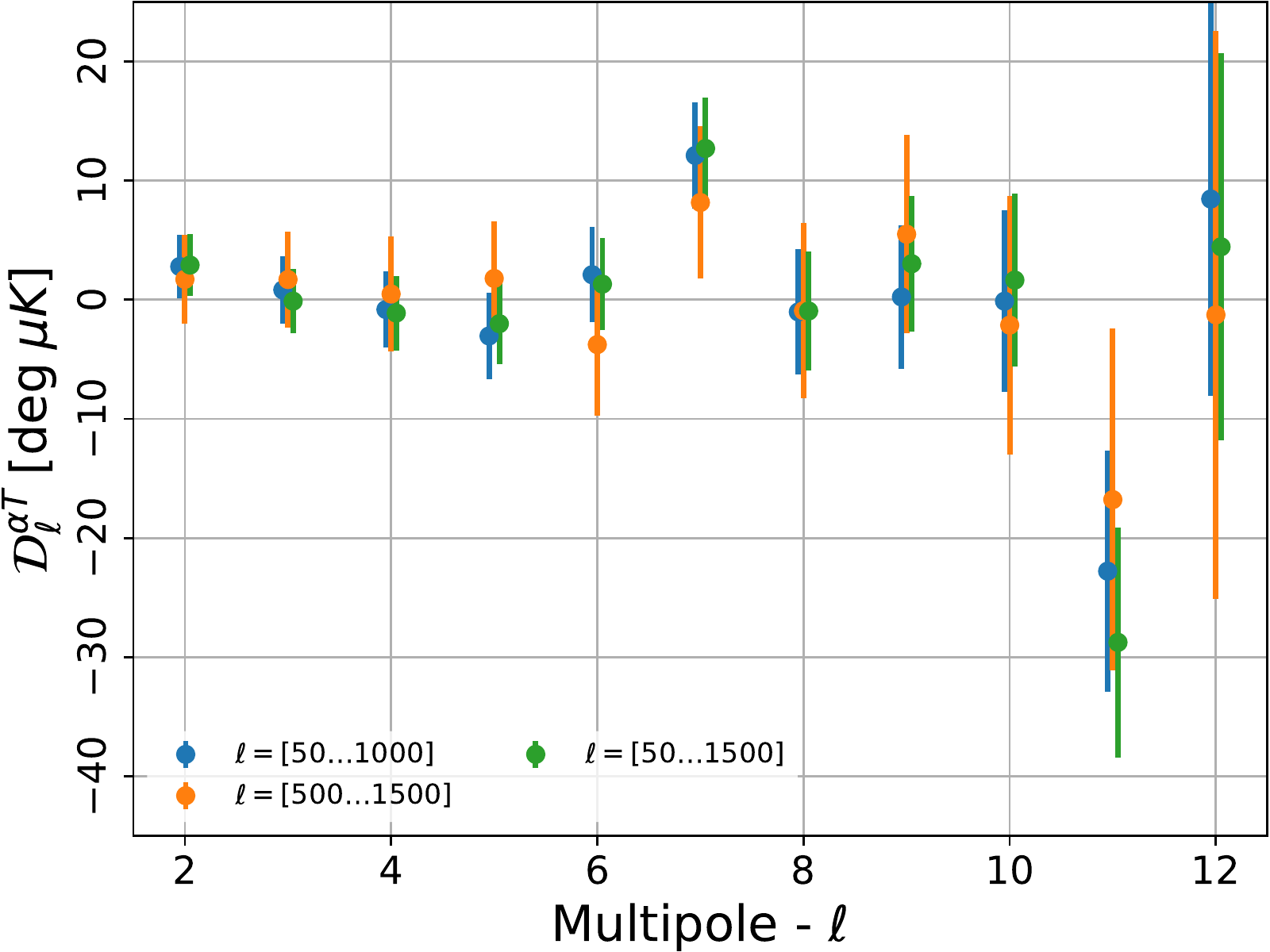}
\includegraphics[width=.49 \textwidth]{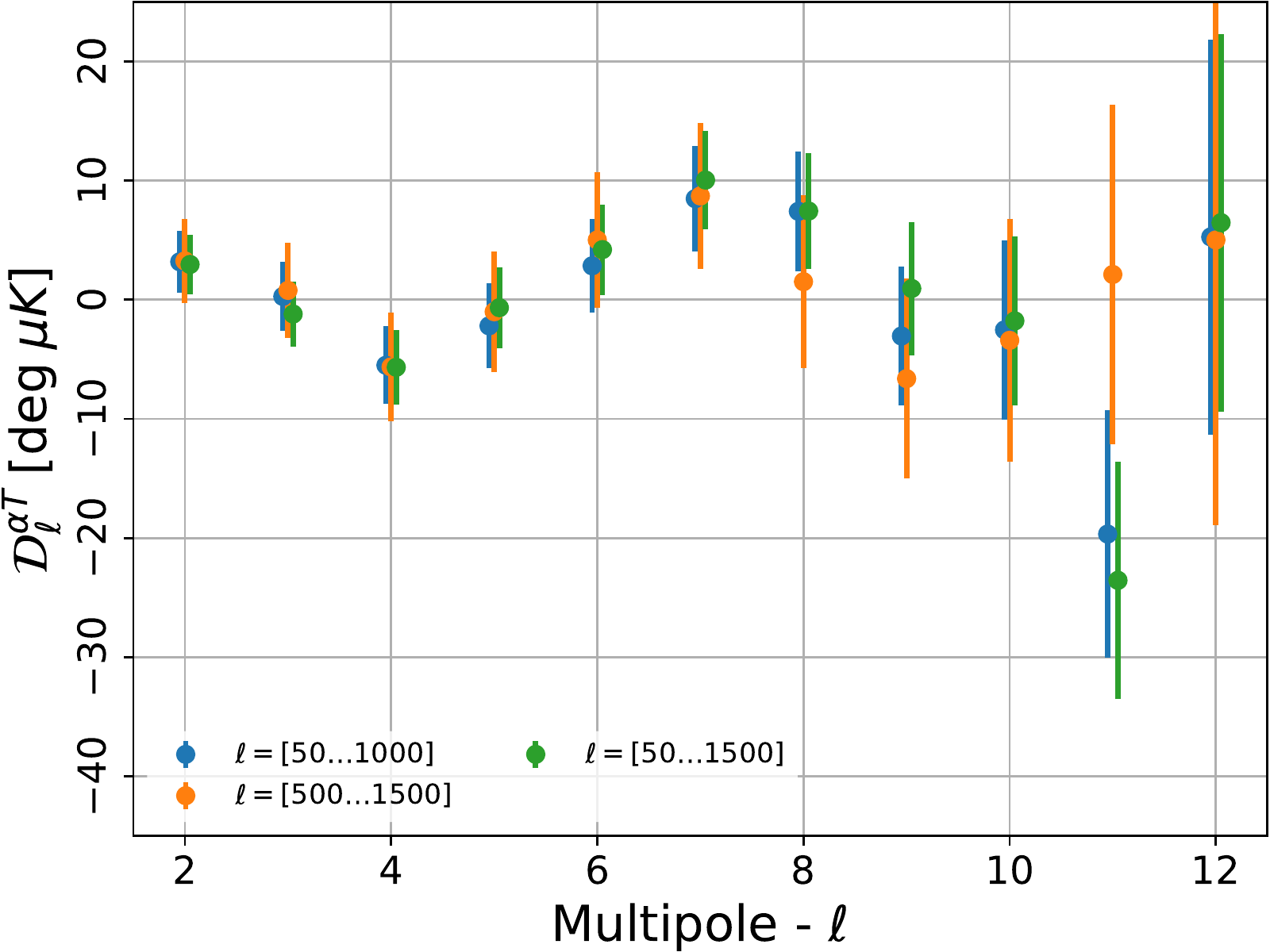}
\caption{Stability of $C_{\ell}^{\alpha T}$ we obtained from \smica\ versus the range of multipoles considered in the pipeline used to build the maps of $\alpha(\hat n)$. Left panel HM, right panel OE.
}
\label{fig:CelltTalpharobustness}
\end{figure}

In Figure \ref{fig:CelltTalpharobustness} we test the stability of the estimates of $C_{\ell}^{\alpha T}$ we obtained from \smica\ versus the range of multipoles considered in the pipeline used to build the maps of $\alpha(\hat n)$.
Left panel refers to HM and right panel to OE. 
This verifies that the stability is very good and the estimates typically shift within $\sim$1$\sigma$ C.L. depending on the range of multipoles considered.

\subsection{Pixel-based likelihood approach}
\label{Pixel_based_likelihood_approach}

It is customary to provide constraints on the anisotropic birefringence effect through the amplitude $A^{\alpha \alpha}$ of the quantity $\mathcal{D}_\ell^{\alpha \alpha} \equiv \ell (\ell + 1) C_{\ell}^{\alpha \alpha} / 2 \pi$ assuming a scale invariant 
APS, see e.g.~\cite{Pogosian:2019jbt}. That is, we look for a constant amplitude $A^{\alpha \alpha}$ to model $\mathcal{D}_\ell^{\alpha \alpha} $ over a given range of multipoles. 
Similarly, we can provide bounds on the cross-correlation between CMB temperature anisotropy and anisotropic birefringence fitting $\ell (\ell +1) C_{\ell}^{\alpha T}/ 2 \pi$ in the harmonic range $[2-12]$ with a constant amplitude, $A^{\alpha T}$. In this section we set joint constraints on $A^{\alpha \alpha}$ and $A^{\alpha T}$ adopting a pixel-based likelihood approach.

Assuming that CMB temperature and $\alpha$-anisotropies are Gaussian distributed, we build the following pixel-based likelihood function,
\be
{\cal L}(A^{\alpha \alpha},A^{\alpha T}) = {(\mbox{det} C )}^{-1/2} (2 \pi)^{-N^{tot}_{pix}/2} \exp{( x^t C^{-1} x/2)} \, ,
\label{likelihood}
\ee
where the covariance $C=C(A^{\alpha \alpha},A^{\alpha T})$ is defined as
\be
   C(A^{\alpha \alpha},A^{\alpha T}) =
  \left( {\begin{array}{cc}
   C^{TT} & C^{T \alpha}(A^{\alpha T})\\
   \left[ C^{T \alpha} (A^{\alpha T})\right]^t & C^{\alpha \alpha} (A^{\alpha \alpha}) \\
  \end{array} } \right) \, ,
  \label{pixelcov}
\ee
and the blocks entering Eq.~(\ref{pixelcov}) are given by
\be 
C^{TT}_{\imath \jmath} &=& \sum_{\ell=2}^{\ell_{max}} {{2 \ell + 1}\over{4 \pi}} P_{\ell} (\hat \imath \cdot \hat \jmath) C^{TT}_{\ell}+ N^{TT}_{\imath \jmath} \, ,  \\ 
C^{\alpha \alpha}_{\imath \jmath} &=& S_{\imath \jmath}(A^{\alpha \alpha}) + N^{\alpha \alpha}_{\imath \jmath} \, , \\
C^{T \alpha}_{\imath \jmath} &=& S_{\imath \jmath}(A^{\alpha T}) \, , \\
\left( C^{T \alpha}_{\imath \jmath} \right)^t &=& C^{T \alpha}_{\jmath \imath} \, , 
\ee
with $P_{\ell}$ being the Legendre polynomials, $C_{\ell}^{TT}$ the APS of the CMB temperature anisotropies 
and where the matrix $S_{\imath \jmath}(A^{X})$ is given by
\be
S_{\imath \jmath}(A^{X}) = A^{X} \sum_{\ell=\ell^{X}}^{\ell_{max}} {{2 \ell + 1}\over{2 \ell (\ell +1)}} P_{\ell} (\hat \imath \cdot \hat \jmath)
\, ,
\label{covariancedef}
\ee
with $X$ standing for ``$\alpha \alpha$'' or ``$\alpha T$'', and where the noise matrices $N^{\alpha \alpha}_{\imath \jmath}$, $N^{TT}_{\imath \jmath}$ are taken to be diagonal. 
In particular
$N^{TT}_{\imath \jmath} = 0.25 \, \mu K^2$ and
\be
N^{\alpha \alpha}_{\imath \jmath} = \sigma_{\imath}^2 \, \delta_{\imath \jmath}
\, ,
\ee
where $\sigma_{\imath}^2$ are evaluated from FFP10 simulations.
Note also that the signal matrix for CMB temperature and for the cross-correlation takes contribution from the multipole $\ell^{\alpha T}=2$ while the signal matrix for the birefringence angles starts from $\ell^{\alpha}=1$.
The vector $x$ in Eq.~(\ref{likelihood}) is defined as 
\be 
x= \left( T(\hat n), \alpha(\hat n) \right) \, ,
\ee 
where $T(\hat n)$ is the CMB temperature anisotropy map and $\alpha(\hat n)$ is the map of the birefringence angle.
Moreover, $N^{tot}_{pix}=N^{T}_{pix} + N^{\alpha}_{pix}$ where $N^{T}_{pix}$ is the number of observed pixels in the CMB temperature anisotropy map
and $N^{\alpha}_{pix}$ is the number of observed pixels in the birefringence map.
The indexes $\imath, \jmath$ run over the observed pixels.

\subsubsection{Validation with simulations}
\label{validationlike}

We start sampling Eq.~(\ref{likelihood}) on $(A^{\alpha \alpha},A^{\alpha T})$ for each of the $\alpha(\hat n)$ realisations constructed from FFP10 simulations considering the baseline configuration, and build a {\it total} likelihood given by the product of each of the sampled 2D-likelihoods. Technically this is performed marginalising over the monopole and the dipole of the CMB and over the monopole of the birefringence map. 
If the region around the peak of the total likelihood contains the model described by the simulations, i.e. no birefringence effect, then we can safely state that the considered likelihood function is validated.

\begin{figure}[t]
\centering
\includegraphics[width=.49\textwidth]{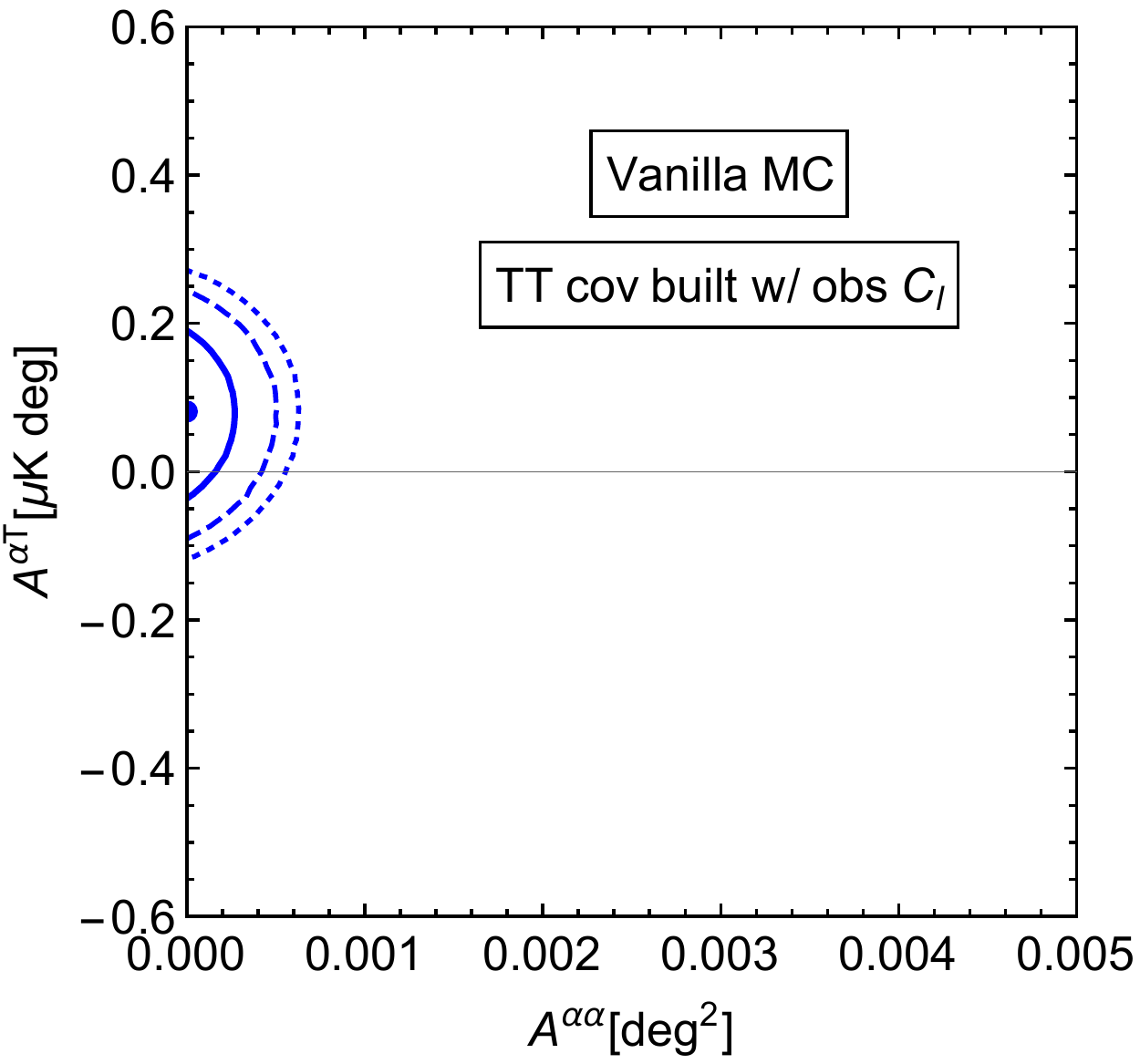}
\caption{2D posterior distribution function of $A^{\alpha \alpha}$ and $A^{\alpha T}$ obtained from vanilla simulations.
Solid, dashed and dotted curves stand for $68$, $95$ and $99\%$ C.L..}
\label{fig:VanillaValidation}
\end{figure}

\begin{figure}[t]
\centering
\includegraphics[width=.49\textwidth]{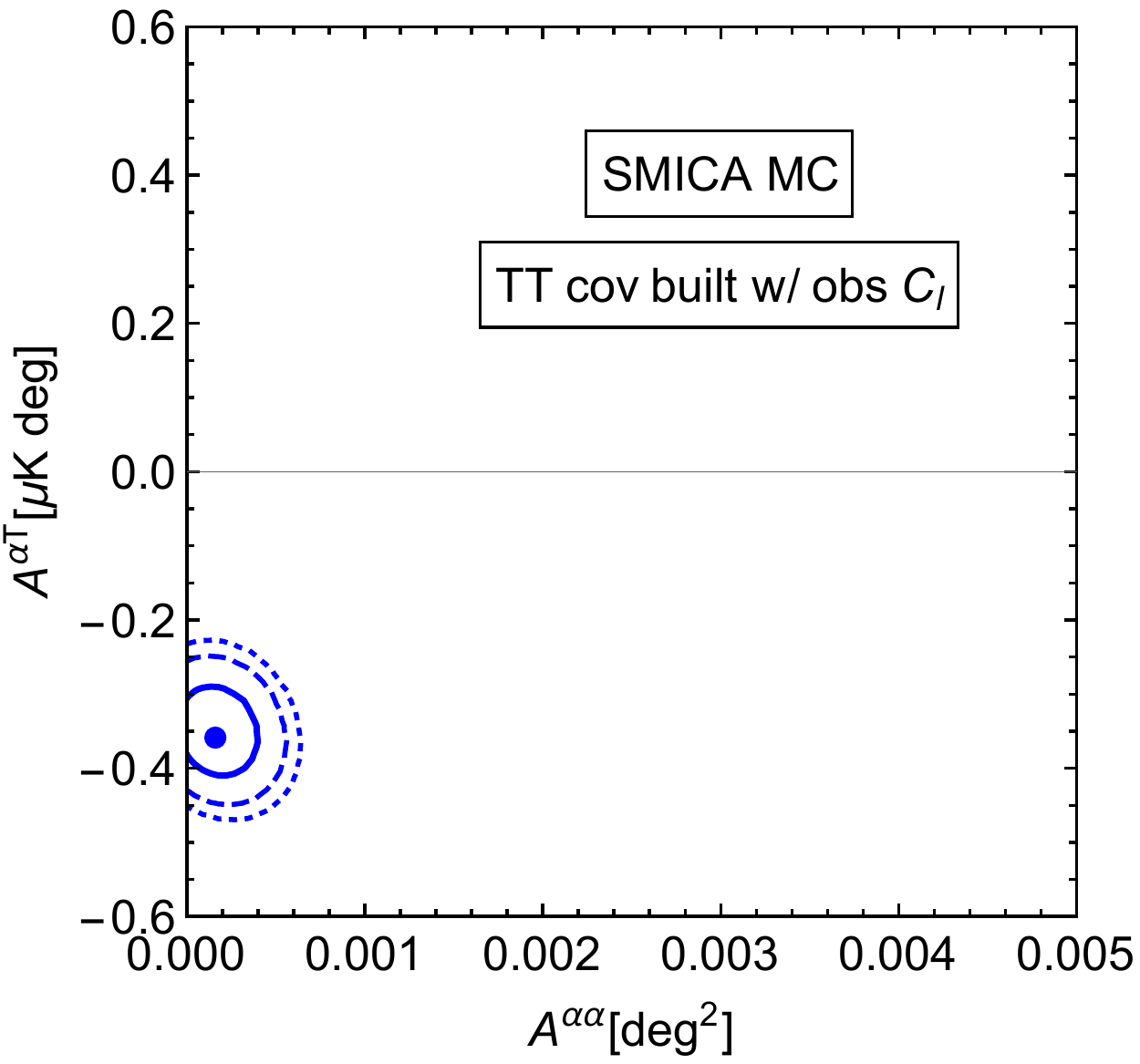}
\includegraphics[width=.49\textwidth]{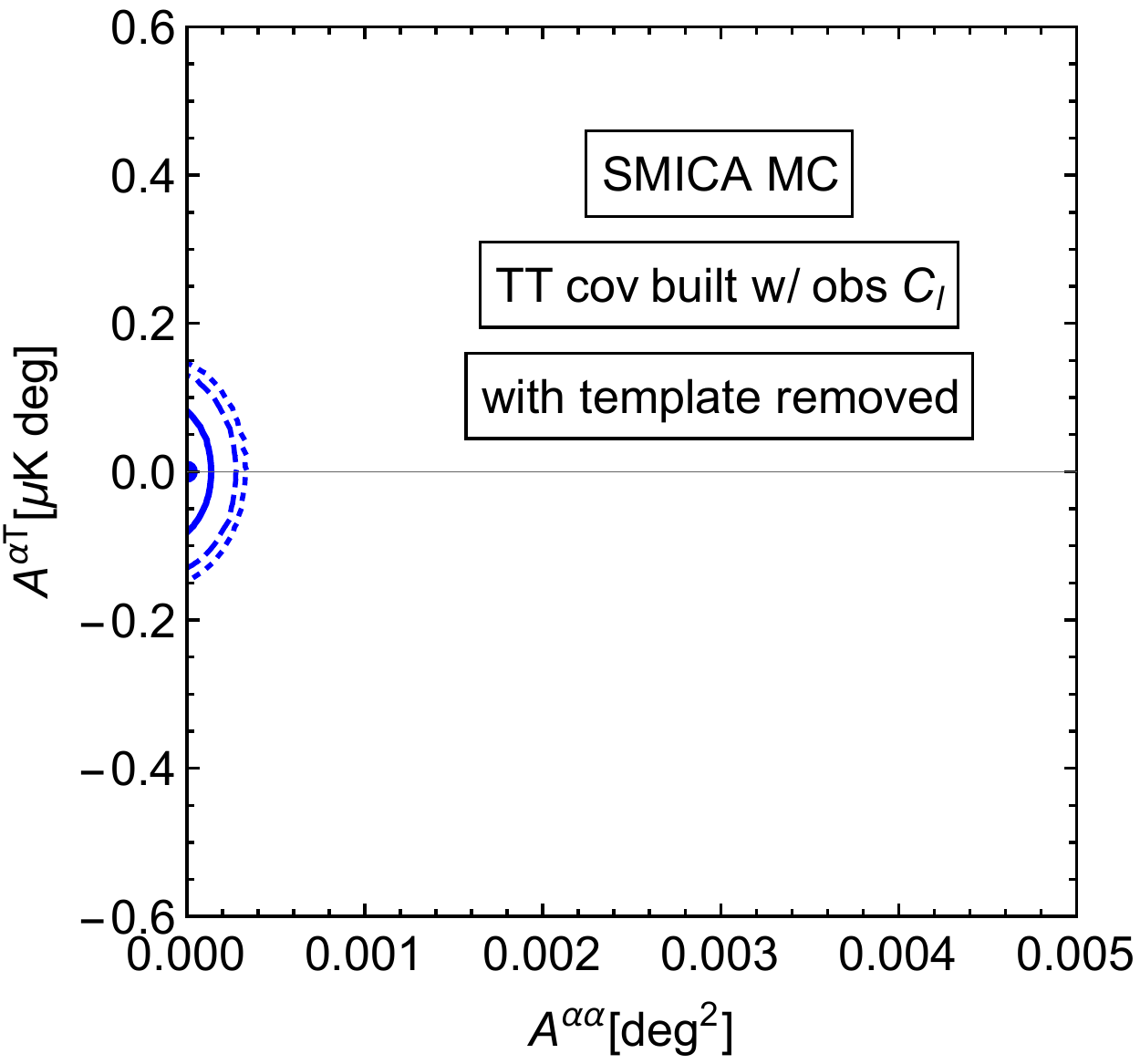}
\caption{2D posterior distribution function of $A^{\alpha \alpha}$ and $A^{\alpha T}$ obtained from HM \smica\ simulations. Left panel without removing the FFP10 template and right panel removing it from simulations.
Solid, dashed and dotted curves stand for $68$, $95$ and $99\%$ C.L..}
\label{fig:SMICAValidation}
\end{figure}

\begin{figure}[t]
\centering
\includegraphics[width=.49\textwidth]{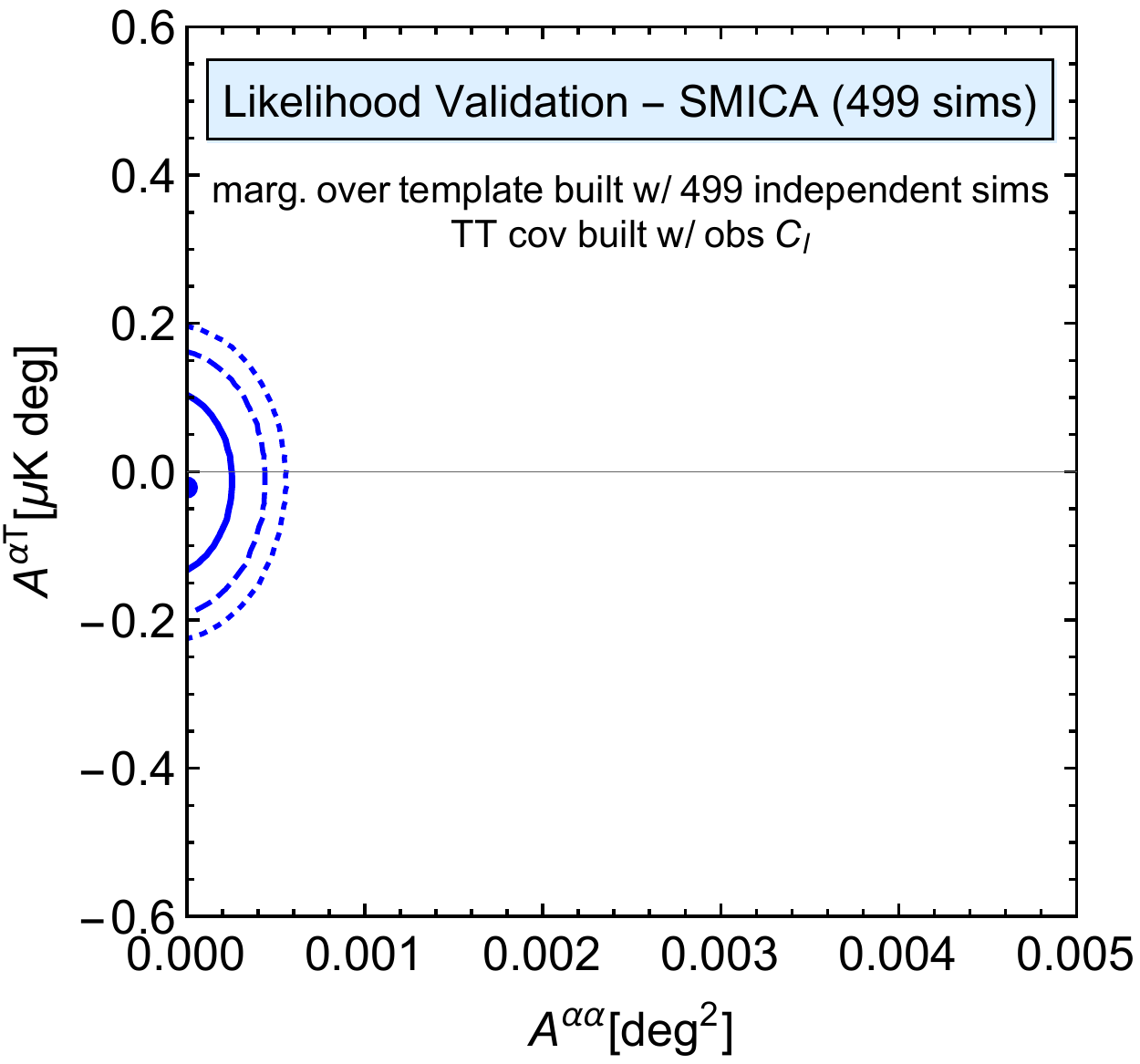}
\includegraphics[width=.49\textwidth]{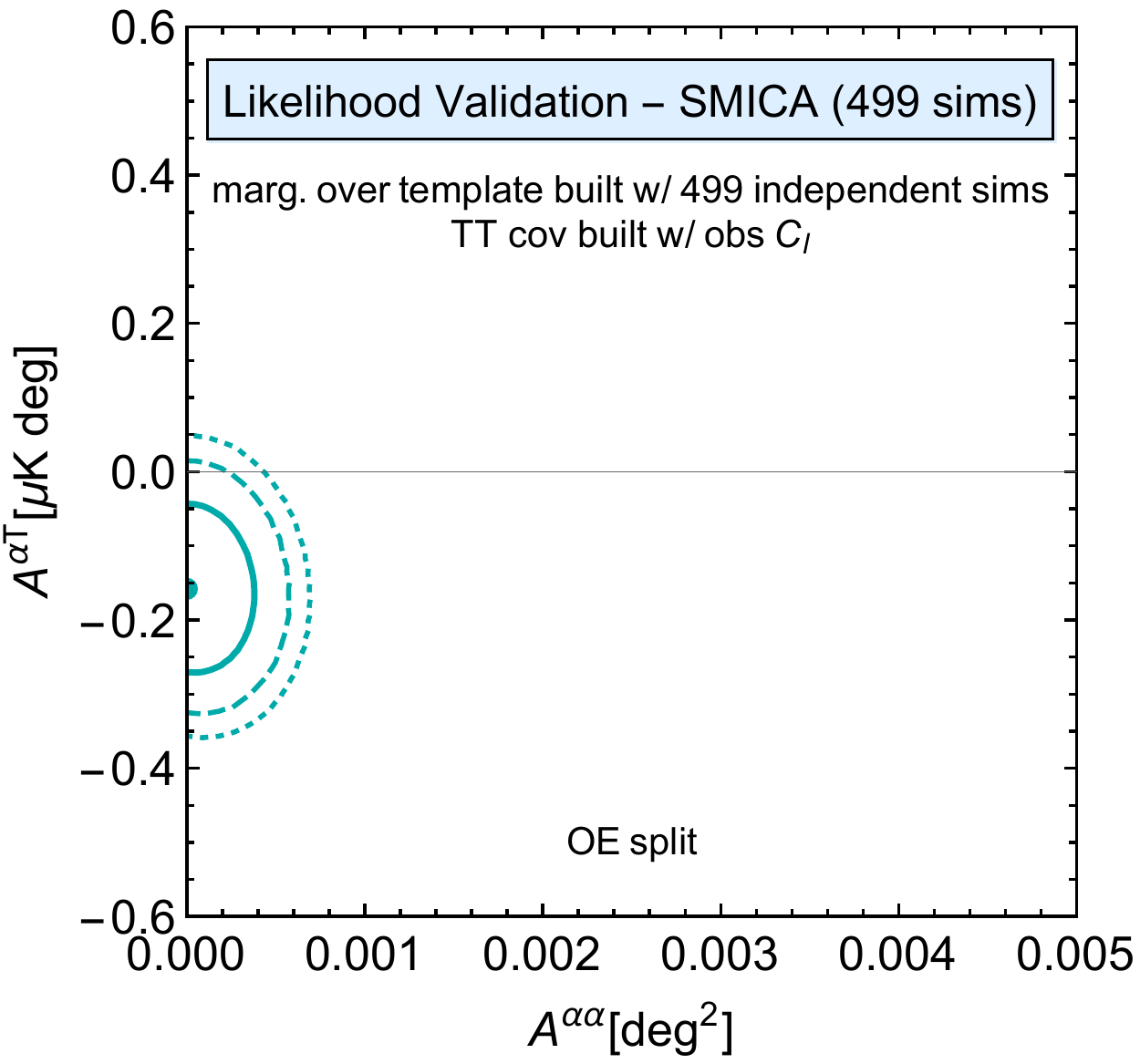}
\caption{2D posterior distribution function of $A^{\alpha \alpha}$ and $A^{\alpha T}$ obtained from HM (left panel) and OE (right panel) 499 \smica\ simulations.
The marginalisation over the FFP10 template, built with 499 further independent simulations, has been performed.
Solid, dashed and dotted curves stand for $68$, $95$ and $99\%$ C.L..}
\label{fig:SmicaValidation2}
\end{figure}

Employing the vanilla simulations, already used above in Section \ref{further}, we derive Figure \ref{fig:VanillaValidation}.
Since the point $(0,0)$ in the $(A^{\alpha \alpha},A^{\alpha T})$-plane, is well within the $68\%$ contour (solid curve) we find a nice validation of the considered likelihood function given in Eq.~(\ref{likelihood}).
When we take into account HM \smica\ simulations we find the total likelihood shown in left panel of Figure \ref{fig:SMICAValidation}.
In this case the point (0,0) is outside the region defined by the $99\%$ contour (dotted curve).
This means that in this case the likelihood is not formally validated (specifically for $A^{\alpha T}$) even if, as it will be clear later, the impact of this bias on the data is not large and, if not corrected, can be quantified as about 1/5 of the standard statistical deviation of $A^{\alpha T}$. 
In the right panel of Figure \ref{fig:SMICAValidation} we show the HM \smica\ case when we remove the FFP10 template from the simulations.
This can be seen as a sanity check, since the FFP10 template is exactly the average map and therefore the peak of the total likelihood has to be on top of $(0,0)$ in the 2D plane $(A^{\alpha \alpha},A^{\alpha T})$ by construction.
For a fair validation one should divide the set of simulations in two parts: one of those has to be used to build the FFP10 template and the other one to build the total likelihood using the FFP10 template. 
In this way the simulations used to validate the likelihood are independent from those adopted to build the template.
In Figure \ref{fig:SmicaValidation2} we show exactly this approach for the HM (left panel) and OE (right panel) \smica\ case where half of the simulations are used to build the FFP10 template and the other half are employed to build the total likelihood where a marginalisation over the FFP10 template is taken into account.
In both cases, the point $(0,0)$ is well within expected statistical fluctuations, represented by solid, dashed and dotted contours, providing a validation of the method.
Analogous plots can be obtained for the other cases that we do not report here for sake of brevity.

\subsubsection{Results on data}
\label{likeondata}

\begin{figure}[t]
\centering
\includegraphics[width=.49\textwidth]{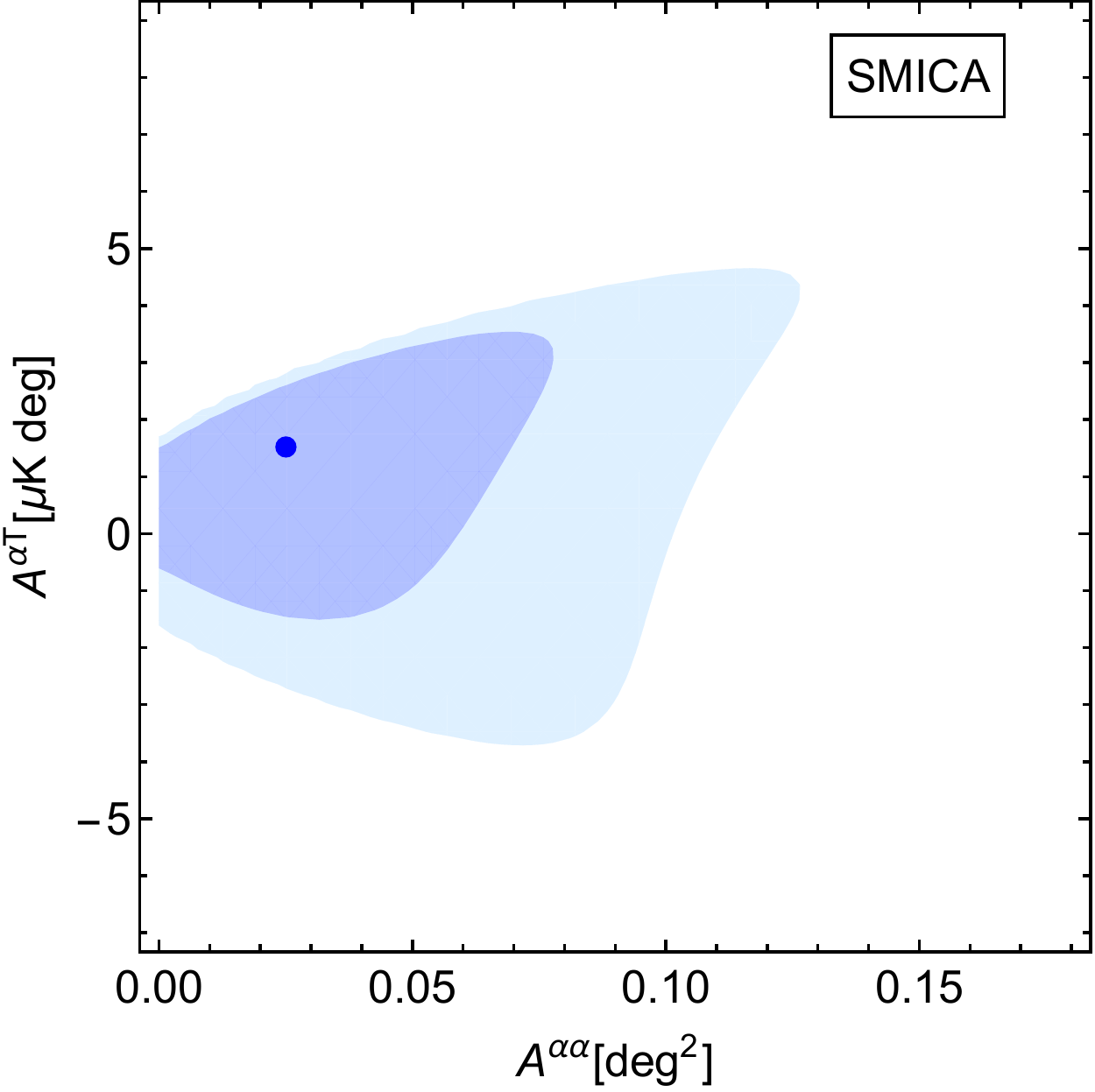}
\includegraphics[width=.49\textwidth]{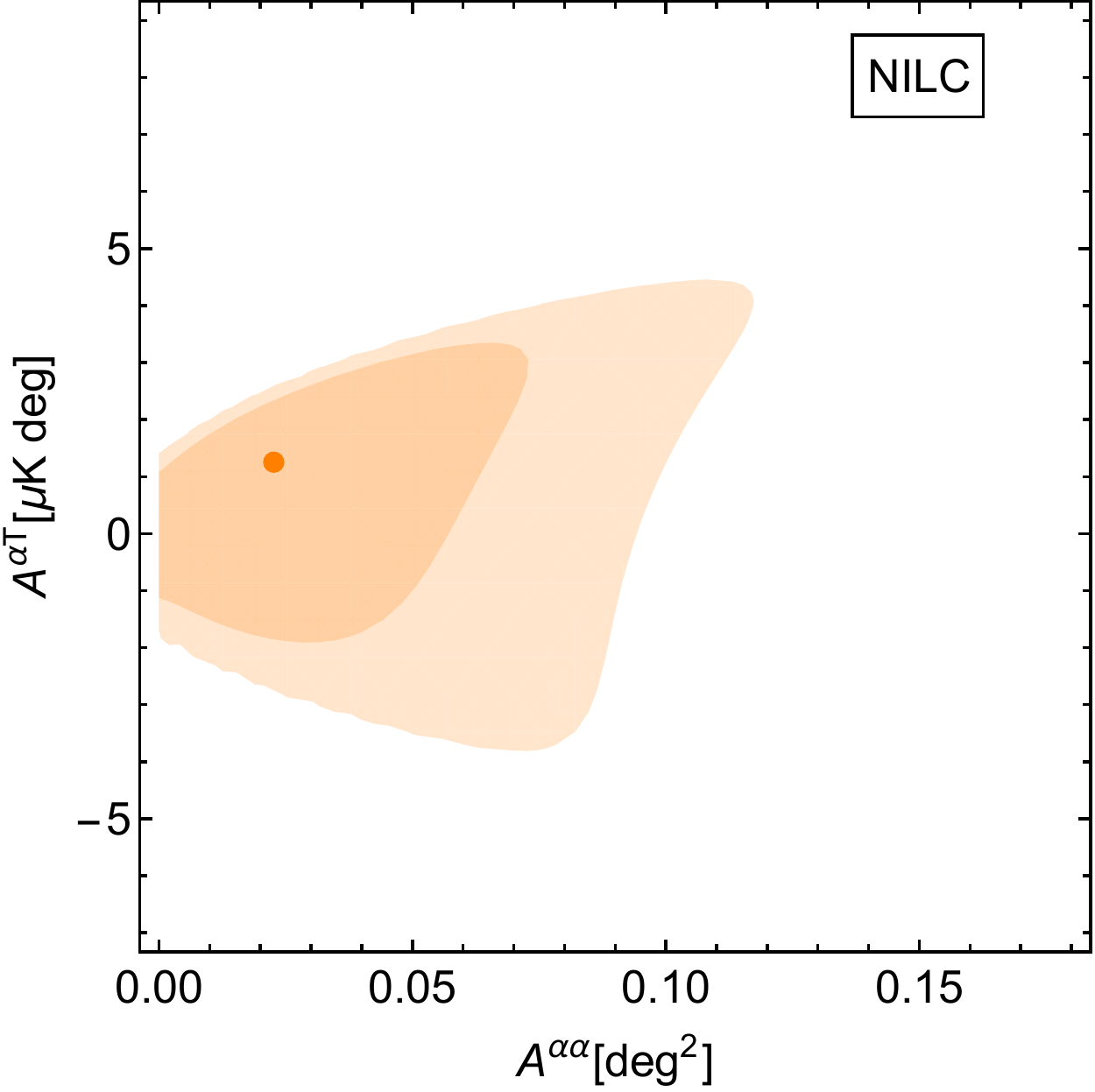}
\includegraphics[width=.49\textwidth]{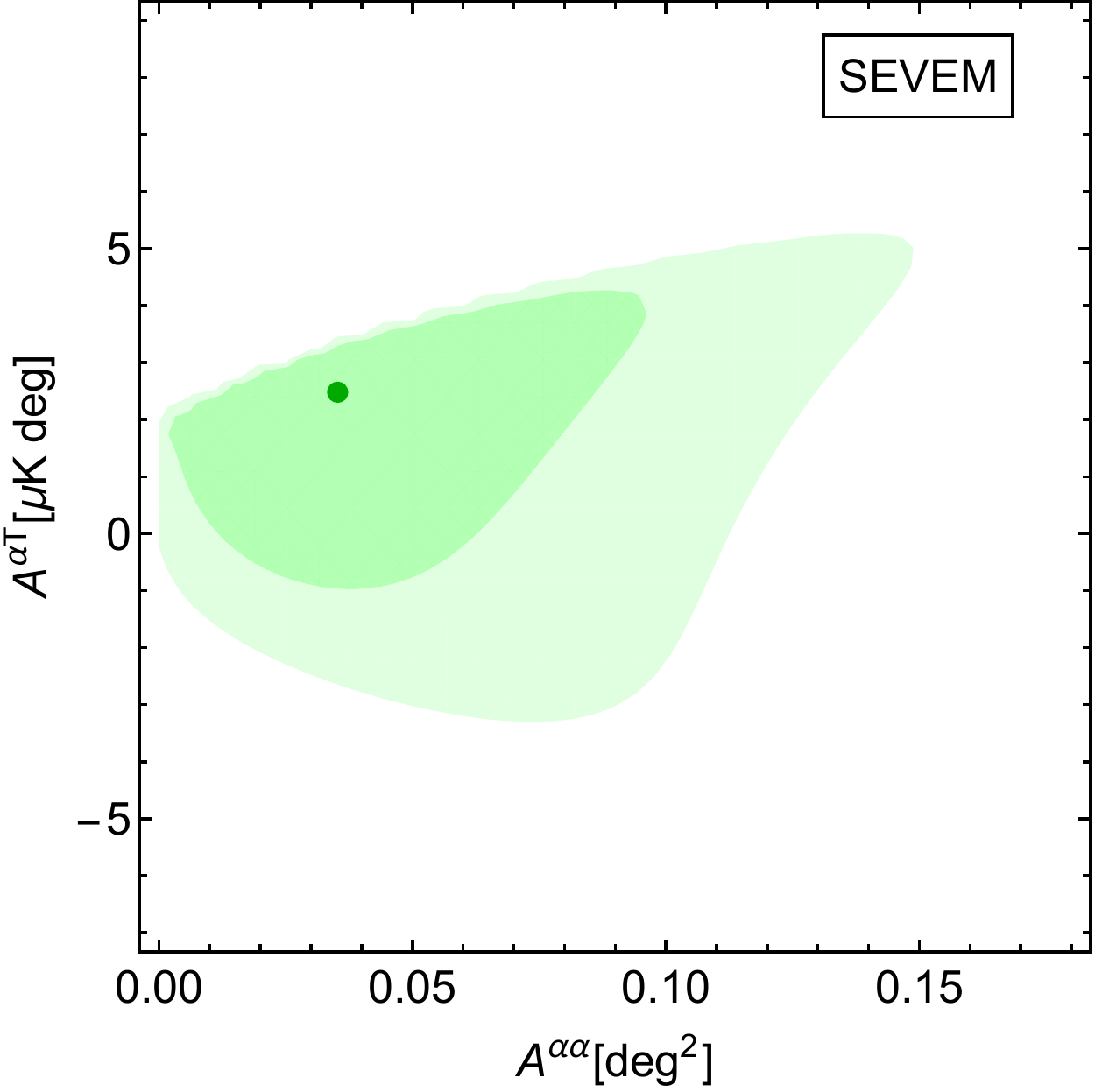}
\includegraphics[width=.49\textwidth]{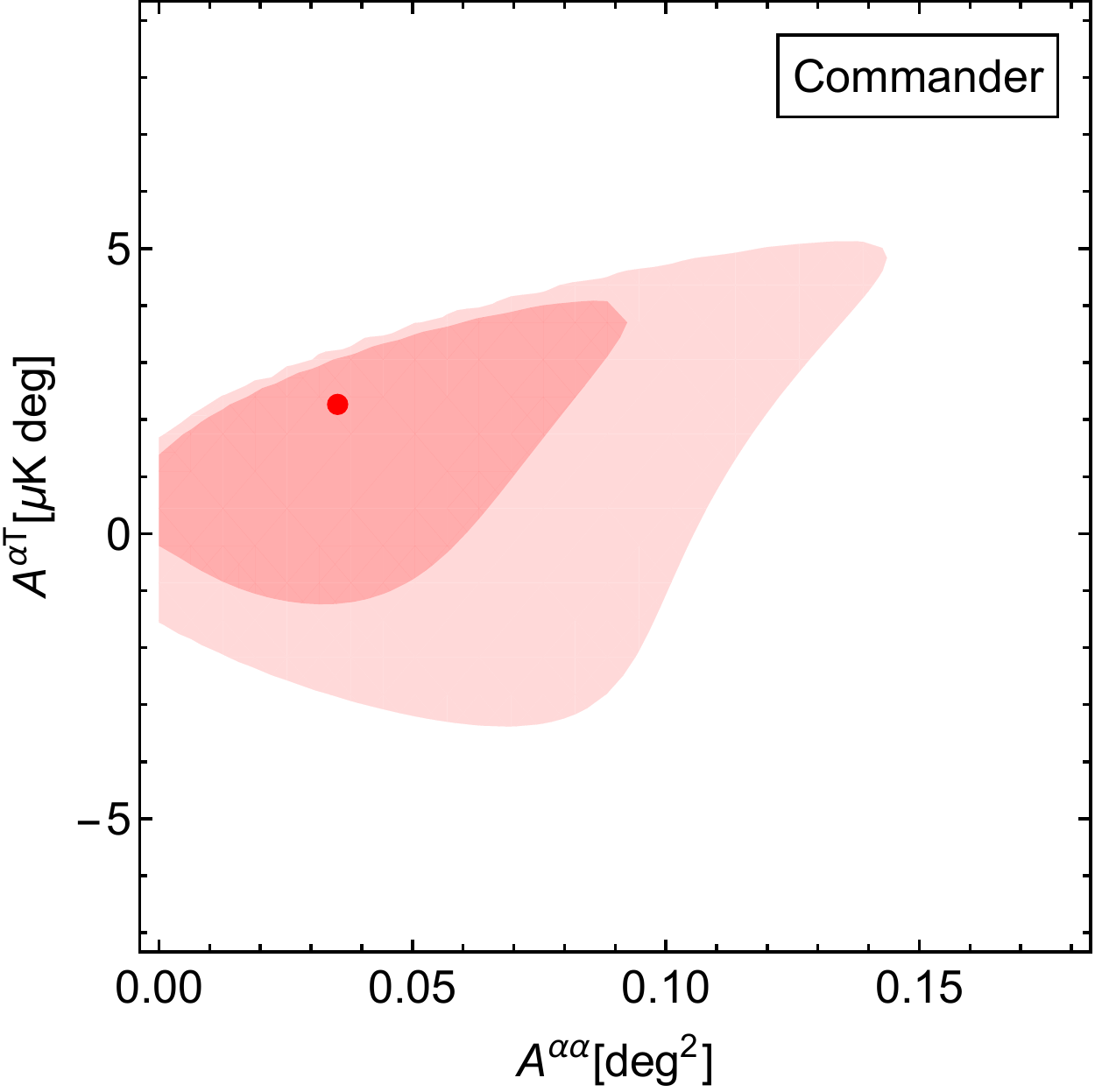}
\caption{Two-dimensional contour plot for $A^{\alpha \alpha}$ and $A^{\alpha T}$, obtained from \smica\ (upper left), \nilc\ (upper right), \sevem\ (lower left), \commander\ (lower right) HM data in the baseline configuration.
In each panel the darker (lighter) area covers the $68\%$ ($95\%$) confidence region.}
\label{fig:2dlikelihood}
\end{figure}

\begin{figure}[t]
\centering
\includegraphics[width=.49\textwidth]{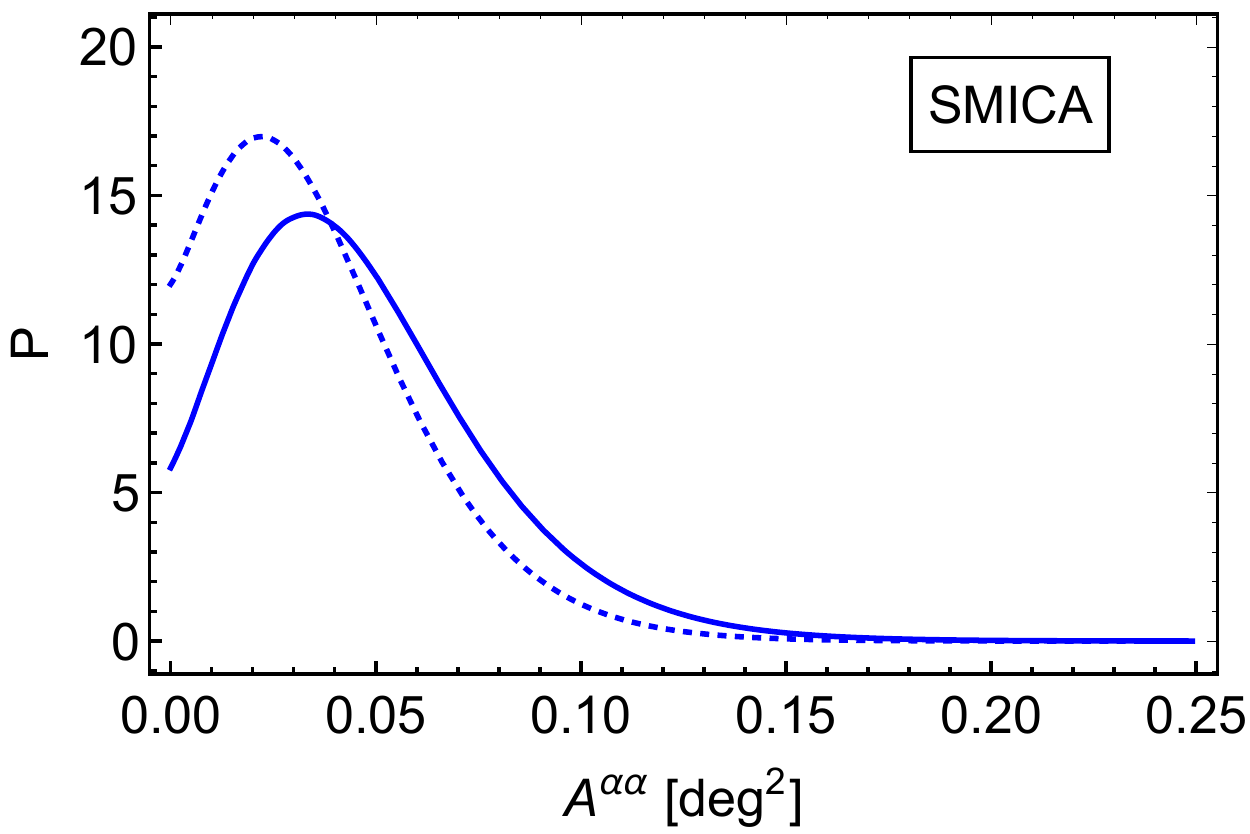}
\includegraphics[width=.49\textwidth]{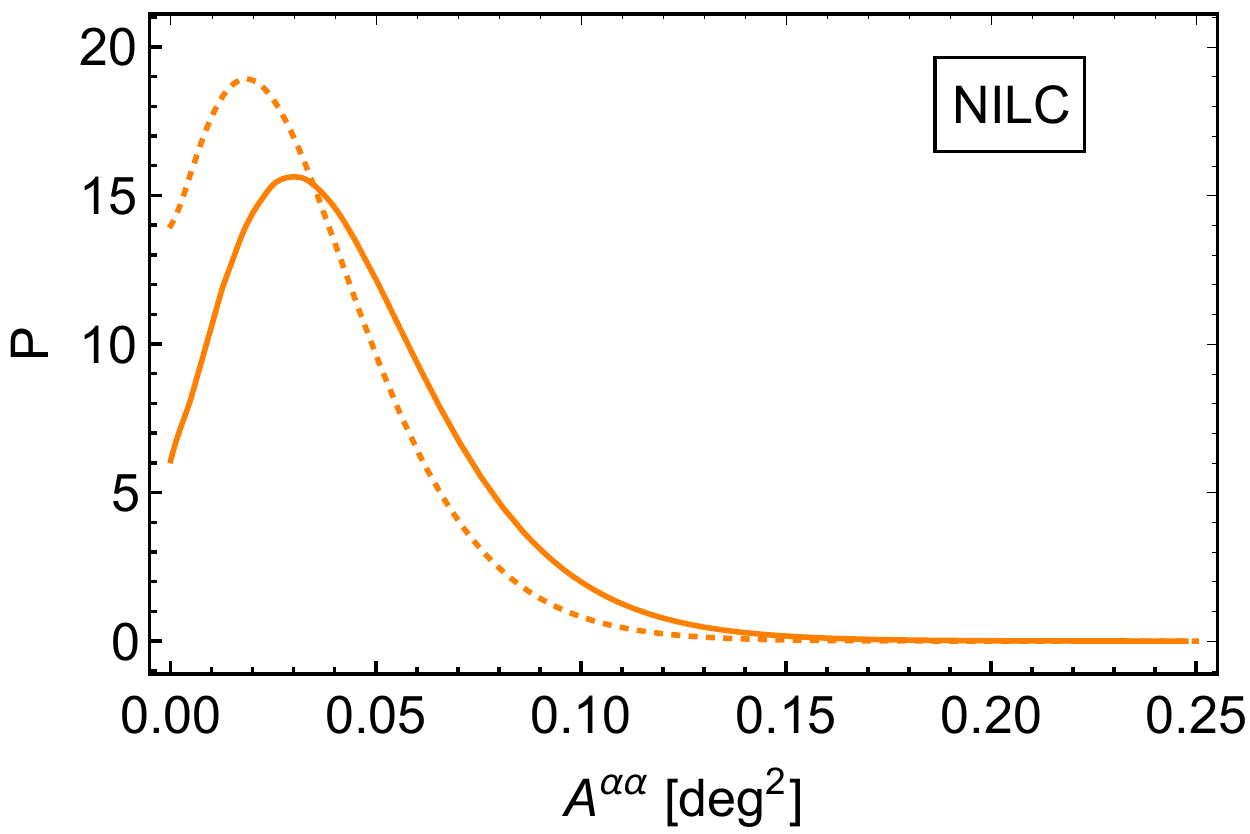}
\includegraphics[width=.49\textwidth]{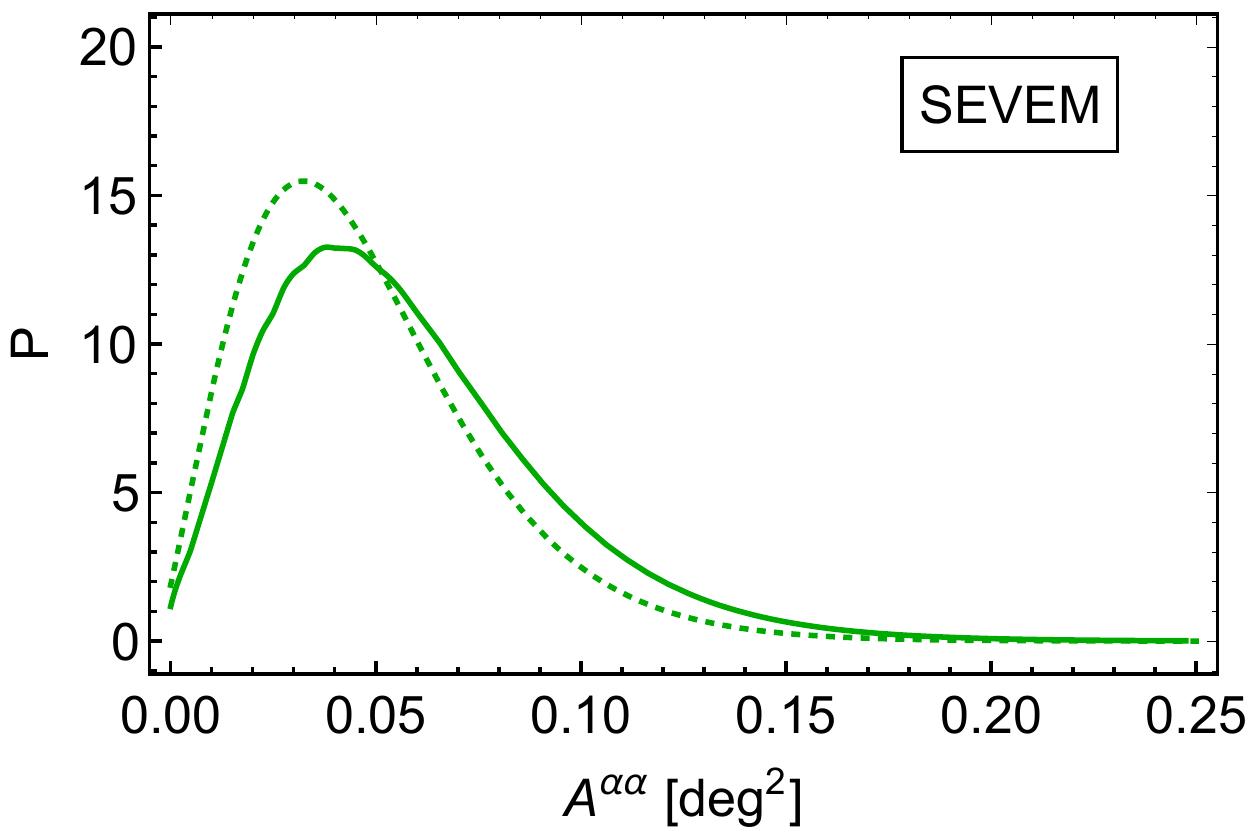}
\includegraphics[width=.49\textwidth]{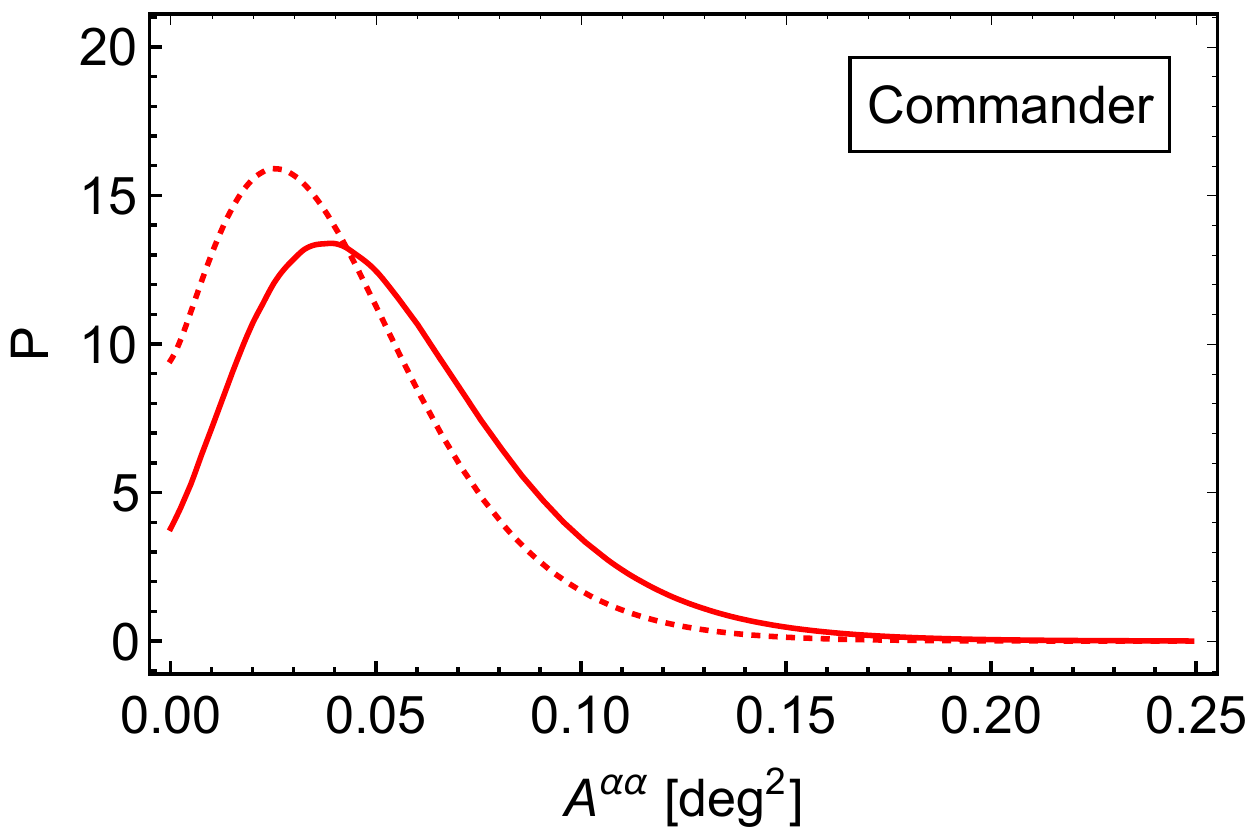}
\caption{Posterior distribution function of $A^{\alpha \alpha}$ obtained from \smica\ (upper left), \nilc\ (upper right), \sevem\ (lower left), \commander\ (lower right) HM data in the baseline configuration.
In each panel the solid curves are computed marginalising the 2D distribution function over $A^{\alpha T}$ and the dashed ones slicing the 2D distribution function at $A^{\alpha T}=0$. }
\label{fig:Aalphamarginalised}
\end{figure}

\begin{figure}[t]
\centering
\includegraphics[width=.49\textwidth]{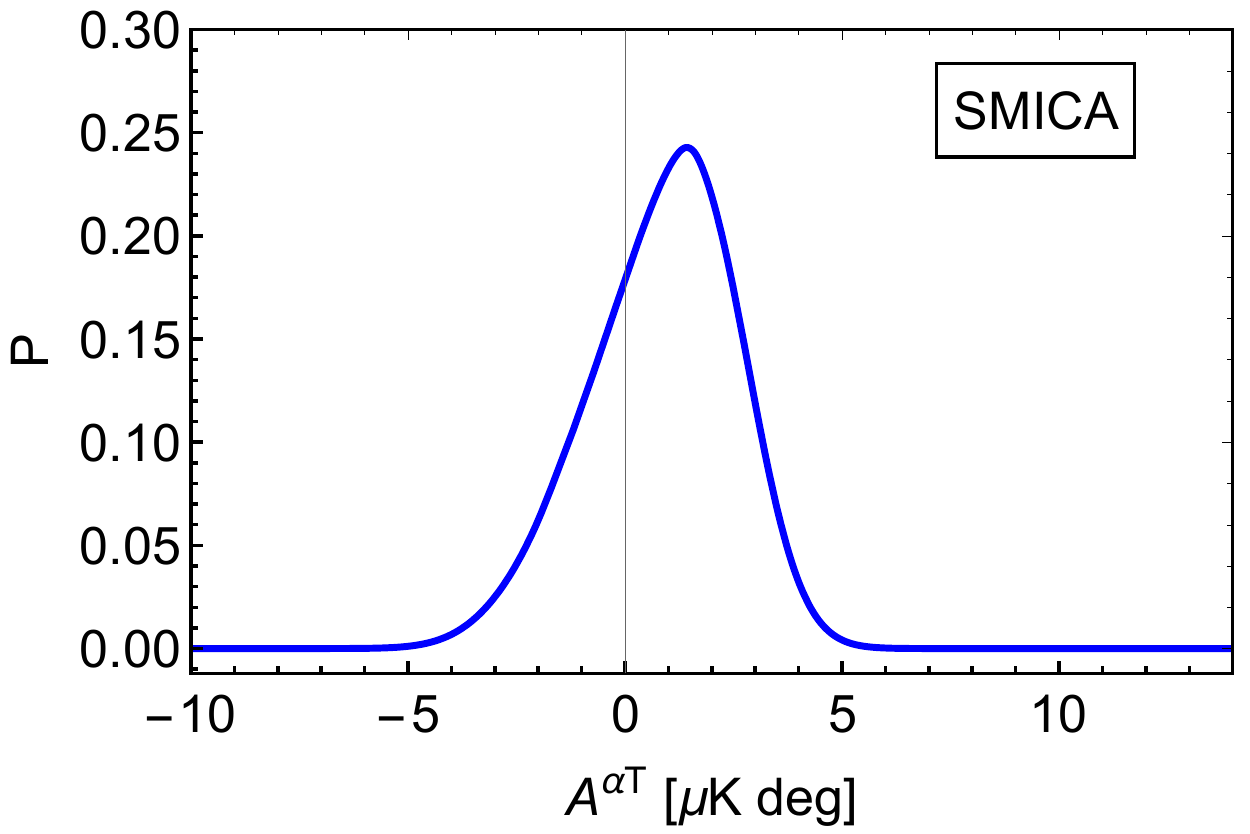}
\includegraphics[width=.49\textwidth]{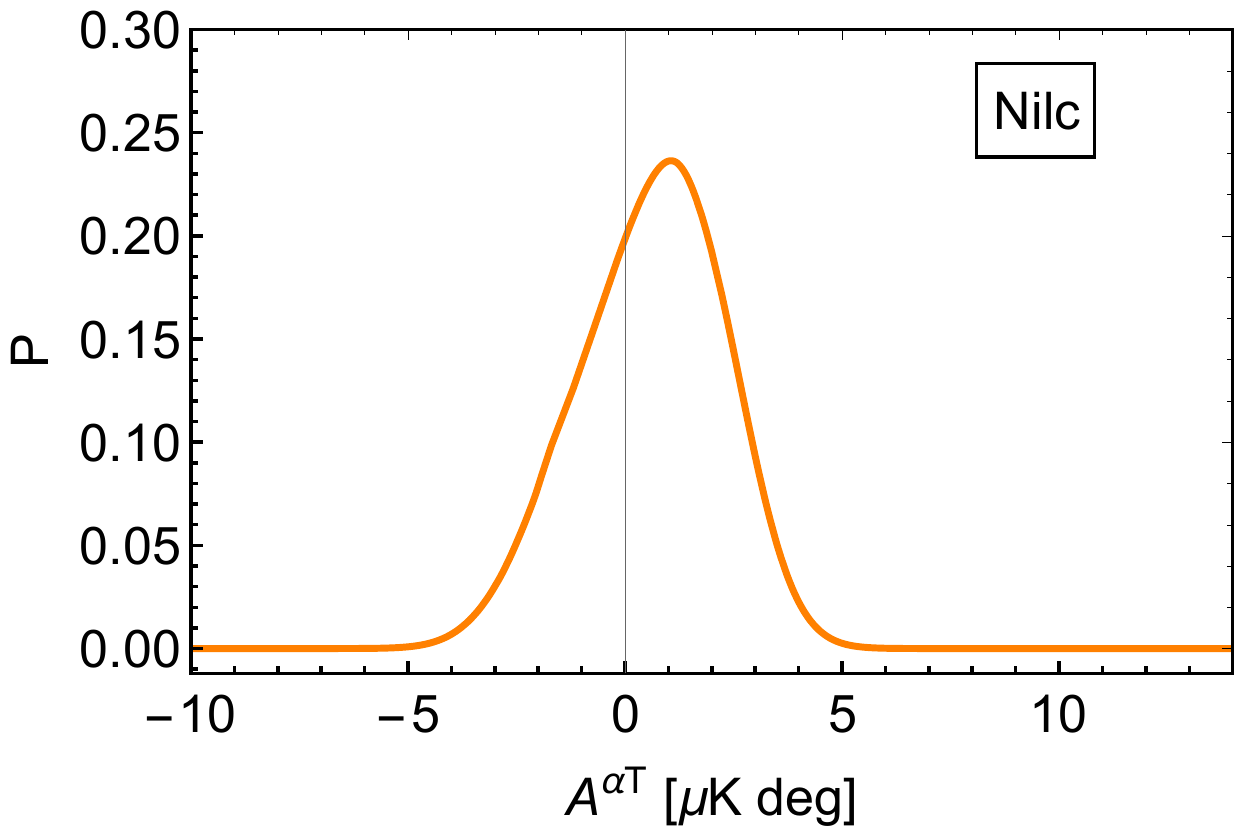}
\includegraphics[width=.49\textwidth]{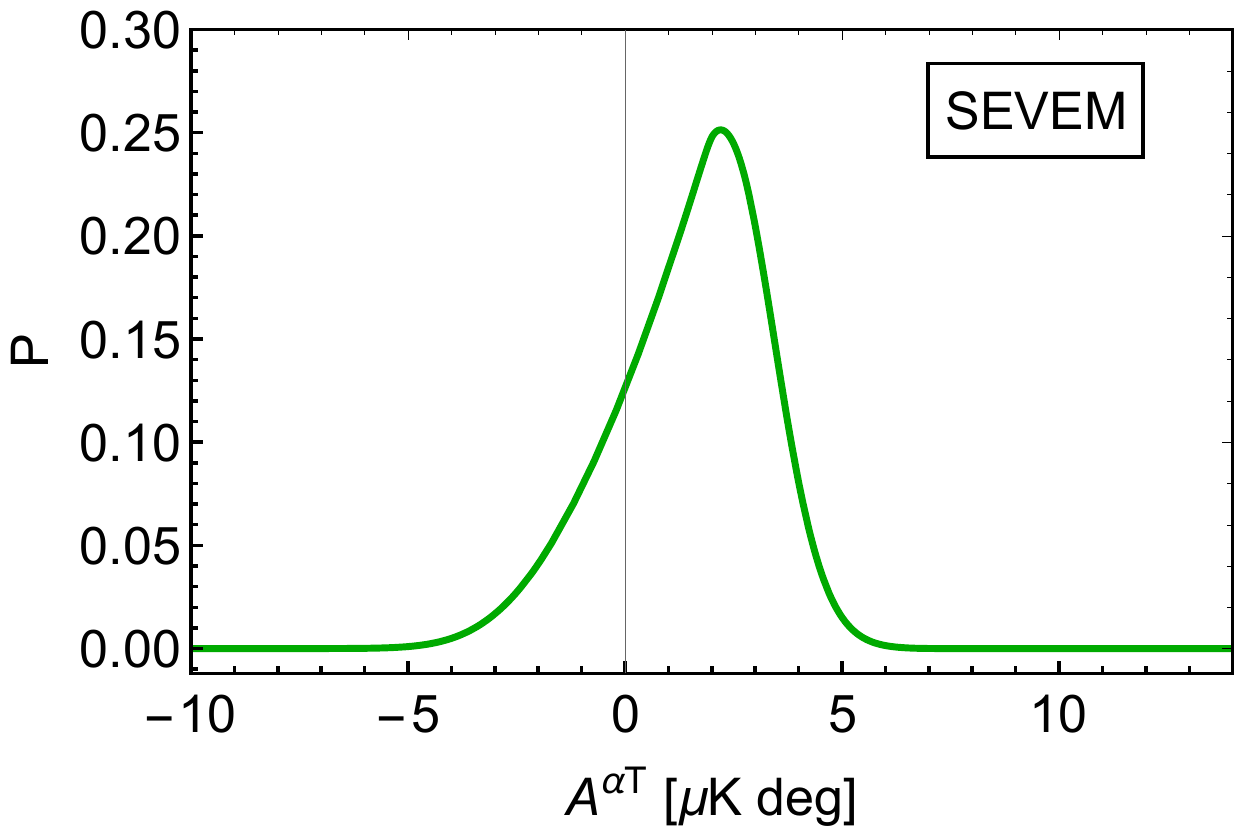}
\includegraphics[width=.49\textwidth]{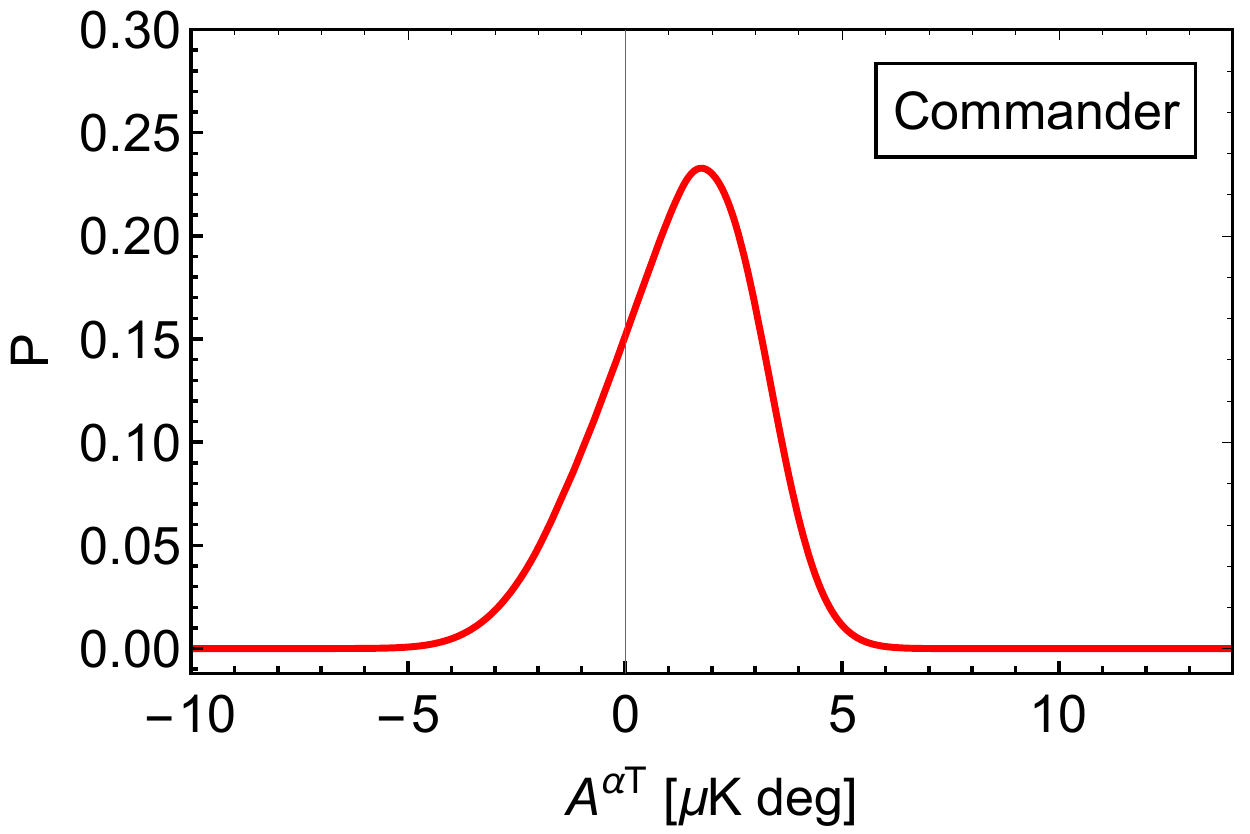}
\caption{Posterior distribution function of $A^{\alpha T}$ obtained from \smica\ (upper left), \nilc\ (upper right), \sevem\ (lower left), \commander\ (lower right) HM data in the baseline configuration.
In each panel the solid curves are computed marginalising the 2D distribution function over $A^{\alpha \alpha}$. }
\label{fig:AalphaTmarginalised}
\end{figure}

In Figure \ref{fig:2dlikelihood} we show the 2D contour plot of $A^{\alpha \alpha}$ and $A^{\alpha T}$ for HM data of \smica, \sevem, \nilc\ and  \commander.
Marginalising the 2D posterior distribution function over $A^{\alpha T}$ we obtain the posterior distribution function of $A^{\alpha \alpha}$ shown in Figure \ref{fig:Aalphamarginalised} (solid curves).
In the same Figure, dashed curves are computed slicing the 2D posterior distribution function at $A^{\alpha T}=0$. 
The confidence levels for $A^{\alpha \alpha}$ are reported in Table \ref{tab:table2new}.
\begin{table}[h!]
  \begin{center}
    \caption{Limits of $A^{\alpha \alpha}$ at $95\%$ C.L. as derived through the 2D pixel-based likelihood approach marginalising over $A^{\alpha T}$ (upper part of the Table) or slicing at $A^{\alpha T}=0$ (lower part of the Table). 
    In the upper part we provide the best fit for $A^{\alpha \alpha}$ as obtained from the 2D contour-plot, whereas in the lower part we give the most likely value of $A^{\alpha \alpha}$ when we slice the 2D probability distribution function at $A^{\alpha T}=0$.  
    Units are $[\mbox{deg}^2]$.}
    \label{tab:table2new}
    \begin{tabular}{lcc}
     \hline
    \hline
      Data/ Method  & Best fit of $A^{\alpha \alpha}$ & Limit(s) at $95 \%$ C.L. \\
      \hline
      HM \commander & 0.035 & 0.114  \\
      HM \smica & 0.025 & 0.104 \\
      HM \nilc & 0.023 & 0.096 \\
      HM \sevem & 0.035 & 0.122 \\
       \hline
      OE \commander & 0.053 & [0.008, 0.135]  \\
      OE \smica & 0.045 & [0.008, 0.135]  \\
      \hline
      \hline
       \end{tabular}
    \begin{tabular}{lccc} 
      Data/ Method  & $\phantom{a/b}$ $A^{\alpha \alpha}$ sliced & Limit(s) at $95 \%$ C.L.\\
      \hline
      HM \commander & 0.026 & 0.092 \\
      HM \smica & 0.023 & 0.085 \\
      HM \nilc & 0.019 & 0.078 \\
      HM \sevem & 0.033 & 0.103 \\
       \hline
      OE \commander & 0.040 & 0.131  \\
      OE \smica & 0.043 & [0.003, 0.112] \\
      \hline
       \end{tabular} \\
      \end{center}
\end{table}
If we instead marginalise over $A^{\alpha \alpha}$ we obtain the posterior distribution function of $A^{\alpha T}$, displayed in Figure \ref{fig:AalphaTmarginalised}.
The confidence levels for $A^{\alpha T}$ are reported in Table \ref{tab:table3new}. 
\begin{table}[h!]
  \begin{center}
    \caption{Limits of $A^{\alpha T}$ at $68\%, 95\%$ and $99\%$ C.L. as derived through the 2D pixel-based likelihood approach marginalising over $A^{\alpha \alpha}$. Units are $[\mu \mbox{K} \cdot \mbox{deg}]$.}
    \label{tab:table3new}
    \begin{tabular}{lccccc}
     \hline
    \hline
      Data/ Method  & Best fit $A^{\alpha T}$ & $68 \%$ C.L.& $95 \%$ C.L. & $99 \%$ C.L  \\ 
      \hline
      HM \commander & 2.25 & [-0.21, 3.26] & [-2.32, 4.36]  & [-3.53, 5.08]   \\
      HM \smica & 1.50 & [-0.57, 2.78] & [-2.60, 3.91]  & [-3.74, 4.61]  \\
      HM \nilc & 1.25 & [-0.88, 2.50] & [-2.74, 3.66]  & [-3.76, 4.39]  \\
      HM \sevem & 2.50 &[0.18, 3.55] & [-2.20, 4.55]  & [-3.55, 5.25]  \\
      \hline
      OE \commander & 2.75 & [-0.60, 3.75] & [-3.10, 4.76]  & [-4.20, 5.44]  \\
      OE \smica & 1.00 & [-1.63, 2.86] & [-3.65, 4.21]  & [-4.65, 5.03]  \\
      \hline
      \hline
       \end{tabular}
    \\
  \end{center}
\end{table}

\begin{figure}[t]
\centering
\includegraphics[width=.49\textwidth]{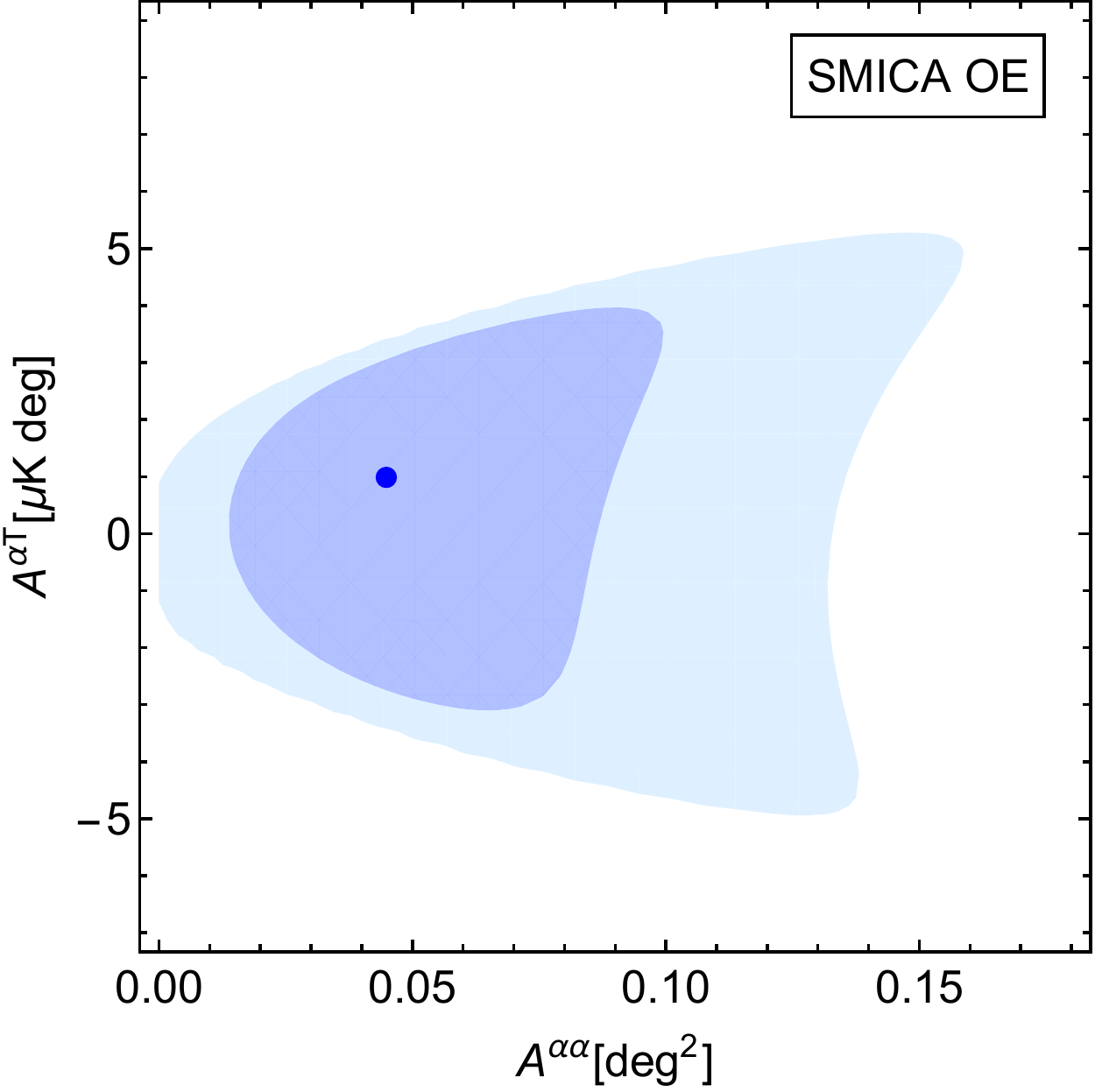}
\includegraphics[width=.49\textwidth]{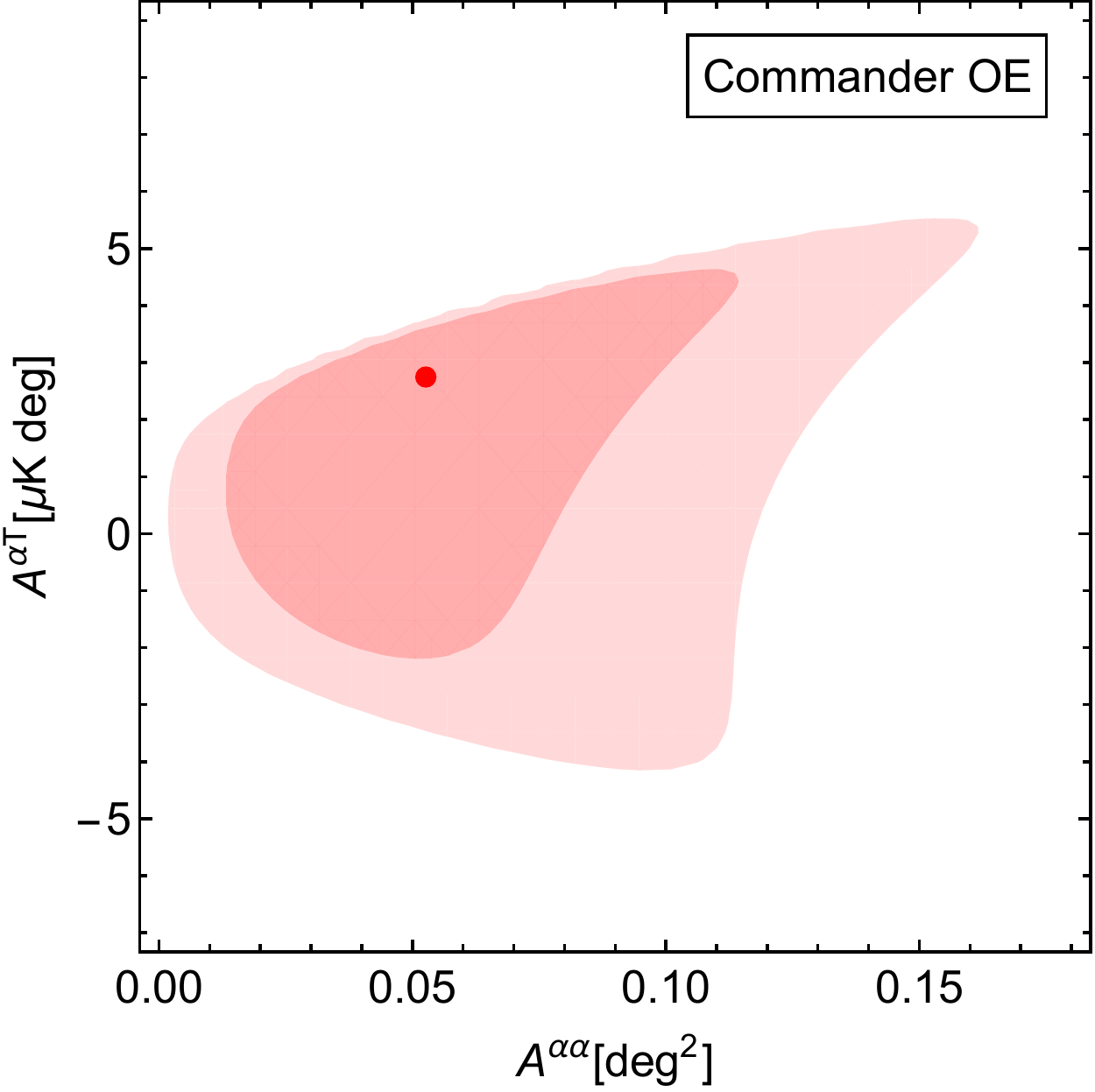}
\caption{Two-dimensional contour plot for $A^{\alpha \alpha}$ and $A^{\alpha T}$, obtained from \smica\ (left), \commander\ (right) OE data in the baseline configuration.
In each panel the darker (lighter) area covers the $68\%$ ($95\%$) confidence region.}
\label{fig:2dlikelihood_OE}
\end{figure}

\begin{figure}[t]
\centering
\includegraphics[width=.49\textwidth]{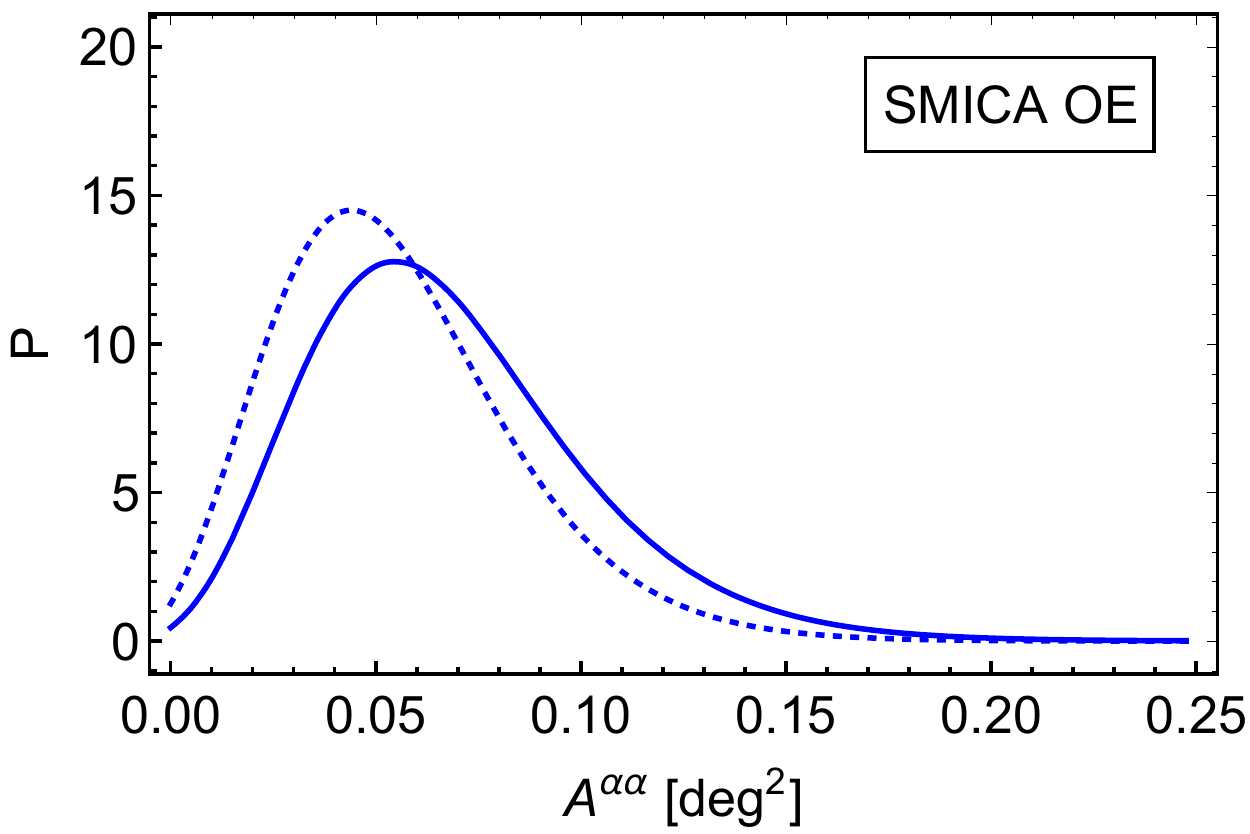}
\includegraphics[width=.49\textwidth]{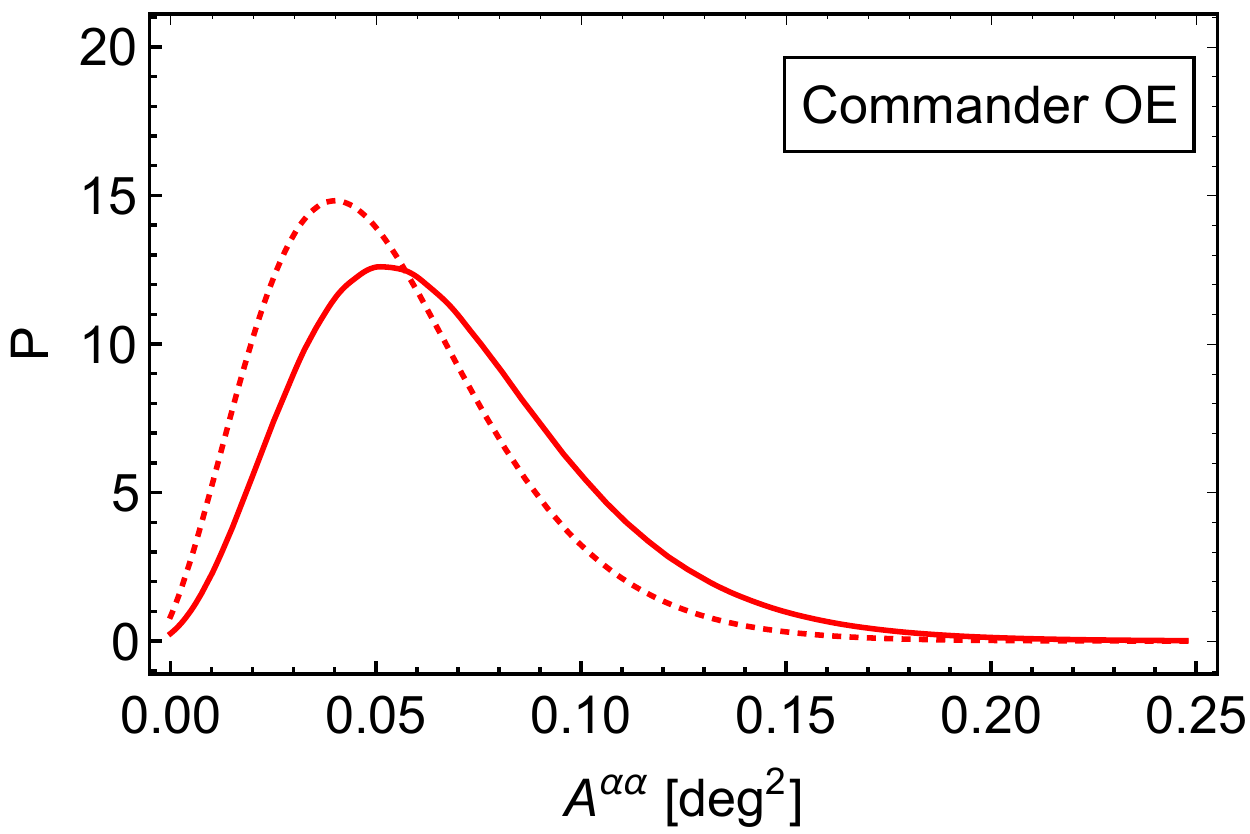}
\caption{Posterior distribution function of $A^{\alpha \alpha}$ obtained from \smica\ (left), \commander\ (right) OE data in the baseline configuration.
In each panel the solid curves are computed marginalising the 2D distribution function over $A^{\alpha T}$ and the dashed ones slicing the 2D distribution function at $A^{\alpha T}=0$. }
\label{fig:Aalphamarginalised_OE}
\end{figure}

\begin{figure}[t]
\centering
\includegraphics[width=.49\textwidth]{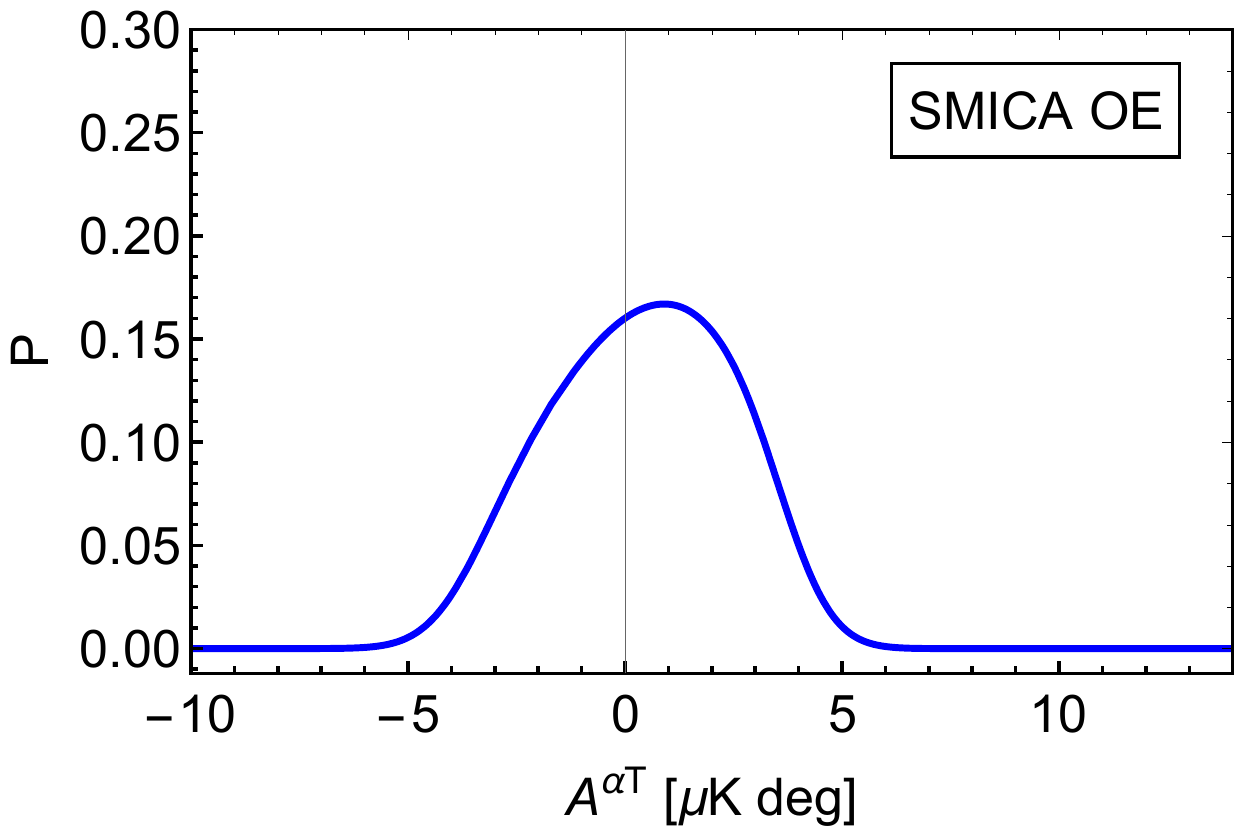}
\includegraphics[width=.49\textwidth]{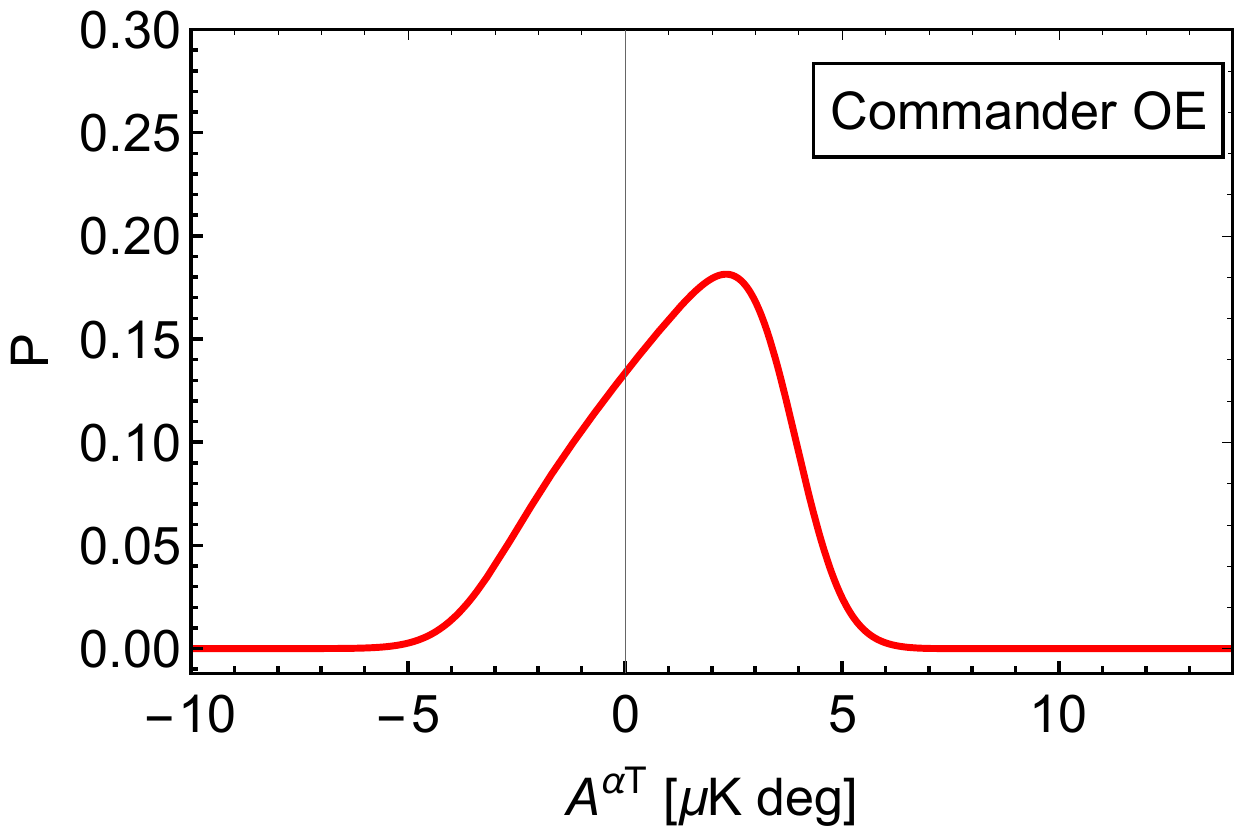}
\caption{Posterior distribution function of $A^{\alpha T}$ obtained from \smica\ (left), \commander\ (right) OE data in the baseline configuration.
In each panel the solid curves are computed marginalising the 2D distribution function over $A^{\alpha \alpha}$. }
\label{fig:AalphaTmarginalised_OE}
\end{figure}

Similarly in Figure \ref{fig:2dlikelihood_OE} we show the 2D contour plots of $A^{\alpha \alpha}$ and $A^{\alpha T}$ for the OE data of \smica\ and  \commander.
Marginalising the 2D posterior distribution function of Figure \ref{fig:2dlikelihood_OE} over $A^{\alpha T}$ we obtain the posterior distribution function of $A^{\alpha \alpha}$ shown as solid curves in Figure \ref{fig:Aalphamarginalised_OE}.
In the same Figure, dashed curves are obtained slicing  the 2D posterior distribution at $A^{\alpha T}=0$. 
If we instead marginalise over $A^{\alpha \alpha}$ we obtain the posterior distribution function of $A^{\alpha T}$ given as solid curves in Figure \ref{fig:AalphaTmarginalised_OE}.
The confidence levels for $A^{\alpha \alpha}$ and $A^{\alpha T}$ corresponding to the OE splits, are reported in Table \ref{tab:table2new} and \ref{tab:table3new}.

We note that $A^{\alpha \alpha}$ is typically compatible with zero at $\sim$1$\sigma$ considering the HM data and at $\sim$2$\sigma$ for the OE data splits. 
For what concerns $A^{\alpha T}$ both the data splits are well compatible with no correlation at $\sim$1$\sigma$ level, however the OE data display slightly larger uncertainties
than HM.
Comparing the best-fits of $A^{\alpha \alpha}$ obtained from HM and OE \commander\ or \smica\ data we note a very good compatibility: the half-distance between the two is at most of the order of $0.01 \, [\mbox{deg}^2]$
representing the 10$\%$ of the statistical bound at $95\%$ C.L.. 
Similarly for $A^{\alpha T}$ we evaluate an half-distance of the order of $0.25 \, [\mu \mbox{K} \cdot \mbox{deg}]$, which is assessed to be at most the 15$\%$ of the statistical uncertainty at $68\%$ C.L..

\begin{figure}[t]
\centering
\includegraphics[width=.49\textwidth]{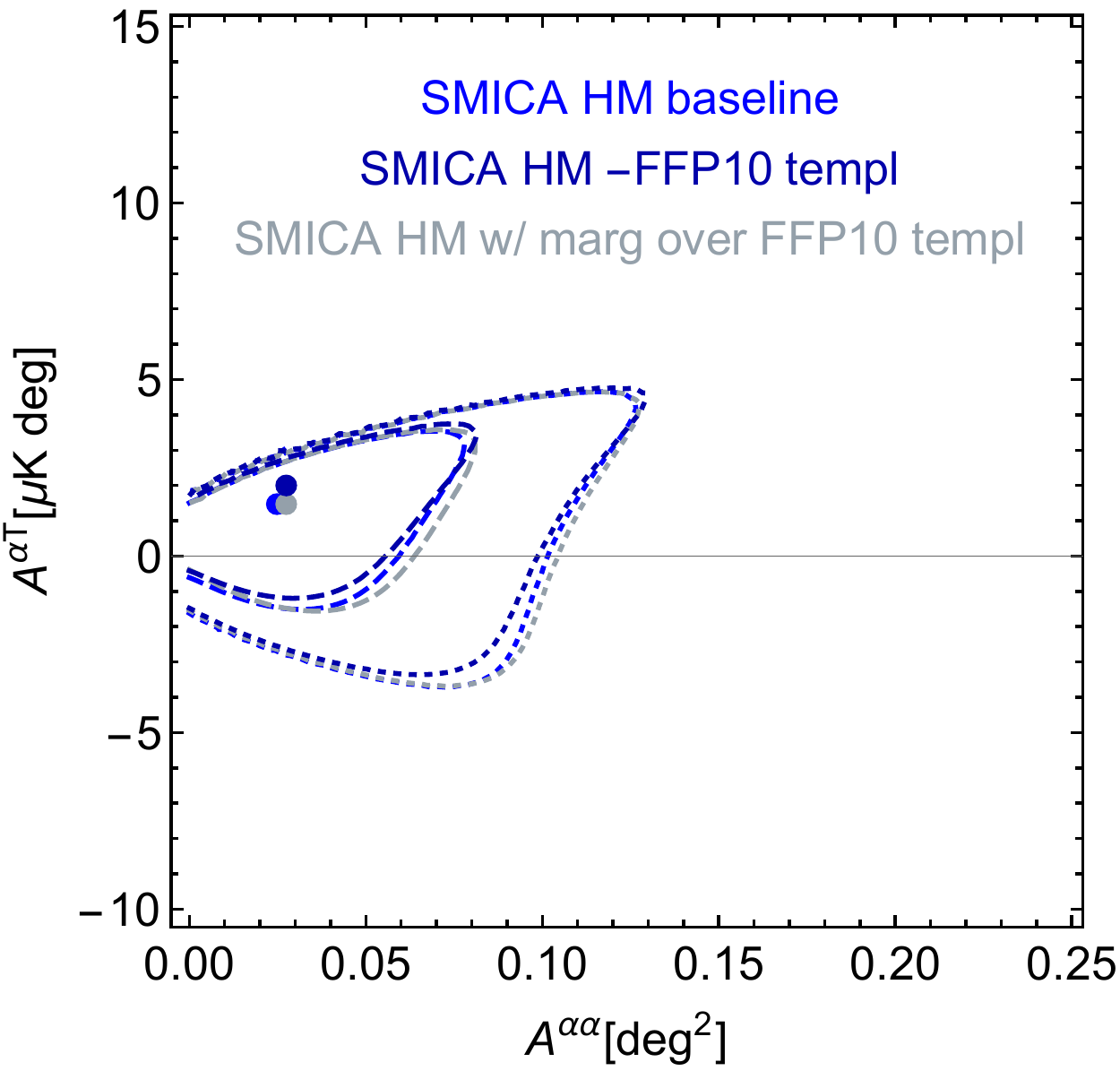}
\caption{Two-dimensional contour plot for $A^{\alpha \alpha}$ and $A^{\alpha T}$, obtained from \smica\ HM data in the baseline configuration (blue contours), 
in the baseline configuration with the removal of the FFP10 template from the map of $\alpha$ (dark blue contours) and in the baseline configuration with the marginalisation over the FFP10 template (gray contours). 
Dashed and dotted contours define the 68$\%$ and 95$\%$ confidence regions.}
\label{fig:2dposteriosSMICAtest}
\end{figure}

We have also verified that the impact of the removal of the FFP10 template on the 2D contour for $A^{\alpha \alpha}$ and $A^{\alpha T}$ is very small. In Figure \ref{fig:2dposteriosSMICAtest} we provide 
the case for \smica, where we compare for the HM case, the baseline configuration (blue contours), the baseline configuration with the removal of the FFP10 template from the map of $\alpha$ (dark blue contours)
and the baseline configuration with the marginalisation over the FFP10 template (gray contours). Dashed and dotted contours define the 68$\%$ and 95$\%$ confidence regions.

\section{Conclusions}
\label{conclusions}

We have constrained the anisotropic birefringence effect and its cross-correlation with the CMB temperature map from {\it Planck} 2018 data building maps of $\alpha=\alpha(\hat n)$ at low resolution ($N_{side}=4$). 
This probes the angular scales larger than $\sim$15 degrees. 

First of all, let us note that the CMB {\it Planck} dataset is the only one available which is sensitive to such large-angle $C_{\ell}^{\alpha \alpha}$, see Figure \ref{fig:spectrabirefringencewithtoterror}. 
While also WMAP has been used to provide similar constraints, results are strongly limited by its lower sensitivity as compared to \planck\, see~\cite{Gluscevic:2012me}.
To improve the robusteness of our results, we have considered all the component separation methods adopted by \planck\ in the HM data split, different sky fractions, various multipole ranges, two pipelines based on TB and the more efficient EB spectra.
Moreover, while our results are based on the HM data split, we have taken into account for consistency also the OE split.

A few considerations are in order:
\begin{enumerate}

\item These $C_{\ell}^{\alpha \alpha}$ at the lowest $\ell$ allow to efficiently constraint $A^{\alpha \alpha}$, i.e. the scale-invariant angular spectrum of $\alpha$-anisotropies, as defined in Section \ref{Pixel_based_likelihood_approach}. 
Our constraint on $A^{\alpha \alpha}$ reads
\be 
A^{\alpha \alpha}_{\mbox{\smica}}  <  0.104  \, \mbox{deg$^2$ at } 95 \% \mbox{C.L.} \, ,
\label{Aalphaconstraintmarg}
\ee 
which is obtained through a pixel based likelihood approach and a marginalisation over $A^{\alpha T}$.
If we set $A^{\alpha T}=0$ we obtain a ``slice'' of the likelihood function and in this case our constraint on $A^{\alpha \alpha}$ is tighter and reads
\be 
A^{\alpha \alpha}_{\mbox{\smica}}  <  0.085 \, \mbox{deg$^2$ at } 95 \% \mbox{C.L.} \, ,
\label{Aalphaconstraint}
\ee 
which is at the same level of precision 
of the Bicep2/Keck Array \cite{Array:2017rlf} and {\it Planck} 2015 data \cite{Contreras:2017sgi} and a factor $2-3$ worse than recent ACTPol \cite{Namikawa:2020ffr}
and SPTpol results \cite{Bianchini:2020osu} that probe however complementary angular scales.
Constraints given in Eq.~(\ref{Aalphaconstraintmarg}) and Eq.~(\ref{Aalphaconstraint}) automatically take into account a cosmic variance contribution, i.e. a variance based on the model itself.
It is interesting to note that this contribution is dominating the error budget. 
In fact, when we drop out this term adopting a $\chi^2$-minimisation (see Appendix \ref{chi2minimisationAalpha}) 
we find 
\begin{equation}
A^{\alpha \alpha}_{\mbox{\smica}}= -0.0006 \pm 0.0044  \, [\mbox{deg}^2] \mbox{ at } 68 \% \mbox{C.L.}
\, ,
\end{equation}
which is roughly a factor of order $10$ better than Eq.~(\ref{Aalphaconstraint}). 
In Figure \ref{fig:uncertaintyAmplitude} we report our constraints on $A^{\alpha \alpha}$ along with all the previous ones, namely WMAP-7 \cite{Gluscevic:2012me}, Polarbear \cite{Ade:2015cao}, Bicep2/Keck Array \cite{Array:2017rlf}, {\it Planck} 2015 \cite{Contreras:2017sgi}, ACTPol \cite{Namikawa:2020ffr} and SPTpol \cite{Bianchini:2020osu}. Note that previous {\it Planck} constrains  \cite{Contreras:2017sgi} were based on 2015 data and also on a different approach.
\begin{figure}[t]
\centering
\includegraphics[width=.75\textwidth]{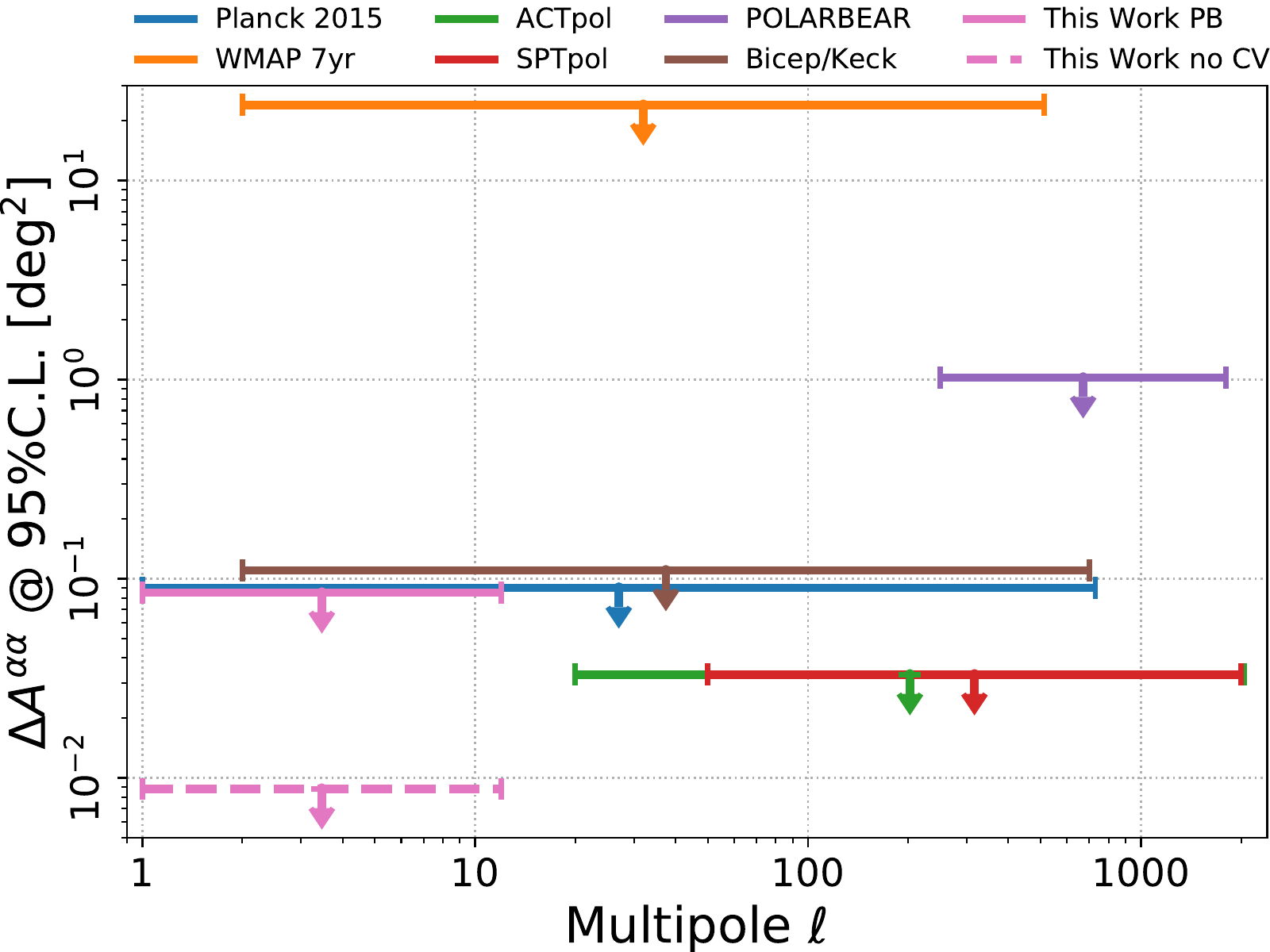}
\caption{Uncertainty at $95\%$ C.L. on the amplitude of the scale invariant spectrum of the anisotropic birefringence angle, $A^{\alpha \alpha}$, from several CMB experiments, namely WMAP \cite{Gluscevic:2012me}, Polarbear \cite{Ade:2015cao}, Bicep2/Keck Array \cite{Array:2017rlf}, {\it Planck} 2015 \cite{Contreras:2017sgi}, ACTPol \cite{Namikawa:2020ffr}, SPTpol \cite{Bianchini:2020osu} and from {\it Planck} 2018 as evaluated in this paper. 
In solid pink we report the error of $A^{\alpha \alpha}$ as obtained through our pixel based likelihood approach slicing at $A^{\alpha T}=0$ and in dashed pink we instead provide the uncertainty found through a $\chi^2$-minimisation. The latter uncertainty is much smaller than the former because of a cosmic variance contribution which is automatically present in the pixel-based likelihood case whereas it is not taken into account through the $\chi^2$-minimisation, see also the text. The WMAP value has been obtained extrapolating at $95\%$ C.L. the bound $A^{\alpha \alpha} < 12$ [deg$^2$] at $68\%$ C.L..}
\label{fig:uncertaintyAmplitude}
\end{figure}

\item We can recast the bound on $A^{\alpha \alpha}$ in terms of primordial stochastic magnetic fields according to \cite{De:2013dra,Pogosian:2013dya}:
\be
B_{\mbox{\tiny 1Mpc}}  = 2.1 \times 10^2 \; {\mbox{nG}} \left({\nu \over 30 \mbox{GHz}}\right)^2 (A^{\alpha \alpha})^{1/2} \, ,
\label{pmf}
\ee
where $\nu$ is the frequency of observation of the CMB expressed in GHz, and where the scale invariant amplitude, $A^{\alpha \alpha}$, has to be given in $[rad]^2$.
To convert our constraints from $A^{\alpha \alpha}$ to $B_{\mbox{\tiny 1Mpc}}$, we suppose that the considered CMB signal is coming only from the most sensitive {\it Planck} channel, i.e. the $143$ GHz one. 
This channel is the one which weights 
more\footnote{We note that a fluctuation of the $\sim 5\%$ around the chosen frequency, impacts on the calculated bound as $\sim 10\%$.} in the component separation methods \cite{Akrami:2018mcd}.
Therefore we translate the bound of Eq.~(\ref{Aalphaconstraintmarg}) as
\be
B_{\mbox{\tiny 1Mpc}} < 26.9 \, \mbox{nG} \, ,
\ee
and the one of Eq.~(\ref{Aalphaconstraint}) as
\be
B_{\mbox{\tiny 1Mpc}} < 24.3 \, \mbox{nG} \, ,
\ee
both at 95$\%$ C.L..
The obtained bound is worse than the one published by the SPT collaboration, i.e. $B_{\mbox{\tiny 1Mpc}} < 17 \, \mbox{nG}$ at 95$\%$ C.L. \cite{Bianchini:2020osu}, but better than what found by {\sc Polarbear} which reads $B_{\mbox{\tiny 1Mpc}} < 93 \, \mbox{nG}$ at 95$\%$ C.L. \cite{Ade:2015cao}. Both constraints are based on the CMB anisotropies four-point correlation function. Previous \planck\ constraints on primordial magnetic fields through Faraday rotation are much weaker because based only on the BB two-point correlation functions of the 70~GHz channel \cite{Ade:2015cva}.
Forecasts on the capability of future experiments to constrain primordial magnetic fields through the four-point correlation function can be found in \cite{Pogosian:2018vfr}.

\item We have evaluated, for the first time to our knowledge, the cross-correlation between anisotropic birefringence and CMB temperature at these large angular scales. 
We have evaluated the $C_{\ell}^{\alpha T}$ from $\ell=2$ to $\ell=12$, see Figure \ref{fig:CelltTalpha}. 
We find very good compatibility with null effect. Moreover, our results are proven to be stable with respect to the method of component separation adopted among the four provided by \planck, the sky fraction, the multipole range, 
the considered data splitting in the pipeline, either HM or OE.

Considering a constant $A^{\alpha T}$ fitting $\ell (\ell +1) C_{\ell}^{\alpha T}/ 2 \pi$ in the harmonic range $[2-12]$ we obtain 
\be
A_{\smica}^{\alpha T}=1.50^{+2.78}_{-2.07} \, \mbox{[$\mu$K$\cdot$deg] at } 68 \% \mbox{C.L.}
\, .
\ee
This is found adopting a pixel based likelihood approach and marginalising over $A^{\alpha \alpha}$. 
As for the corresponding $A^{\alpha \alpha}$ analysis, here the uncertainty already include a cosmic variance term.
If we remove the latter, see Appendix \ref{alphaT}, we find:
\be
A_{\smica}^{\alpha T} = 0.90 \pm 1.09 \, \mbox{[deg$\cdot \mu$K] at } 68\%  \mbox{C.L.} \, ,
\ee
which has an uncertainty roughly $50\%$ smaller than the total error.
It is interesting to note that the constraint of $A^{\alpha T}$ seems to be more significant with respect to $A^{\alpha \alpha}$ 
for some early dark energy models recently proposed to alleviate the $H_{0}$ tension \cite{Capparelli:2019rtn}.
In particular, from Figure 1 of  \cite{Capparelli:2019rtn}, it is possibile to extrapolate that the models, the authors of  \cite{Capparelli:2019rtn} consider, provide $A^{\alpha T}_{(model)} \sim$ $-$4.5 and +5 [deg$\cdot \mu$K]
at the lowest $\ell$. Comparing them to the distribution probability of $A^{\alpha T}$ estimated here, we find that these values are at $\sim$3$\sigma$ for \smica\ HM and $\sim$2$.5\sigma$ for \smica\ OE.
The same proposed models are instead well compatible, below $\sim$1$\sigma$, with our and previous constraints of $A^{\alpha \alpha}$ since they predict a value $\sim$0.03 [deg$^2$].

\end{enumerate}

\begin{figure}[t]
\centering
\includegraphics[width=.70\textwidth]{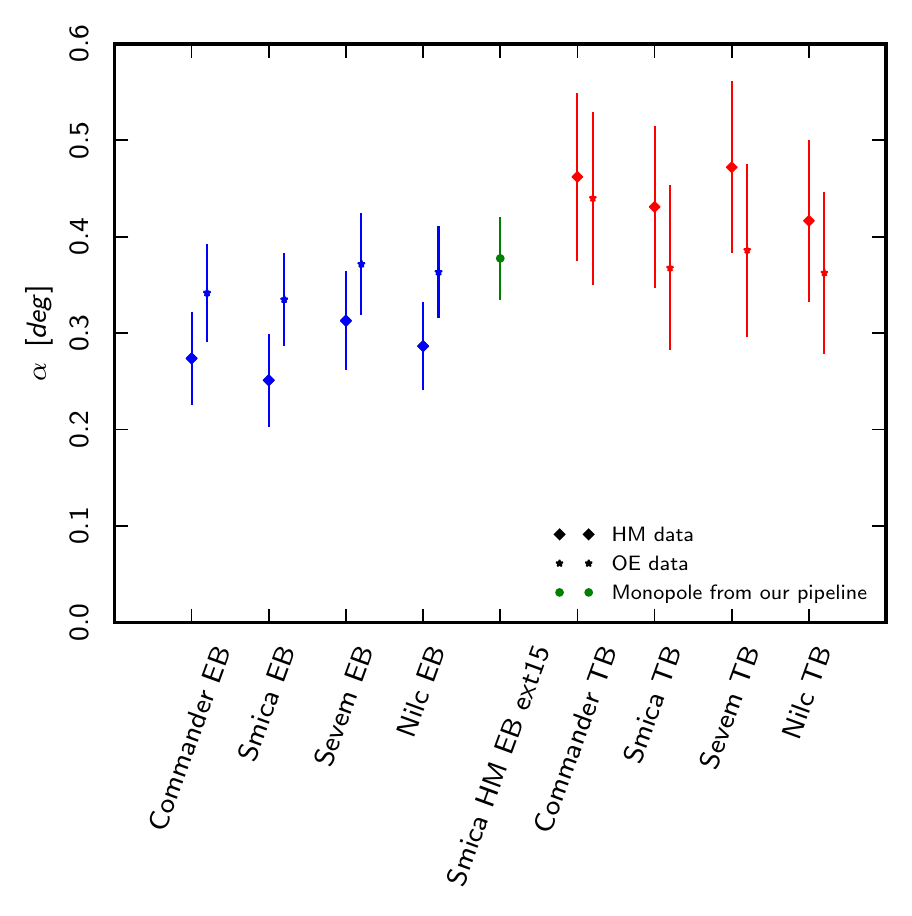}
\caption{Isotropic birefringence angle from {\it Planck} 2018 data, for all the component separation methods and for the $D_{\ell}^{EB}$ (blue) and $D_{\ell}^{TB}$-based pipeline (red).
Only statistical uncertainties shown. Diamonds (stars) refer to HM (OE) splits. In green we show the amplitude of the monopole estimated from our EB pipeline. }
\label{fig:isotropicbire}
\end{figure}

We end this section commenting on the isotropic birefringence effect, i.e. the monopole term of the map of $\alpha(\hat n)$, which has been also constrained and found to be fully compatible with previous {\it Planck} estimates \cite{Aghanim:2016fhp}. 
See Figure \ref{fig:isotropicbire} where the isotropic birefringence angle $\alpha$ is estimated for all the CMB component separation solutions and for both EB- and TB-pipeline. 
For isotropic birefringence, we find that the compatibility with the null hypothesis is reached only when accounting for the systematic uncertainty in the instrumental polarisation angle \cite{Aghanim:2016fhp}. Further details can be found in Appendix \ref{isotropicbire}. See also \cite{Gruppuso:2016fwu,Gruppuso:2015xza} for other isotropic birefringence angle estimates based on {\it Planck} 2015 data.

As a final remark, we note that, in principle, the method considered here could be used to estimate the APS of the birefringence anisotropies also at higher multipoles, i.e. beyond the maximum multipole $\ell=12$ assumed in this work. However, the related computational cost would be considerable: assuming the immediately higher \texttt{HEALPix} resolution $N_{side}=8$, required to reach $\ell=24$, one would need to apply our procedure $f_{sky}\cdot$768 times, compared to $f_{sky}\cdot$192 as done here (with $f_{sky}$ being the observed sky fraction). Moreover, one would need to handle $f_{sky}\cdot$768 different Monte Carlo sets, i.e. one for each observed low resolution pixel. Hence, it is clear that this method is not so suitable to quickly reach higher multipoles, due to the a nasty $N_{side}^2$ scaling. The easiest way to reach higher multipoles would be to employ a different estimator, such as the one applied in \cite{Gluscevic:2012me}, which exploits the birefringence induced correlations among CMB modes at different multipoles. The latter approach, typically called MCE (Mode-Coupling estimator), is explained in detail in \cite{Gluscevic:2009mm}.
To assess what is the most efficient technique from a statistical point of view in order to analyse \planck\ data is not trivial, and a fair comparison would be possible only after a full implementation of the MCE approach. However we note that in \cite{Gluscevic:2009mm} a nice forecast of the MCE performances on \planck\ data is given. By comparing such a forecast at a reference multipole, say $\ell=10$, with the brute-force method employed here, we provide an interesting first attempt of such an evaluation: in the present paper we find an uncertainty of $\sim$$0.008$ deg$^2$ while in \cite{Gluscevic:2009mm} the authors provide $\sim$$0.02$ deg$^2$ \footnote{In fact the forecast reads $\sim 0.04$ deg$^2$ but was derived assuming an observation time roughly half of what Planck has obtained in practice.}. It is nice to show that the brute force approach yields an error in-line with what is expected by the MCE method, despite real-world complications of the actual analysis, including a proper noise assessment which takes component separation in account. A proper comparison between the MCE method and our approach is deferred to future work.

\acknowledgments
This work is based on observations obtained with Planck (http://www.esa.int/Planck), an ESA science mission with instruments and contributions directly funded by ESA Member States, NASA, and Canada.
We acknowledge the use of computing facilities at NERSC and those provided by the INFN theory group (I.S.\ InDark) at CINECA. 
Some of the results in this paper have been derived using the \texttt{HEALPix} \cite{Gorski:2004by} package.
We acknowledge the financial support from the INFN InDark project and from the COSMOS network (www.cosmosnet.it) through the ASI (Italian Space Agency) Grants 2016-24-H.0 and 2016-24-H.1-2018.
We thank Daniela Paoletti for useful comments.

\appendix

\section{$\chi^2$-minimisation for $A^{\alpha \alpha}$}
\label{chi2minimisationAalpha}

\begin{figure}[t]
\centering
\includegraphics[width=.49 \textwidth]{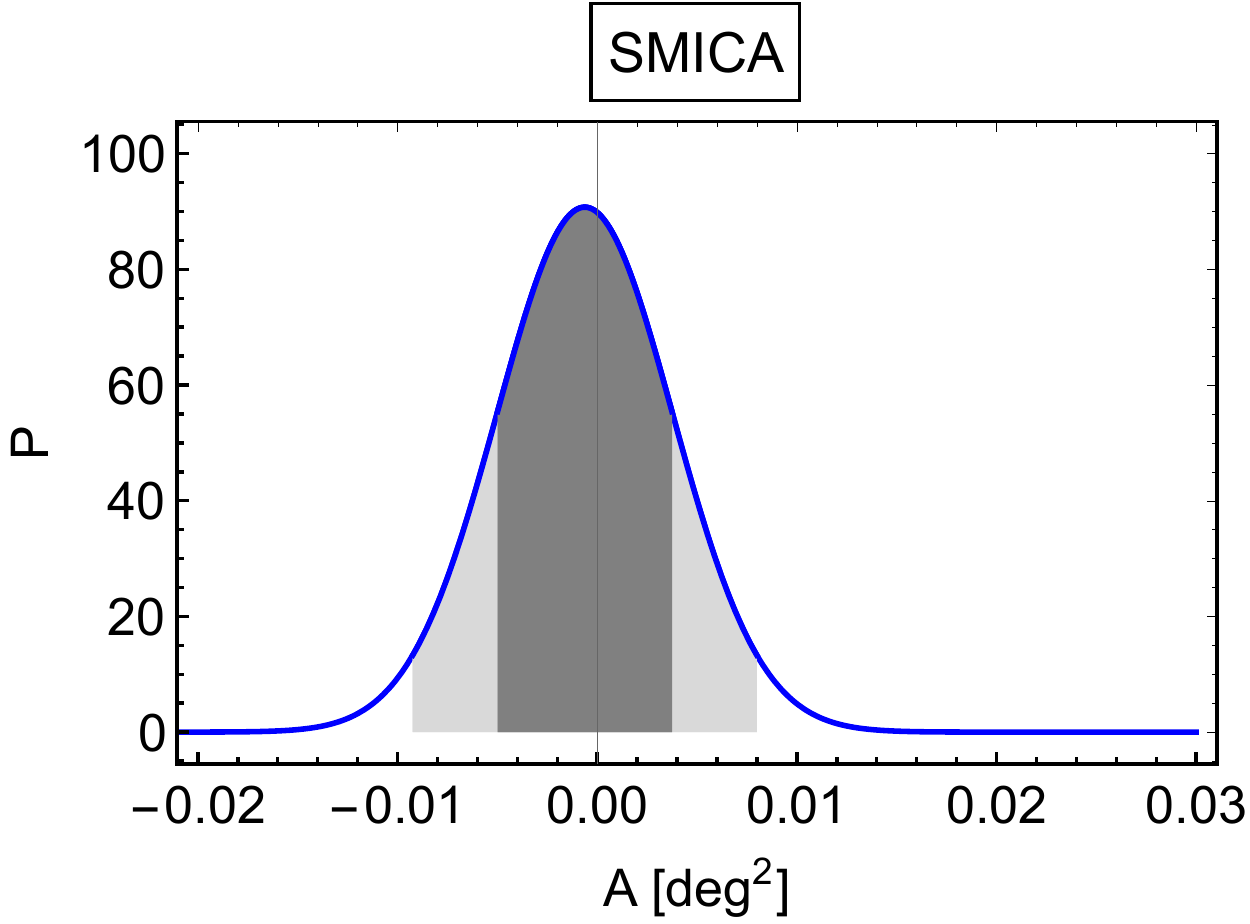}
\includegraphics[width=.49 \textwidth]{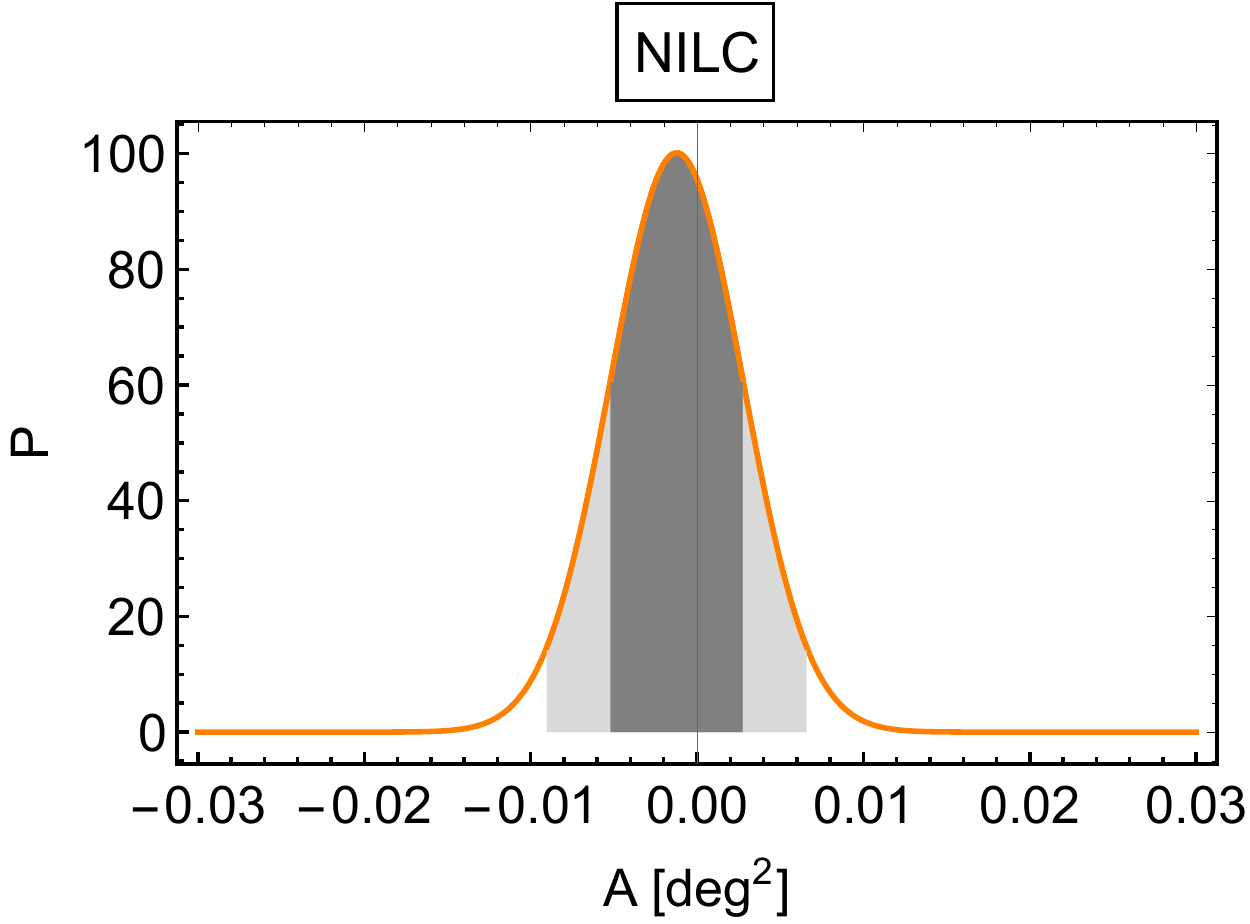}
\includegraphics[width=.49 \textwidth]{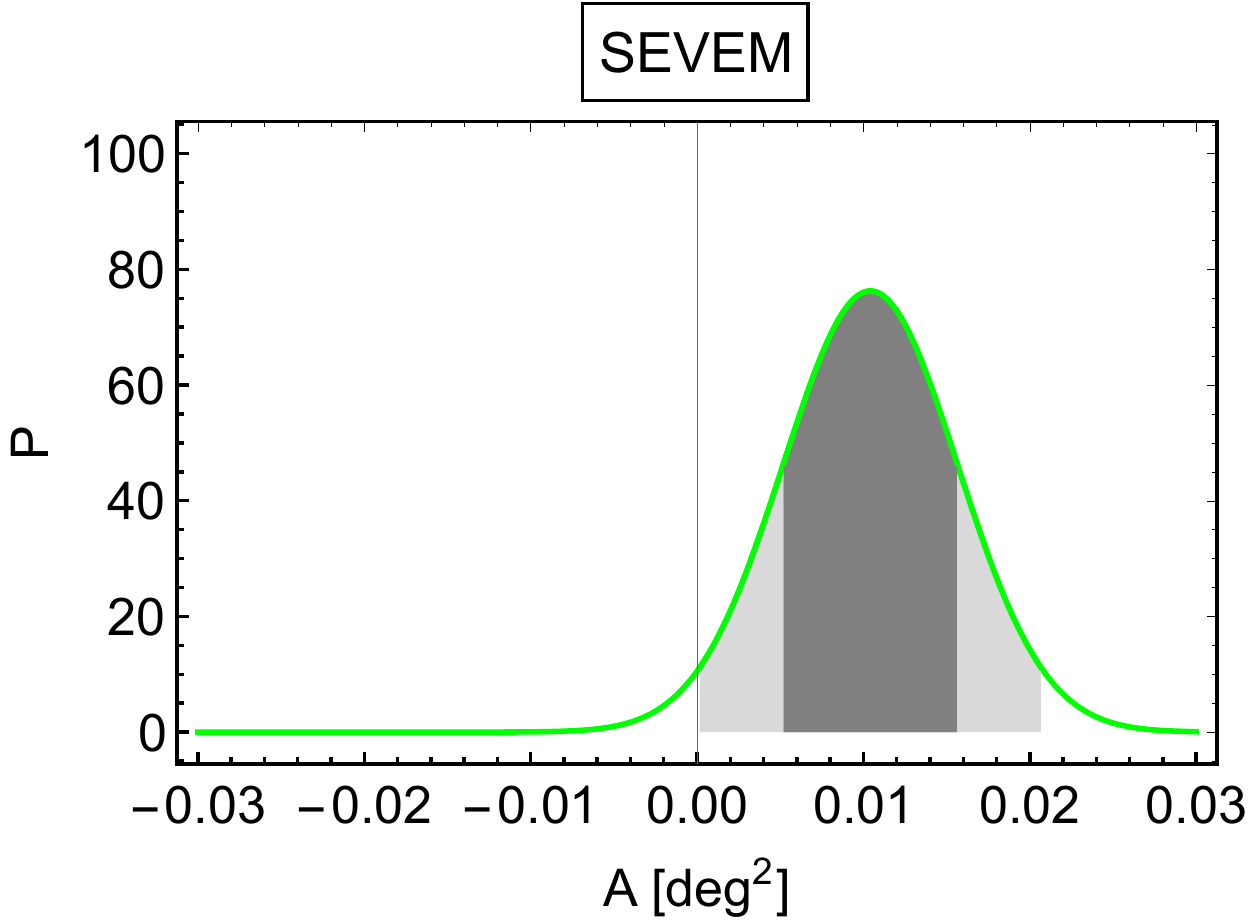}
\includegraphics[width=.49 \textwidth]{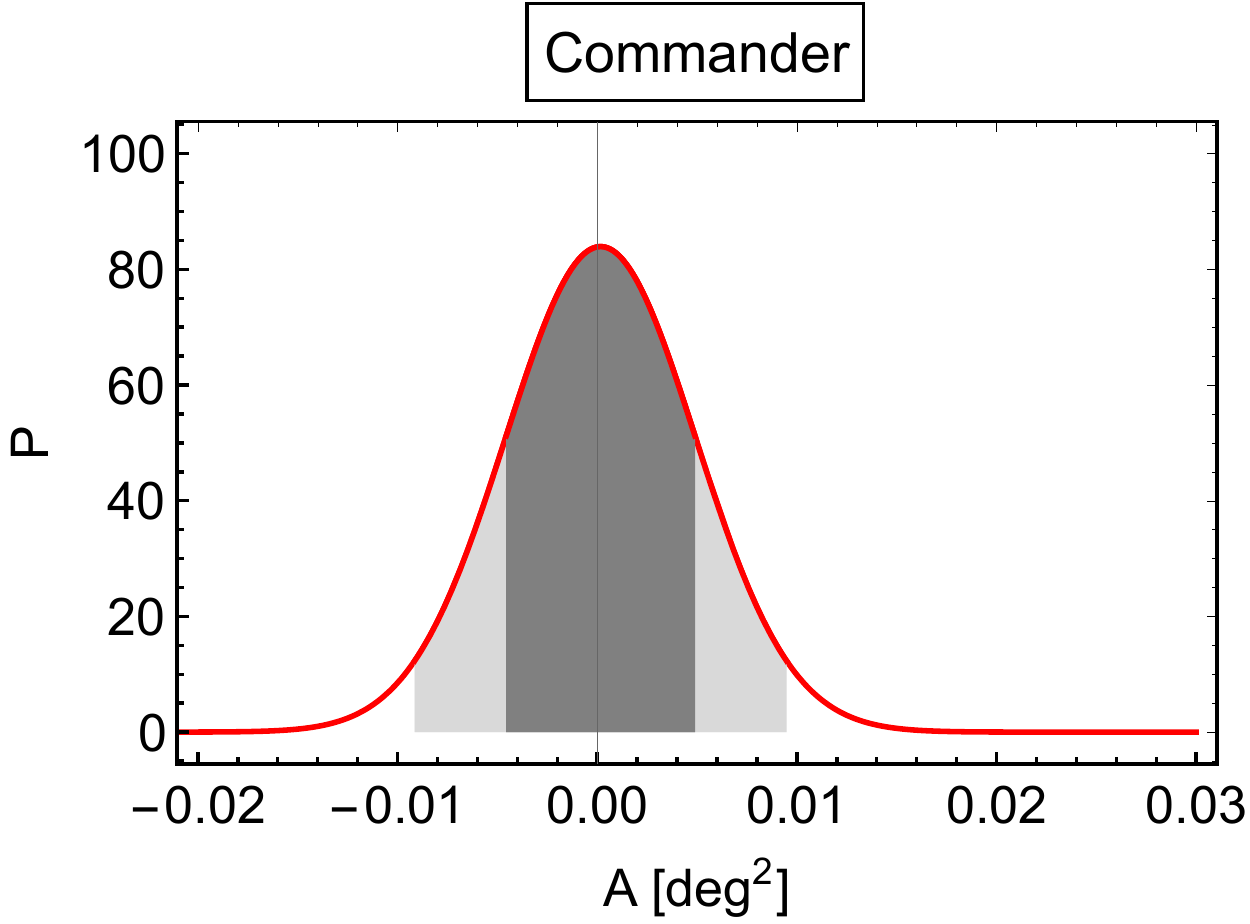}
\caption{Amplitude $A^{\alpha \alpha}$ of the scale-invariant angular spectrum of $\alpha$-anisotropies expressed in $[\mbox{deg}^2]$. 
Each panel shows the probability distribution function derived from Eq.~(\ref{chiquadroarmonico2}), see text for details.  
All four component separation methods are shown, for the HM data split in the baseline configuration. The dark (light) shaded portions represent the $68\%$ ($95\%$) C.L.\ intervals.
}
\label{fig:Aalpha}
\end{figure}

We estimate here $A^{\alpha \alpha}$ through a $\chi^2$-minimisation, as for example done in \cite{Array:2017rlf}.
The best fit value for $A^{\alpha \alpha}$ is obtained by minimising the following $\chi^2 (A^{\alpha \alpha})$:
\begin{equation}
\chi^2(A^{\alpha \alpha}) = \sum_{\ell,\ell^{\prime}=1}^{12} (\mathcal{D}_\ell^\alpha - A^{\alpha \alpha} ) \langle \mathcal{D}_{\ell}^{\alpha} \mathcal{D}_{\ell^{\prime}}^{\alpha} \rangle^{-1} (\mathcal{D}_{\ell^{\prime}}^\alpha - A^{\alpha \alpha} ) \, ,
\label{chiquadroarmonico2}
\end{equation}
where the covariance $ \langle \mathcal{D}_{\ell}^{\alpha} \mathcal{D}_{\ell^{\prime}}^{\alpha} \rangle$ has been evaluated using the FFP10 simulations. In Figure \ref{fig:Aalpha} we show the area-normalised probability distribution function $\propto \exp{\left [-\chi^2(A^{\alpha \alpha})/2 \right ]}$  in the baseline configuration for all four component separation methods. The dark (light) shaded portions represent the $68\%$ ($95\%$) C.L.\ intervals. As it can be seen, good compatibility with the null hypothesis is achieved for all methods.
\begin{table}[h!]
  \begin{center}
    \caption{Best fit estimates of $A^{\alpha \alpha}$ with uncertainties as derived by minimising Eq.~\ref{chiquadroarmonico2}. Upper part for the HM and lower part for the OE data split. Units are $[\mbox{deg}]^2$.}
    \label{tab:table1}
    \begin{tabular}{lcc}
     \hline
      Method & Best fit $A^{\alpha \alpha}$ & Stat.\ uncert.$^{\dagger}$ \\
      \hline
      \commander\ HM & 0.0002 & 0.0048 \\
      \smica\ HM & -0.0006 & 0.0044 \\
      \nilc\ HM & -0.0012 & 0.0040 \\
      \sevem\ HM &  0.0104 & 0.0052 \\
      \hline
      \commander\ OE & -0.0004 & 0.0051 \\
      \smica\ OE & -0.0042 & 0.0046 \\
       \hline
       \end{tabular} \\
          \footnotesize
     $\! \! \! \! \! \! \! \! \! \! \! \! \! \! \! \! \! \! \! \! \! \! \! \! \! \! \! \! \! \! \! \! \! \! \! \! \! \! \! \! \! \! \! \! \! \! \! \! \! \! \! \! \! \! \! \! \! \! \! \! \! \! \! \! \! \! \! \! \! \! \! \! \! \! \! \! \! \! 
     \! \! \! \! \! \! \! \! \! \! \! \! \! \! \! \! \! \! \! \! \! \! \! \! \! \! \! \! \! \! \! \! \! \! \! \! \! \! \! \! \! \! \! \! \! \! \! \! \! \! \! \! \! \! \! \! \! \! \! \! \! \! ^{\dagger}$ $68\%$ C.L.\\
  \end{center}
\end{table}
In Table \ref{tab:table1} we report the best fit values of $A^{\alpha \alpha}$, obtained by minimising Eq.~(\ref{chiquadroarmonico2}), along with the statistical uncertainty at $68\%$ C.L.. 
All methods provide very-well consistent estimates, with the possible mild exception of \sevem\ which is in any case within $\sim$2$\sigma$ from the others.

It is interesting to note that the uncertainties presented here are markedly more stringent, by a factor $\sim$$10$, with respect to what given in Section \ref{Pixel_based_likelihood_approach}.
The difference between the two approaches is given by the presence of a cosmic variance term in the pixel-based likelihood approach which turns out to dominate the contribution at the largest scales.

\section{$\chi^2$-minimisation for $A^{\alpha T}$}
\label{alphaT}

\begin{figure}[t]
\centering
\includegraphics[width=.43 \textwidth]{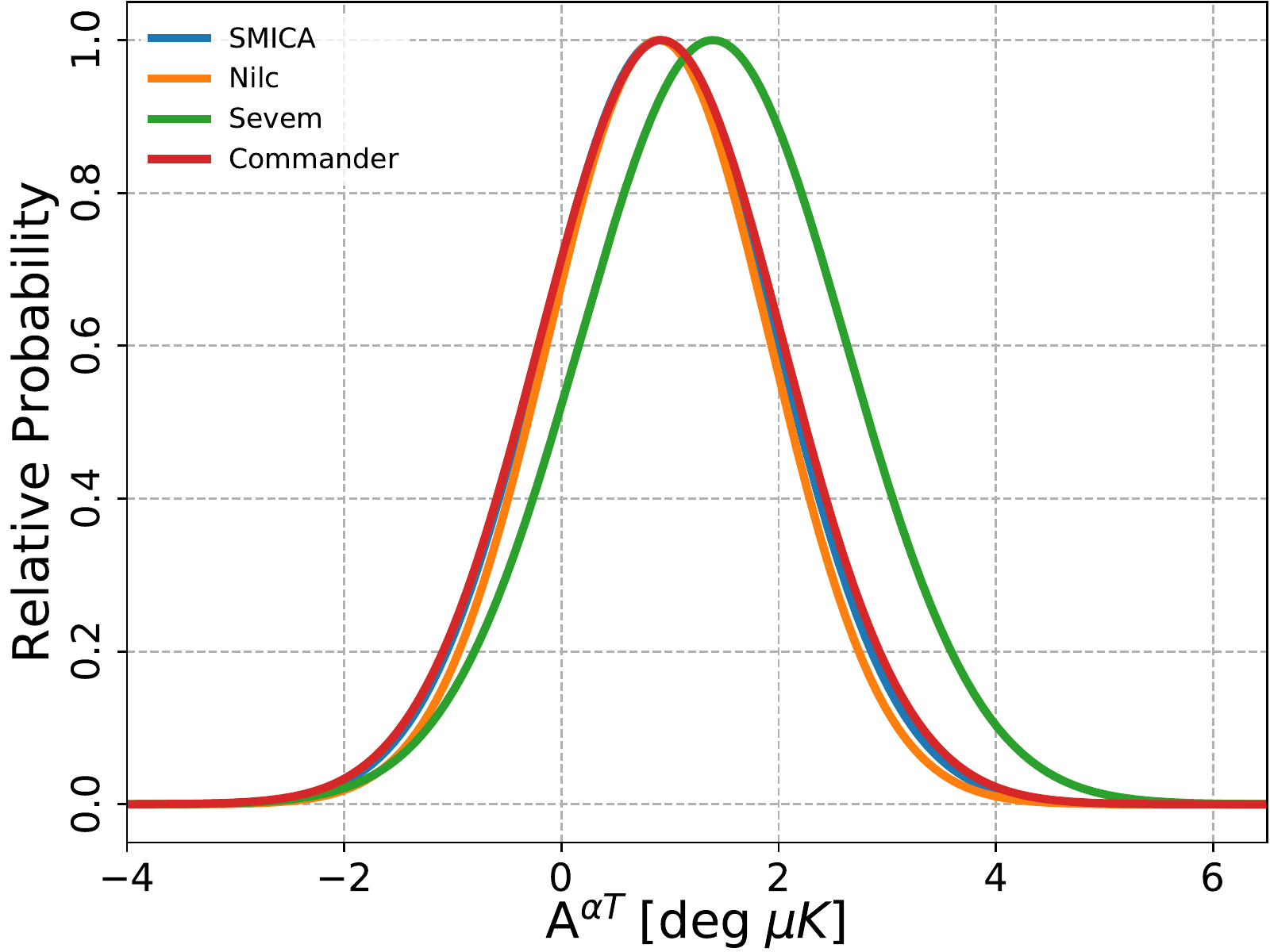}
\includegraphics[width=.43 \textwidth]{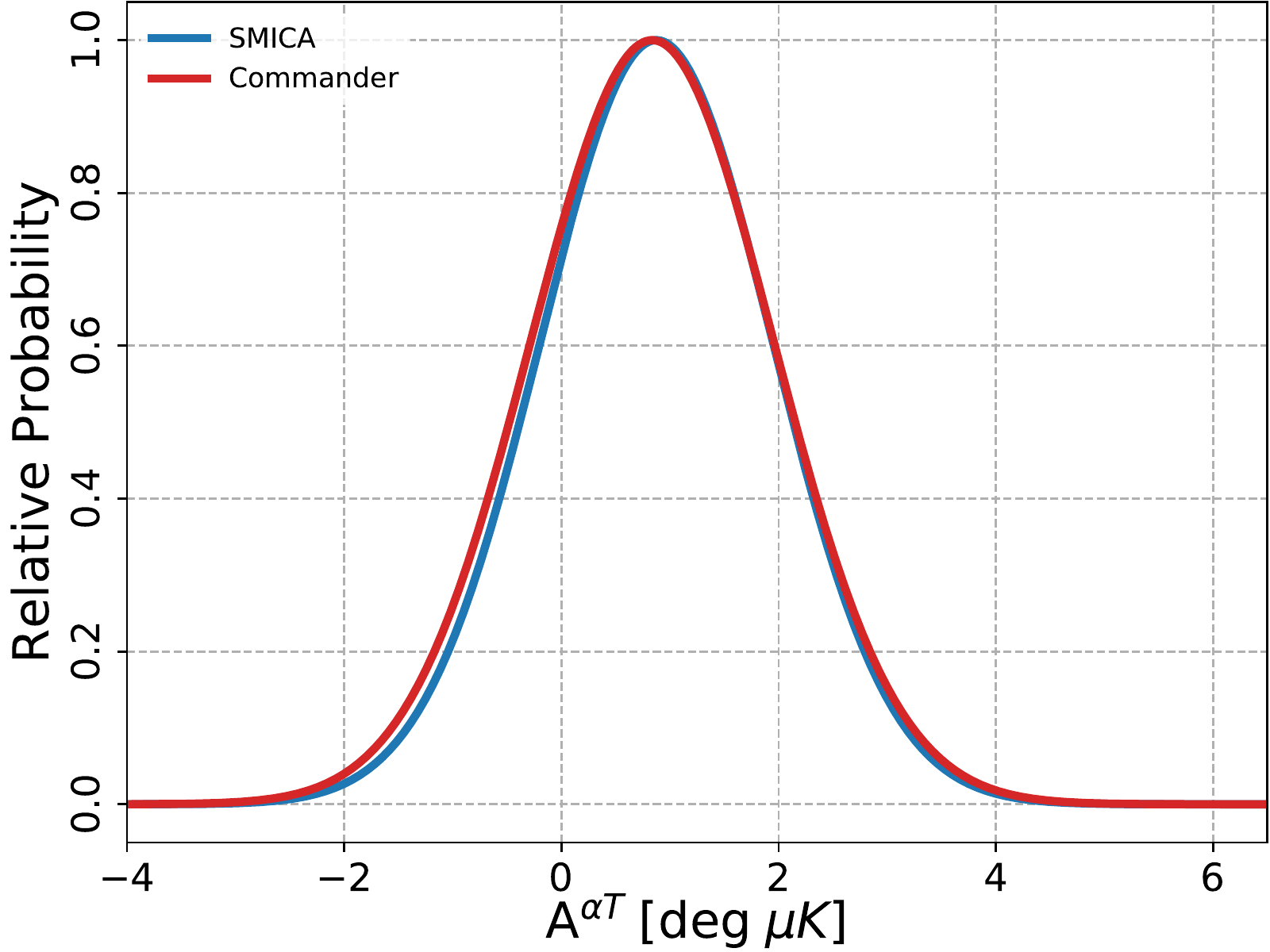}
\caption{Posterior distribution function of $A^{\alpha T}$ for all the component separation methods. Left panel HM, right panel OE. See also Section \ref{alphaT}.
}
\label{fig:A_Talpha}
\end{figure}

We provide the compatibility with null effect also fitting the spectrum of the cross-correla\- -tion $\ell (\ell +1) C_{\ell}^{\alpha T}/ 2 \pi$ in the harmonic range $[2-12]$ with a constant amplitude, $A^{\alpha T}$.
As for Section \ref{chi2minimisationAalpha} this is done here performing a minimisation of a $\chi^2$.
The posterior distribution functions for $A^{\alpha T}$ are shown in Figure \ref{fig:A_Talpha} where the left panel is for the HM pipeline and the right panel for the OE pipeline.
All the estimated $A^{\alpha T}$ are compatible with zero and in particular we find
\be
A^{\alpha T} &=& 0.899 \pm 1.089 \, \mbox{[deg$\cdot \mu$K] for \smica}  \, , \\
A^{\alpha T} &=& 0.897 \pm 1.026 \, \mbox{[deg$\cdot \mu$K] for \nilc} \, , \\
A^{\alpha T} &=& 1.394 \pm 1.223 \, \mbox{[deg$\cdot \mu$K] for \sevem} \, , \\
A^{\alpha T} &=& 0.918 \pm 1.119 \, \mbox{[deg$\cdot \mu$K] for \commander} \, ,
\ee
for the HM pipeline and
\be
A^{\alpha T} &=& 0.874 \pm 1.069 \, \mbox{[deg$\cdot \mu$K] for \smica} \, \\
A^{\alpha T} &=& 0.835 \pm 1.117 \, \mbox{[deg$\cdot \mu$K] for \commander} \, .
\ee
for the OE pipeline, where the uncertainties are given at $68 \%$ C.L..
The uncertainty here is smaller than what given in Table \ref{fig:AalphaTmarginalised} because of a contribution of the cosmic variance term for the signal part of $\alpha$ which is not taken into account in the minimisation process while is automatically accounted for in the pixel-based likelihood approach.

\section{Monopole vs.\ purely isotropic birefringence. Results and consistency tests}
\label{isotropicbire}

We can also provide direct constraints to  isotropic birefringence by minimising Eq.~(\ref{chisq}) over the whole available sky (i.e. considering the masks shown in Figure \ref{fig:common_masks}). 
This analysis closely mimics the harmonic pipeline employed by the \planck\ collaboration for the 2015 release \cite{Aghanim:2016fhp} but using
FFP10 simulations and \planck\ 2018 data for all components separation methods. 
As also done in \cite{Aghanim:2016fhp} we have considered the harmonic range $[51-1511]$. Results are shown in Figure \ref{fig:isotropicbire} where only statistical uncertainties are quoted.
It is interesting to note the very good compatibility among all the component separation solutions for both the HM and OE data splits and for the $D_{\ell}^{EB}$- and $D_{\ell}^{TB}$-based pipelines.
Note also that the isotropic birefringence angle found on the ``full sky'' analysis, is well consistent 
with the estimate derived from the maps of $\alpha(\hat n)$ which are used for the anisotropic birefringence analysis: in Figure  \ref{fig:isotropicbire} we show in green the estimate derived from the monopole which has been obtained from the \smica\ HM data split in the baseline configuration.

In addition our findings for the isotropic birefringence angle are well consistent with the {\it Planck 2015} estimates, 
whose compatibility with the null hypothesis can be obtained only accounting for a systematic uncertainty in the orientation of the {\it Planck}'s polarisation-sensitive bolometers \cite{Aghanim:2016fhp}.

\end{document}